\documentclass[useAMS,usenatbib]{mn2e}
\usepackage{times}
\usepackage{natbib}
\usepackage[dvips]{graphicx}
\usepackage{subfigure}
\usepackage{multirow}
\title[An MHD Gadget for cosmological simulations]
{An MHD Gadget for cosmological simulations}
\author[K. Dolag, F. Stasyszyn]
{K.~Dolag$^{1}$\thanks{E-mail: kdolag@mpa-garching.mpg.de},
F. Stasyszyn$^{1}$,$^{2}$\\
$^{1}$ Max-Planck-Institut f\"ur Astrophysik, Garching, Germany\\
$^{2}$ Instituto de Astronom\'{\i}a Te\'{o}rica y Experimental IATE (UNC-CONICET), 
Observatorio Astron\'{o}mico C\'{o}rdoba, Francisco N. Laprida 922, C\'{o}rdoba, Argentina}

\begin{document}

\date{Accepted ???. Received ???; in original form ???}

\pagerange{\pageref{firstpage}--\pageref{lastpage}} \pubyear{0000}

\maketitle

\label{firstpage}

\begin{abstract}
Various radio observations have shown that the hot atmospheres of
galaxy clusters are magnetized. However, our understanding of the
origin of these magnetic fields, their implications on structure
formation and their interplay with the dynamics of the cluster
atmosphere, especially in the centers of galaxy clusters, is still
very limited. In preparation for the upcoming new generation of
radio telescopes (like EVLA, LWA, LOFAR and SKA), a huge effort is
being made to learn more about cosmological magnetic fields from
the observational perspective. Here we present the implementation
of magneto-hydrodynamics in the cosmological SPH code GADGET
\citep{springel2001,springel2005}. We discuss the details of the
implementation and various schemes to suppress numerical
instabilities as well as regularization schemes, in the context of
cosmological simulations. The performance of the SPH-MHD code is
demonstrated in various one and two dimensional test problems,
which we performed with a fully, three dimensional setup to test the
code under realistic circumstances. Comparing solutions
obtained using ATHENA \citep{2008arXiv0804.0402S}, we find
excellent agreement with our SPH-MHD implementation. Finally we
apply our SPH-MHD implementation to galaxy cluster formation within
a large, cosmological box. Performing a resolution study we
demonstrate the robustness of the predicted shape of the magnetic
field profiles in galaxy clusters,  which is in good agreement
with previous studies.
\end{abstract}



\section{Introduction} \label{sec:intro}

Magnetic fields have been detected in galaxy clusters by radio
observations, via the Faraday Rotation Signal of the magnetized
cluster atmosphere towards polarized radio sources in or behind
clusters \citep[see][for recent reviews]{2002ARA&A..40..319C,2004IJMPD..13.1549G}
and from diffuse synchrotron emission of the cluster atmosphere  
\citep[see][ for recent reviews]{2004IJMPD..13.1549G,2008SSRv..134...93F}.
Our understanding of the origin of cosmological magnetic
fields is particularly limited. But their evolution and possible
implications for structure formation are also not yet fully understood. 
In addition their interplay with the large-scale structure formation
processes, as well as their link to additional dynamics within the
cluster atmosphere is unclear, especially their role in the cool
core regions and the influence of these regions on the evolution of
the magnetic fields.

The upcoming, new generation of radio 
telescopes (like EVLA, LWA, LOFAR and SKA) will dramatically increase 
the volume of observational data relevant for our understanding
of cosmological seed magnetic fields in the near future. 
To investigate the general 
characteristics of the magnetic fields in and beyond galaxy clusters 
at the level required for a meaningful comparison to current and 
forthcoming observations, numerical simulations are mandatory.
Non-radiative simulations of galaxy clusters within a cosmological
environment which follow the evolution of a primordial magnetic
seed field have been performed using Smooth-Particle-Hydrodynamics
(SPH) codes
\citep{1999A&A...348..351D,2002A&A...387..383D,dolag2005b}  as
well as Adaptive Mesh Refinement (AMR) codes
\citep{2005ApJ...631L..21B,2008A&A...482L..13D}. Although these
simulations are based on quite different numerical techniques they
show good agreement in the predicted properties of the
magnetic fields in galaxy clusters. When radiative cooling is included,
strong amplification of the magnetic fields inside the cool-core
region of clusters is found \citep{2008A&A...482L..13D}, in good
agreement with previous work \citep{2000cucg.confE..75D}.
Cosmological, magneto-hydrodynamical simulations were also
performed using finite-volume and finite-difference methods. Such
simulations are used to either follow a primordial magnetic field
\citep{2008ApJS..174....1L} or the creation of magnetic fields in
shocks through the so-called Biermann battery effect
\citep{1997ApJ...480..481K,Ryu..1998}, on which a subsequent
turbulent dynamo may operate. The latter predict magnetic field
strength in filaments with somewhat higher values \citep[e.g. see
][]{2004PhRvD..70d3007S} than predicted by simulations which
start from a primordial magnetic seed field, but are in
line with predictions of magnetic field values from turbulence
\citep{2008Sci...320..909R}. Therefore further investigations are
needed to clarify the structure, evolution and origin of magnetic
fields in the largest structures of the Universe, their
observational signatures, as well as their interplay with other
processes acting in galaxy clusters and the large scale
structure.

The complexity of galaxy clusters comes principly from their
hierarchical build up within the large-scale structure of the
Universe. In order to study their formation it is necessary, to
follow a large volume of the Universe. However, one must also
describe cosmic structures down to relatively small scales, thus
spanning 5 to 6 orders of magnitudes in size. The complexity of
the cluster atmosphere reflects the infall of thousands of smaller
objects and their subsequent destruction or survival within the
cluster potential. Being the source of shocks and turbulence,
these processes directly act on the magnetic field causing
re-distribution and amplification. Therefore realistic modelling
of these processes critically depends on the ability of the
simulation to resolve and follow correctly this dynamics in galaxy
clusters.

Starting from a well-established cosmological n-body 
smoothed particle hydrodynamic (SPH) code GADGET 
\citep{springel2001,springel2005} we present here 
the implementation of magneto-hydrodynamics, which 
allows us to explore the full size and dynamical range of
state of the art cosmological simulations. GADGET also allows us to turn
on the treatment of many additional physical processes which are
of interest for structure formation and make interesting
links with the treatment of magnetic fields for future studies.
This includes thermal conduction
\citep{2004MNRAS.351..423J,2004ApJ...606L..97D}, physical
viscosity \citep{2006MNRAS.371.1025S}, cooling and star-formation
\citep{springel2003}, detailed modelling of the stellar population
and chemical enrichment
\citep{2004MNRAS.349L..19T,2007MNRAS.382.1050T} and a self
consistent treatment of cosmic rays
\citep{2007A&A...473...41E,2007MNRAS.378..385P}. The MHD
implementation presented here is fully compatible with all these
extensions, but here we want to focus on non-radiative
simulations. 
All such processes are expected to increased the complexity 
and lead to interplay with the evolution of the magnetic 
field. This would make it impossible to critically check the numerical 
effects caused by the different SPH-MHD implementations and therefore
we will ignore such additional processes in this work.

The paper is structured as follows: In section 2 we present the
details of the numerical implementation, whereas in section 3 we
present various code validation tests, all performed in fully
three dimensional setups. In section 4 we present the formation of a
galaxy cluster as an example for a cosmological application before we
present our conclusions in section 5. In addition we present a
convergence test for the code  in the appendix.

\section{SPH-MHD Implementation}
We have implemented the MHD equations in the cosmological SPH code
{\small GADGET} \citep{springel2001,springel2005}. In this section we
present the relevant details of this implementation. While developing
the MHD implementation made use of {\small GADGET-1}
\citep{springel2001} and {\small GADGET-2} \citep{springel2005}, all
simulations presented in this paper are based on the most recent
version of the code, {\small GADGET-3} (Springel, in prep). Note that
the implementation therefore is fully parallelized and benefits from
many optimizations within the general parts of the code, especially
the calculation of self gravity and optimization in data structures as
well as work-load balancing. Therefore, this implementation is an
ideal tool to follow the evolution of magnetic fields and allows 
us to explore dynamical ranges of more than 5 orders of magnitude within 
cosmological simulations.

During the last years, many general improvements in the 
implementation of the SPH method have been made. Examples include 
are more modern formulations of the artificial viscosity 
\citep{1997JCoPh..136....298S}, the introduction of self-consistent 
correction terms from varying smoothing length \citep{springel02,2002MNRAS.335..843M}
or the continuous definition of the smoothing length \citep{springel02}.
All these improvements not only increased the accuracy and stability 
of the underlying SPH formulation but also improved, directly or indirectly,
the accuracy and stability of any MHD implementation. The main, indirect 
benefits of these improvements are the self-consistent treatment of the 
magnetic waves within the formulation of the artificial 
viscosity \citep{2004MNRAS.348..123P},
the possibility to drop the viscosity limiter (see equation 10 and 
discussion in section 2.3) and the inclusion of the correction terms
from varying smoothing length in both the induction equation and 
the Lorenz force term \citep{2004MNRAS.348..139P}. In the remaining 
sections we will discuss our SPH-MHD implementation in detail.

\subsection{SPH implementation in GADGET} \label{sec:code}

The basic idea of SPH is to discretize the fluid in mass elements
($m_i$), represented by particles at positions $\vec x_i$
\citep{1977AJ.....82.1013L,1977MNRAS.181..375G}.
To build continuous fluid quantities,
one starts with a general definition of a kernel smoothing method.
The most frequently used kernel $W(|\vec x|,h)$ is
the $B_2$-Spline \citep{1985A&A...149..135M}, which can be written
as
\begin{equation}
   W(x,h)=\frac{8}{\pi h^3}\left\{\begin{array}{ll}
      1 - 6 \left(\frac{x}{h}\right)^2 + 6 \left(\frac{x}{h}\right)^3 \;\;& 0 \le \frac{x}{h} \le 0.5 \\
      2 \left(1 - \frac{x}{h}\right)^3                              & 0.5 \le \frac{x}{h} \le 1 \\
      0                                                             & 1 \le \frac{x}{h} \\
   \end{array} \right. , \label{SPH:kern}
\end{equation}
It is worth stressing that, contrary to other SPH implementation,
{\small GADGET} uses the notation in which the kernel $W(x,h)$
reaches zero at $x/h=1$ and not at $x/h=2$. The density $\rho_i$
at each particle position $\vec x_i$ can be estimated via
\begin{equation}
   \left<\rho_i\right> = \sum_j m_j W({\vec x_i} - {\vec x_j},h_i),
   \label{SPH:density}
\end{equation}
where the smoothing length $h_i$ is defined by solving the
equation
\begin{equation}
\frac{4\pi}{3}h_i^3\rho_i = N m_i.
\label{eqn:hsml}
\end{equation}
A typical value for $N$ is in the range of 32-64, which correspond
to the number of neighbors which are traditionally chosen in SPH
implementations.

In {\small GADGET}, the equation of motion for the SPH particles
are implemented based on a derivation from the fluid Lagrangian
\citep{springel02} and take the form
\begin{equation}
\left(\frac{\mathrm{ d}\vec{v}_i}{\mathrm{ d}t}\right)^{(\mathrm{hyd})} = - \sum_{j=1}^{N}m_j
\left[ f_i^\mathrm{ co
}\frac{P_i}{\rho_i^2}\vec{\nabla}_i W_i+f_j^\mathrm{ co
}\frac{P_j}{\rho_j^2}\vec{\nabla}_i W_j)\right].
\end{equation}
The coefficients $f_i$ are defined by
\begin{equation}
f_i^\mathrm{ co}=\left[1+\frac{h_i}{3\rho_i}\frac{\partial
\rho_i}{\partial h_i }\right]^{-1},
\end{equation}
and reflect the full, self-consistent correction terms arising
from varying the particle smoothing length. The abbreviation
$W_i=W(|\vec{r}_i-\vec{r}_j|,h_i)$ and
$W_j=W(|\vec{r}_i-\vec{r}_j|,h_j)$ are the two
kernels of the interacting particles. The
pressure of each particle is given by $P_i=A_i\rho_i^\gamma$,
where the entropic function $A_i$ stays constant for each particle
in the absence of shocks or other sources of heat.

To capture shocks properly, artificial viscosity is needed.
Therefore, in {\small GADGET} the viscous force is implemented
as
\begin{equation}
  \left(\frac{\mathrm{ d}v_i}{\mathrm{ d}t}\right)^{(\mathrm{visc})} = -
\sum_{j=1}^{N}m_j\Pi_{ij}\nabla_i\bar{W}_{ij},
\end{equation}
where $\Pi_{ij}\ge0$ is non-zero only when particles approach each
other in physical space. The viscosity generates entropy at a rate
\begin{equation}
  \frac{\mathrm{ d}A_i}{\mathrm{ d}t} =
\frac{1}{2}\frac{\gamma-1}{\rho_i^{\gamma-1}}
\sum_{j=1}^{N}m_j\Pi_{ij}\vec{v}_{ij}\cdot\nabla_i\bar{W}_{ij},
\label{eqn:dt_entropy}
\end{equation}
Here, the symbol $\bar{W}_{ij}$ denotes the arithmetic mean of the
two kernels $W_i$ and $W_j$.

For the parameterization of the artificial viscosity, starting with
version 2 of {\small
GADGET}, a formulation proposed by
\cite{1997JCoPh..136....298S} based on an analogy with Riemann
solutions of compressible gas dynamics, is used. In this case, the
resulting viscosity term can be written as
\begin{equation}
\Pi_{ij}=\frac{-0.5\alpha v_{ij}^\mathrm{ sig}
\mu_{ij}}{\rho_{ij}}f_{ij}^\mathrm{ shear} \label{eqn:visc2}
\end{equation}
for $\vec{r}_{ij}\cdot\vec{v}_{ij}\le 0$ and $\Pi_{ij} = 0$
otherwise, i.e.~the pair-wise viscosity is only non-zero if the
particles are approaching each other. Here
$\mu_{ij}=\vec{v}_{ij}\cdot\vec{r}_{ij}/|\vec{r}_{ij}|$ is the
relative velocity projected onto the separation vector and the
signal velocity is estimated as
\begin{equation}
v_{ij}^\mathrm{ sig} = c_i + c_j - \beta\mu_{ij},
\end{equation}
with $c_i=\sqrt{\gamma P_i/\rho_i}$ denoting the sound velocity.
In {\small GADGET-2} the values $\alpha=1$ and $\beta=3$ are
commonly used for the dimensionless parameters within the
artificial viscosity. Here we have also included a
viscosity-limiter $f_{ij}^\mathrm{ shear}=(f_i^\mathrm{
shear}+f_j^\mathrm{ shear})/2$,
which is often used to suppress the viscosity locally in regions
of strong shear flows, as measured by
\begin{equation}
f_{i}^\mathrm{ shear} =
\frac{|\left<\vec{\nabla}\cdot\vec{v}\right>_i|}{|\left<\vec{\nabla}\cdot\vec{v}\right>_i|
+ |\left<\vec{\nabla}\times\vec{v}\right >_i|+\sigma_i},
\end{equation}
which can help to avoid spurious angular momentum and vorticity
transport in gas disks
\citep{1995JCoPh.121..357B,1996IAUS..171..259S},
with the common choice
$\sigma_i=0.0001 c_i/h_i$.

This also leads naturally to a Courant-like hydrodynamical
time-step
\begin{equation}
\Delta t_i^\mathrm{ (hyd)}=\frac{C_\mathrm{ courant}h_i}{\mathrm{
max}_j(v_{ij}^\mathrm{ sig}) }, \label{eqn:timestep}
\end{equation}
where $C_\mathrm{ courant}$ is a numerical constant, typically choosen
to be in the range $0.15-0.2$.

\subsection{Co-moving variables and integration}

The equations of motion are integrated using a leap-frog
integration making use of a kick-drift-kick scheme. Within this
scheme, all the pre-factors due to the cosmological background
expansion are taken into account within the calculation of the
kick- and drift-factors \citep[see][]{1997astro.ph.10043Q,springel2005}. 
For the integration
of the entropy within a cosmological simulation, a factor
$(Ha^2)^{-1} = \frac{\mathrm{ d}t}{\mathrm{ d}a}$ is present in
equation \ref{eqn:dt_entropy} to take into account that the
internal time variable in {\small GADGET} is the expansion
parameter $a$. The formulation of the MHD equations within
{\small GADGET} has to be be adapted accordingly to this choice
of variables.

\subsection{Magnetic signal velocity}

A natural generalization of the signal velocity
$v_{ij}^\mathrm{ sig}$ in the framework of MHD is to replace the sound velocity $c_i$ by
the fastest magnetic wave as suggested by \citet{2004MNRAS.348..123P}.
Therefore the sound velocity $c_i$ gets replaced by
\begin{eqnarray}
v_i &=&
\frac{1}{\sqrt{2}}\left[\left(c_i^2+\frac{B_i^2}{\mu_0\rho_a}\right)
\cdot \right.
 \nonumber \\ & &\left.\sqrt{\left(c_i^2+\frac{B_i^2}{\mu_0\rho_i}\right)^2 -
4\frac{c_i^2(\vec{B}\cdot\vec{r}_{ij}/|\vec{r}_{ij}|)^2}{\mu_0\rho_i}}
\right]^{0.5}.
\end{eqnarray}
As this new definition of the signal velocity also enters
the time-step criteria (\ref{eqn:timestep}), no extra time-step
criteria due to the magnetic field has to be defined. We note that
we still see improvements in the solution to the test problems, if
we choose more conservative settings within the Courant
condition. Therefore we generally use $C_\mathrm{ courant}=0.075$,
which is half the value usually used in pure hydrodynamical
problems. Different authors also propose to use different values
for $\alpha$ and $\beta$ within the artificial viscosity
definition (\ref{eqn:visc2}). Whereas typically $\alpha=1$ is chosen,
\citet{1997JCoPh..136....298S} proposed to
use $\beta=3$. \citet{2004MNRAS.348..123P,2004MNRAS.348..139P} propose to use
$\beta=2$ or $\beta=1$ respectively. We find slight improvements
in our test problems when using  $\alpha=2$ and $\beta=1.5$, which
we use throughout this paper. We also note, that the viscosity
suppression switch $f_{ij}^{shear}$ was introduced based on an
earlier realization of the artificial viscosity and it is not clear
if it is still needed. As we note significant improvements in our
test problems when neglecting this switch we do not use this
switch throughout this paper. Also for the cosmological
application presented in the last part of this paper, this switch
was always turned off.

\subsection{Induction equation}
The evolution of the magnetic field is given by the induction
equation,
\begin{equation}
\frac{\mathrm{ d}\vec{B}}{\mathrm{ d}t} =
(\vec{B}\cdot\vec{\nabla})\vec{v} -
\vec{B}(\vec{\nabla}\cdot\vec{v}),
\end{equation}
if ohmic dissipation is neglected and the constraint
$\vec{\nabla}\cdot\vec{B}=0$ is used. The SPH equivalent reads
\begin{eqnarray}
\frac{\mathrm{ d}\vec{B}_i}{\mathrm{ d}t} &=& \frac{1}{Ha^2}
\frac{f_i^\mathrm{ co}}{\rho_i} \cdot \nonumber \\
& & \left[\sum_{j=1}^{N}m_j\left[
\vec{B}_i(\vec{v}_{ij}\cdot\vec{\nabla}_i W_i)
-\vec{v}_{ij}(\vec{B}_i\cdot\vec{\nabla}_i W_i)
\right]\right] \nonumber \\ & & - 2\vec{B}_i,
\end{eqnarray}
where $(Ha^2)^{-1} = \frac{\mathrm{ d}t}{\mathrm{ d}a}$ takes into
account that the internal time variable in {\small GADGET} is
the expansion parameter $a$. Note that here, by construction, only the
kernel $W_i$ and its derivative is used. The second term $-2\vec{B}_i$
accounts for the dilution of the frozen in magnetic field due to
cosmic expansion. Both these additions -- the $(Ha^2)^{-1}$ factor 
and the $-2\vec{B}_i$ term -- are only present in the
cosmological simulations and absent for the code evaluation
presented in section \ref{sec:tests}. In component form the
induction equation reads
\begin{eqnarray}
\frac{\mathrm{ d}B_i^k}{\mathrm{ d}t} &=& \frac{1}{Ha^2}
\frac{f_i^\mathrm{ co} }{\rho_i} \left[\sum_{j=1}^{N}m_j
(v^k_{ij}B_i^l - B_i^k v_{ij}^l) \frac{\partial W_i}{\partial
u}\frac{\vec{r}_{ij}^l}{|\vec{r}_{ij}|}\right] \nonumber \\ & & -
2\vec{B}^k_i.
\end{eqnarray}
Note that, as also suggested by \citet{2004MNRAS.348..139P}, we
wrote down the equations including the correction factor
$f_i^\mathrm{ co}$ which reflects the correction terms
($\frac{\mathrm{ d}W}{\mathrm{ d}h}$) arising
from the variable particle smoothing length. Unfortunately it is not possibile to
directly infer the exact form of the correction factors from first
principles for the induction equation. However,
\citet{2004MNRAS.348..139P} showed that, if not chosen in the same
way as for the Lorenz force, inconsistency between the induction
equation and magnetic force results. The effect of these factors in the
induction equation is quite small, but nevertheless one notices
tiny improvements in test problems when they are included.
Therefore, we included them for all applications presented
in this paper.

\subsection{Magnetic force}

The magnetic field acts on the gas via the Lorenz force, which can
be written in a symmetric, conservative form involving the magnetic
stress tensor \citep{1985MNRAS.216..883P}
\begin{equation}
M_i^{kl} = \left( \vec{B}_i^k\vec{B}_i^l - \frac{1}{2}|\vec{B}_i|^2\delta^{kl}\right).
\label{eq:maxtens}
\end{equation}
The magnetic contribution to the acceleration of the $i$-th particle
can therefore be written as
\begin{eqnarray}
\left(\frac{\mathrm{ d}\vec{v}_i}{\mathrm{ d}t}\right)^{(\mathrm{mag})} &=& \frac{a^{3\gamma}}{\mu_0}
\sum_{j=1}^{N}m_j \left[f_i^\mathrm{ co}\frac{M_i}{\rho_i^2}
                       \cdot\vec{\nabla}_i W_i \right.   \nonumber
\\
                       & &+
                       \left.f_j^\mathrm{ co}\frac{M_j}{\rho_j^2}
                       \cdot\vec{\nabla}_j W_j \right].
\end{eqnarray}
Here $a^{3\gamma} = \frac{\mathrm{ d}t}{\mathrm{ d}\eta}$ is needed to
transform the equations to the internal variables for cosmological
simulations and is set to one in all other cases. Also $\mu_0$ has to
be chosen properly as 
\begin{equation}
   \mu_0=\frac{[\mathrm{ TIME}]^2[\mathrm{ LENGTH}]}{4\pi[\mathrm{MASS}]h^2},
\end{equation}
with $h=1$ for non cosmological runs.
The factors $f_i^\mathrm{ co}$ reflect the correction terms
($\frac{\mathrm{ d}W}{\mathrm{ d}h}$) arising
from the variable particle smoothing length as introduced
already \citep[see also][]{2004MNRAS.348..139P}. In
component form the equation reads
\begin{eqnarray}
\left(\frac{\mathrm{ d}\vec{v}_i^k}{\mathrm{ d}t}\right)^{(\mathrm{visc})} &=& \frac{ a^{3\gamma}}{\mu_0}
\sum_{j=1}^{N}m_j\left[f_i^\mathrm{ co}\frac{M_i^{kl}}{\rho_i^2}
                       \frac{\partial W_i}{\partial
u}\frac{\vec{r}_{ij}^l}{|\vec{r}_{ij}|}\right. \nonumber \\
                       & &+ \left.f_j^\mathrm{ co}\frac{M_j^{kl}}{\rho_j^2}
                       \frac{\partial W_j}{\partial u}\frac{\vec{r}_{ij}^l}{|\vec{r}_{ij}|}
                      \right].
\end{eqnarray}
It is well known that this formulation becomes unstable for
situations, in which the magnetic forces are dominating
\citep{1985MNRAS.216..883P}, The reason for this is that the
magnetic stress can become negative, leading to the clumping of
particles. Therefore, some additional measures have to be taken
to suppress the onset of this instability.

\subsection{Instability corrections}

There are several methods proposed in the literature to suppress the onset
of the clumping instability which is caused by the implementation of the
magnetic force. However their performance was found to depend on
the details of the simulation setup. In the next sections we will
briefly discuss the different possibilities in the context of building
up an implementation for cosmological simulations.

\subsubsection{Adding a constant value}

One method to remove the instability was pointed out by
\citet{1985MNRAS.216..883P}, who suggested to calculate the maximum of
the magnetic stress tensor and to subtract it globaly from all
particles. Or similar, as
suggested in \citet{2005MNRAS.364..384P}, to subtract the
contribution of a constant magnetic field. This is simple and
straight forward if there is a strong, external magnetic field contribution
from the initial setup, which can be associated with the term one
subtracts. However, with cosmological simulations in mind, this
approach is not very viable.
 and therefore we did not use this approach.

\subsubsection{Anti clumping term}

\citet{344473} suggested the introduction of an additional term in the
momentum equation which prevents particles from clumping in the
presence of strong magnetic stress. Including this term, equation
(\ref{eq:maxtens}) reads
\begin{equation}
M_i^{kl} = \left( \vec{B}_i^k\vec{B}_i^l -
\frac{1}{2}|\vec{B}_i|^2\delta^{kl} - R_i\vec{B}_i^k\vec{B}_i^l\right),
\end{equation}
where $R_i$ is a steepened kernel which can be defined as
\begin{equation}
R_i = \frac{\epsilon}{2}\left(\frac{W_i}{W_1}\right)^n.
\end{equation}
The modification of the kernel is made so that contributions are
significant only at distances below the average particle
spacing $u_1$, so $W_1$ is defined as $W_1=W(u_1)$. Typical values for
the remaining parameters are $\epsilon=0.8$ and $n=5$. This method was
also used in  \citet{2004MNRAS.348..123P,2004MNRAS.348..139P} where
they found $u_1=1.5h$ to be a good choice for 1D simulations and
switched to $u_1=1.1h$ for 2D simulations. In agreement with
\citet{2005MNRAS.364..384P} we find that in 3D and allowing the
smoothing length to vary, this approach does not help to suppress the
instabilities efficiently. 

\subsubsection{$\mathrm{div}(\vec{B})$ force subtraction}

\citet{2001ApJ...561...82B} suggested explicitly subtracting the
effect of any numerically non-vanishing divergence of $\vec{B}$. Therefore, one
can explicitly subtract the term
\begin{eqnarray}
\left(\frac{\mathrm{ d}\vec{v}_i^k}{\mathrm{ d}t}\right)^{(\mathrm{corr})} &=&
- a^{3\gamma}\frac{1}{\mu_0}\hat\beta\vec{B}_i
\sum_{j=1}^{N}m_j\left[f_i^\mathrm{ co}\frac{\vec{B}_i}{\rho_i^2}
                       \cdot\vec{\nabla}_i W_i \right. \nonumber \\
& & + \left.
                       f_j^\mathrm{ co}\frac{\vec{B}_j}{\rho_j^2}
                       \cdot\vec{\nabla}_j W_j \right]
\end{eqnarray}
from the momentum equation. Here again, $a^{3\gamma} =
\frac{\mathrm{ d}t}{\mathrm{ d}\eta}$ and $\mu_0$ are introduced
to transform the equation to the internal units used. To be
consistent with the other formulations,  we included $f_i^\mathrm{
co}$, which are the $\frac{\mathrm{ d}W}{\mathrm{ d}h}$ terms. In
the original work \citep{2001ApJ...561...82B}, $\hat\beta=1$ was
choosen. In component form this equation reads

\begin{eqnarray}
\left(
\frac{\mathrm{ d}\vec{v}_i^k}{\mathrm{ d}t}
\right)^{(\mathrm{corr})} &=&
- a^{3\gamma}\frac{1}{\mu_0}
\hat\beta\vec{B}_i^k
\sum_{j=1}^{N}m_j
\left[f_i^\mathrm{ co}
\frac{\vec{B}_i^l}{\rho_i^2}
\frac{\partial W_i}{\partial u}
\frac{\vec{r}_{ij}^l}{|\vec{r}_{ij}|}
\right. \nonumber \\
& & +
\left.f_j^\mathrm{ co}
\frac{\vec{B}_j^l}{\rho_j^2}
\frac{\partial W_j}{\partial u}
\frac{\vec{r}_{ij}^l}{|\vec{r}_{ij}|}
\right].
\end{eqnarray}
In principle, this term breaks the momentum conserving form of the
MHD formulation. However, in practice, this seems to be a minor effect.
\citet{2004ApJS..153..447B} argued that stability for linear waves in 2D
can be safely reached even when not subtracting the full term but
choosing $\hat\beta<1$; e.g. they suggested $\hat\beta=0.5$ to
further minimizing the non-conservative
contribution. However, it is not clear if this stays true for 3D
setups and in the non-linear regime.
Additional we do not use a higher order kernel as done in \citet{2004ApJS..153..447B}, 
therefore it is not clear if their conclusions still hold in our case.
As the results in our test problems seem unharmed
by the possible violation of momentum conservation due to the formulation of the correction
factor in the Lorenz force, we keep it in the form suggested in earlier work
\citep[e.g.][]{2001ApJ...561...82B} i.e. $\hat\beta=1$.

In \citet{2006ApJ...652.1306B} a more general formalism to obtain
$\hat{\beta}$ for each particle was introduced with the aim to further
minimize the violation of momentum conservation in the formulation of
the correction terms in the Lorenz force. Unfortunately, in the
light of cosmological simulation, this
seems not to be very practical as it contains a scan for a maximum
value over all particles, which in a cosmological context makes no
sense as there is not a specific single object to which such
characteristics can be tuned to.

In general we find that this correction term significantly improves all
results in our test simulations and effectively suppresses the
onset of the clumping instability. It was also already successfully used
in previous, cosmological applications \citep[e.g.][]{2004JETPL..79..583D,2004APh....22..167R,dolag2005b}.

\subsection{Regularization schemes} \label{sec:reg}
Beside instabilities, noise (e.g. fluctuations of the
magnetic field imprinted by numerical effects when integrating the
induction equation) is a source of errors in SPH-MHD
implementations. The goal of a regularization 
scheme is to obtain a magnetic field which does not show
strong fluctuations below the smoothing length. This can be 
achieved indirectly through improvements in the underlying 
SPH formalism and by reformulation of the the interactions 
to reduce the creation of small irregularities from 
numerical effects (e.g. through particle splitting). 
Alternative approaches are to directly suppress the magnetic field or to 
dissipate small irregularities by introducing artificial dissipation.

\subsubsection{Improvements in the underlying 
SPH formalism}

Here, the entropy conserving formalism \citep{springel02} of the
underlying SPH implementation contributes to a significant
improvement of the MHD formalism compared to previous MHD
implementations in SPH by generally improving the density
estimate and the calculation of derivatives. It has to be noted,
that this is not only due to the $\frac{\mathrm{ d}W}{\mathrm{
d}h}$ terms, but in large part also from the new formalism for
calculation of the smoothing length. As described before, the
smoothing length $h_i$ for each particle is no longer calculated
by counting neighbors within the sphere, but by solving equation
(\ref{eqn:hsml}) for each formal number of neighbors
$N$, there exists only one unambiguous value of $h_i$. Note that,
as this equation is solved iteratively, it is usual to give some
allowed range of $N$, however in our case we can choose the range
smaller than 1 and typically we use $N=64\pm0.1$. 

\subsubsection{Particle splitting}
\citet{2001ApJ...561...82B} developed a scheme to regularize the
interaction of particles in SPH based on a discretization of the smoothing length
$h_i$ by factors of 2. In such cases, interactions between two
particles with different smoothing length can be realized by splitting
the one with the larger smoothing length into $2^n$ (where $n$ is the
dimensionality) particles, placed on a $2^n$ sub-grid. Such split
particles then have the same smoothing length as the particles
with which they interact. Originally this scheme was invented to avoid the
problems induced by a variable smoothing length (before the correction
terms where properly introduced in SPH) and gave good results
in 1D and 2D
\citep[see][]{2001ApJ...561...82B,2004ApJS..153..447B,2006ApJ...652.1306B}.
However, in 3D the resulting change in the number of neighbors is 
quite large when the smoothing length is quantized in 
factors of 2. Therefore the additional
sampling noise for particles can be large. This is especially 
problematic when the lower density is approaching  
 -- but still above -- 
the threshold for doubling the particle smoothing lengths. 
Here particles have a particularly small smoothing length (relative to 
the optimal, unquantized one) and therefore 
have only small numbers of neighbors.
Unfortunately, this effect is much larger than the
gain in accuracy by the regularization, at least when based on
standard SPH formalism, \citep[see][]{2003DelPra}. Also, as the
$\frac{\mathrm{ d}W}{\mathrm{ d}h}$ terms formally take care of all
correction terms induced in the formalism when allowing a variable
smoothing length, this splitting -- and specifically the quantization of the
smoothing length -- is no longer needed. This might be different when
further improving the SPH (and specially the MHD) method. For example
when re-mapping techniqes based on Voronoi tessellation are used
\citep[e.g.][]{2006ApJ...652.1306B}, or special coordinates like
spherical or cylindical are used \citep[e.g.][]{2006Jcap...213.391}.

\subsubsection{Smoothing the magnetic field}

Another method to remove small scale fluctuations and to
regularize the magnetic field
is to smooth the magnetic field periodicly. As suggested by
\citet{2001ApJ...561...82B}, one can calculate a smoothed magnetic
field $\left<\vec{B}_i\right>$ for each particle,
\begin{equation}
   \left<\vec{B}_i\right> = \frac{\sum_j \frac{m_j}{\rho_j} \vec{B}_i W_i}{\sum_j \frac{m_j}{\rho_j} W_i}.
\end{equation}
Then, in periodic intervals, one can calculate a new, regularized
magnetic field by
\begin{equation}
   \vec{B}_i^\mathrm{new} = q\left<\vec{B}_i\right> + (1-q) \vec{B}_i.
\label{eqn:bsmooth}
\end{equation}
Note that this, in principal, acts similar to the mixing process on
resolution scale present in Eulerian schemes. However, introduced in
this way, the amount of mixing (e.g. dissipation) of magnetic field
depends on the frequency with which this procedure is applied and the
value of $q$ chosen. Typically, we set $q$ to one and perform the
smoothing at every 15$^{th}$-20$^{th}$ main time-step. It is worth while to mention that
implemented in this form, total energy is not conserved (as magnetic
field fluctuations on scales smaller than the smoothing length are
just removed) and, as the time-steps depend on the chosen
resolution, this method is even resolution dependent. Never the less
it leads to improvements in the results of our test problems, without
strongly smoothing sharp features. It also works without problem in 3D and
has already been used in cosmological simulations \citep{2004JETPL..79..583D,dolag2005b}.

\begin{figure}
\begin{center}
  \includegraphics[width=0.45\textwidth]{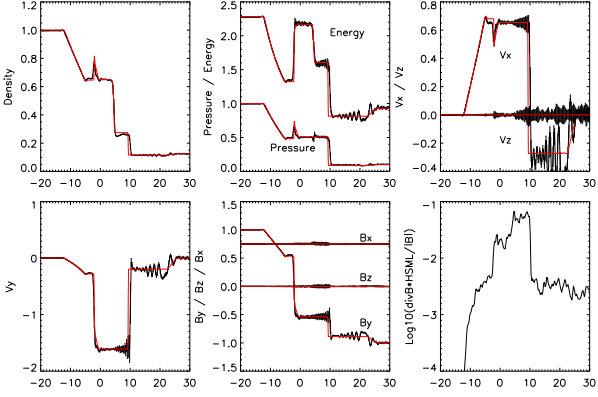}
\end{center}
\caption{Test {\it 5A} at time $t=7$ with the MHD implementation
similar to that used in
\citet{1999A&A...348..351D,2002A&A...387..383D}, but already including
the instability correction due to subtraction of the $\mathrm{
div}(\vec{B})$ term in the force equation and in a fully
three-dimensional setup. Shown in the first row are
the density (left panel), total energy and pressure (middle panel)
and the $x$,$z$ component of the velocity field (right panel). The second row shows
the $y$-component in the velocity field (left panel), the three
components of the magnetic field (middle panel) and the measure of the
$\mathrm{div}(\vec{B})$ error, see equation (\ref{eqn:divb}), in the right panel.
The black lines with error bars show the SPH results, the red lines are
the reference results obtained with Athena in a 1D setup.}
\label{fig:BrioWu1}
\end{figure}

\subsubsection{Artificial magnetic dissipation}

Another possibility to regularize the magnetic field was presented by
\citet{2004MNRAS.348..123P}, who suggested including an artificial
dissipation for the magnetic field, analogous to the artificial
viscosity used in SPH. In
\citet{2004MNRAS.348..123P} it was suggested that the
dissipation terms be constructed based on the magnetic field component perpendicular
to the line joining the interacting particles. However, to better
suppress the small scale fluctuations within the magnetic field
which appear due to numerical effects especially in
multi-dimensional tests, \citet{2004MNRAS.348..139P}
suggested basing the artificial dissipation on the change of the total
magnetic field rather than on the perpendicular field components only. We
also found this to work significantly better in our test cases and
therefore only use the later implementation throughout this paper.
Such an artificial dissipation term can be included in the induction equation as
\begin{eqnarray}
  \left(\frac{\mathrm{ d}\vec{B}_i}{\mathrm{ d}t}\right)^{(\mathrm{ diss})} &=&
  \frac{1}{Ha^2}\frac{\rho_i\alpha_B}{2} \nonumber \\ & &
      \sum_{j=1}^{N}
        \frac{m_jv_{ij}^\mathrm{ sig}}{\hat{\rho}_{ij}^2}\left(\vec{B}_i-\vec{B}_j\right)
        \frac{\vec{r}_{ij}}{|\vec{r}_{ij}|}\cdot\vec{\nabla}_i W_i.
\label{eqn:induction_dis}
\end{eqnarray}
The parameter $\alpha_B$ is used to control the strength of the effect, typical
values are suggested to be around $\alpha_B\sim0.5$. Similar to the
artificial viscosity, this will create entropy at the rate
\begin{eqnarray}
  \left(\frac{\mathrm{ d}A_i}{\mathrm{ d}t}\right)^{(\mathrm{ diss})} &=&
      -\frac{\gamma-1}{\rho_i^{\gamma-1}}
      \frac{\alpha_B}{4\mu_0} \nonumber \\ & &
      \sum_{j=1}^{N}
        \frac{m_jv_{ij}^\mathrm{ sig}}{\hat{\rho}_{ij}^2}\left(\vec{B}_i-\vec{B}_j\right)^2
        \frac{\vec{r}_{ij}}{|\vec{r}_{ij}|}\cdot\vec{\nabla}_i \bar{W}_{ij}.
\label{eqn:entropy_dis}
\end{eqnarray}
The pre-factor $(\gamma-1)/(\rho_i^{\gamma-1})$ properly converts
the dissipation term to a change in entropy.

This method reduces noise significantly. However, depending on the
choice of $\alpha_B$, it can also lead to smearing of sharp features. To
avoid this outside of strong shocks (e.g. where this is needed),
\citet{2005MNRAS.364..384P} proposed evolving $\alpha_B$ for each
particle, similar to the handling of the time dependent viscosity as
suggested by \citet{1997JCoPh..136....41S}. Such, evolution of
$\alpha_B$ for each particle will be followed by integrating
\begin{equation}
\frac{\mathrm{ d} \alpha_B}{\mathrm{ d} t} = - \frac{(\alpha_B-\alpha_B^\mathrm{min})}{\tau} + S,
\end{equation}
where the source term $S$ can be chosen as
\begin{equation}
   S = S_0 \, \mathrm{max}\left(
\frac{|\vec{\nabla}\times\vec{B}|}{\sqrt{\mu_0\rho}}
,
\frac{|\vec{\nabla}\cdot\vec{B}|}{\sqrt{\mu_0\rho}}
\right)
\end{equation}
\citep[see][]{2005MNRAS.364..384P}. The time-scale $\tau$ defines how
fast the dissipation constant decays. Taking the signal velocity,
one can translate this directly into a distance to the shock over which the dissipation
constant decays. A useful choice of $\tau$ can be written as
\begin{equation}
   \tau = \frac{h_i}{C\;v_\mathrm{sig}},
\end{equation}
where the constant $C$ typically is chosen to be around 0.2, allowing
the dissipation constant to decay  within a time that corresponds to
the shock travelling 5 kernel lengths \citep[see][]{2004MNRAS.348..123P},

\begin{figure}
\begin{center}
  \includegraphics[width=0.45\textwidth]{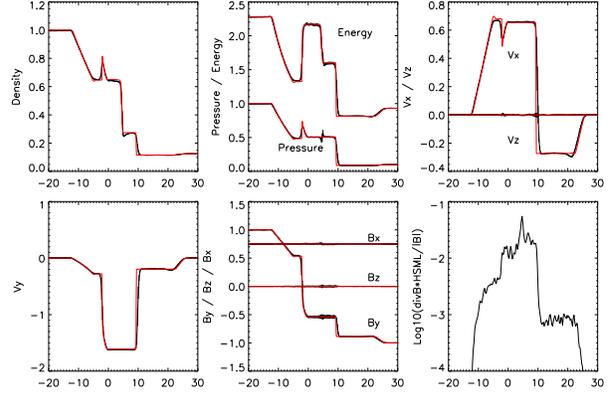}
\end{center}
\caption{As figure \ref{fig:BrioWu1}, but including the
magnetic waves in the signal velocity and turning off the shear
viscosity suppression as explained in the code description. The main
advantage is a significant reduction in the noise, specifically in the
velocity, but also in the magnetic field. Also the $\mathrm{div}(\vec{B})$ errors are
reduced by a factor of $\approx 2$.}
\label{fig:BrioWu2}
\end{figure}

\subsection{Euler potential}

A very elegant way to implement the MHD equations in Lagrangian codes
is the usage of so called {\it Euler potentials} \citep[see][ and
references therein]{2007MNRAS.379..915R}. Two independent variables 
$\alpha$ and $\beta$ are constructed to correspond to an
implicit gauge for the vector potential. They can be
thought of as labels of magnetic field lines and will be advected with
the flow. In this formulation, the magnetic field at any time can be
represented as
\begin{equation}
   \vec{B} = \vec{\nabla}\alpha \times \vec{\nabla}\beta.
\end{equation}
In principle, having obtained the magnetic field, one could also use this
magnetic field in the equation of motion as before. However, 
this would mean that the magnetic force is based on the second 
derivative of a variable. This is usually quite noisy and not
recommendable unless regularization schemes are implemented 
as done by e.g. \citet{2007MNRAS.379..915R}. In addition, 
due to its form, an implementation based on the Euler potential 
cannot be easily investigated in 3D test problems with periodic 
boundary conditions for the resulting magnetic field 
as can be done for the other implementations. A simple 
example here is constant magnetic field, which can be 
represented by a linear function of the {\it Euler potentials}.

Therefore we use this simple description only as a check
in cosmological simulations, to investigate the influence of
$\mathrm{div}(\vec{B})$ driven errors. 
Applied to cosmological simulations the evolution 
of the magnetic field predicted when 
using Euler potentialsis an upper bound on the 
amplification processes in the absence of any dynamo action. 
Therefore Euler potentials are a useful tool 
to check the influence of numerics on the results of 
cosmological simulations where we have no other means to
verify the results.

%

\begin{figure}
\begin{center}
  \includegraphics[width=0.45\textwidth]{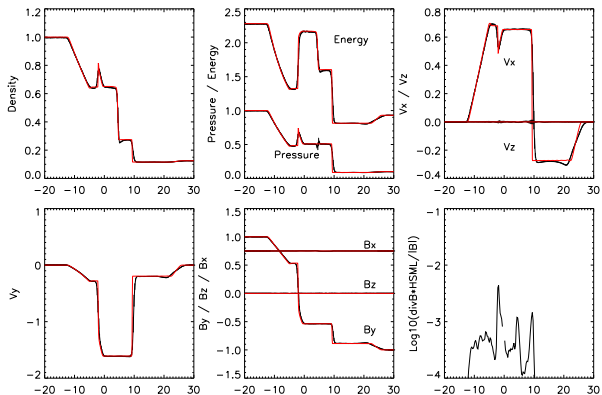}
\end{center}
\caption{As figure \ref{fig:BrioWu2}, but including
the regular smoothing of the magnetic field as a regularization
scheme. This SPH-MHD implementation basically reflects the one used in
\citet{2004JETPL..79..583D,dolag2005b}. The main
advantages are a further, significant reduction in the noise as well as
a strong reduction of the $\mathrm{div}(\vec{B})$ errors by a factor of $\approx 10$.}
\label{fig:BrioWu_bsmooth}
\end{figure}

\section{Test problems} \label{sec:tests}

To test performance of the code and to infer the optimal numerical
settings for the regularization schemes, we performed the series of
shock-tube problems as presented by
\citet{1995ApJ...442..228R}. In particular test {\it 5A}, which is
also used in \citet{BrioWu1988}, was used to show the effects of
different numerical treatments. Additionally we performed several 2D
test cases including the {\it Fast Rotor} test
\citep{Toth,2000ApJ...530..508L,Balsara99}, a {\it Strong Blast}
\citep{2000ApJ...530..508L,Balsara99} and the {\it Orzang-Tang Vortex}
\citep{Orzang,1994JCoPh.115..485D,1991PhFlB...3...29P,2000ApJ...530..508L}.
To obtain results under realistic circumstances, we performed all the
tests by setting up a fully three-dimensional particle
distribution. We also avoid starting from regular grids but used
glass-like \citep{1996clss.conf..349W} initial particle distributions
instead. To obtain such a configuration, the particles are originally
distributed in an random fashion within the volume and then allowed to relax
until they settle in a equilibrium distribution which is quasi force-free 
and homogeneous in density. This is similar to the distribution 
of atoms in an amorphous structure like glass. Compared to a distribution 
of the particles based on a grid, this guarantees that all the kernel averages 
in the SPH formalism sample the kernel in a uniform way rather than multiple
times at the same distances (which furthermore would be fractions of the underlying 
grid spacing). For all tests we used the same
particle masses, independent of the initial density. Therefore,
typical initial particle distributions for the shock-tube tests where
based on $5^3$ particles in low density and $10^3$ particles in
high density regions within unit volume. Usually, these unit volumes
are then replicated 35 times along the $x$-direction each.    For some
test cases with strong (and therefore fast) shocks, we evolved the
simulations longer. In such cases we doubled the simulation setup size
in the $x$-direction.

\begin{figure}
\begin{center}
  \includegraphics[width=0.45\textwidth]{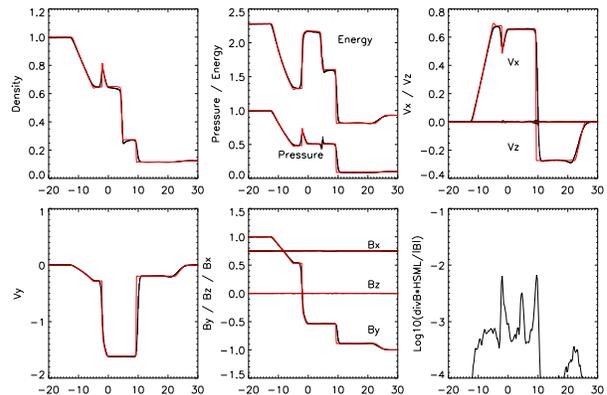}
\end{center}
\caption{As figure \ref{fig:BrioWu2}, but including
artificial magnetic dissipation as a regularization scheme. Similar to
the smoothing of the magnetic field, significant reduction in the noise as well as
a strong reduction of the $\mathrm{div}(\vec{B})$ errors by a factor of
$\approx 10$ is obtained compared to the {\it basic SPH-MHD} implementation.}
\label{fig:BrioWu_dissipation}
\end{figure}

We assume ideal gas (e.g. $\gamma=5/3$) and, as described
before, use an equivalent of 64 neighbors for calculating the SPH
smoothing length. This ensures that, in the low density regions, the SPH
particles get smoothed over a region corresponding to a unit length.
The number of resolution elements corresponding to a unit
length therefore ranges from 1 to 4, depending whether one associates the smoothed
region or the mean inter-particle distance with the effective resolution in
SPH. In general, SPH converges somewhat slower compared
to grid codes when comparing simulations with the same number of grid
cells as SPH particles (see Appendix A for an example).

For the SPH results we usually plot the mean within a 3D slab
corresponding to the smoothing length and (as error bars) the RMS over
the individual particles within this volume. The reference solution
was obtained using Athena \citep{2008arXiv0804.0402S} with typically 10-20 resolution elements
per unit length, depending on the individual test. As one criteria of
the goodness of the SPH simulation result we use the usual measure for
the non-vanishing divergence of the magnetic field,
\begin{equation}
   E_{\nabla\vec{B}} = \mathrm{div}(\vec{B}) \frac{h}{|\vec{B}|}.
\label{eqn:divb}
\end{equation}

\subsection{Shock tube 5A}

The most commonly used MHD shock-tube test is the one used by \citet{BrioWu1988},
e.g. test {\it 5A} in \citet{1995ApJ...442..228R}. The reason for this is
that it involves a shock and a rarefraction 
moving together. Therefore it allows simultaneous testing of the code in different
regimes.

Figure \ref{fig:BrioWu1} shows the result for a code implementation
similar to the first implementation used to study galaxy clusters
\citep[e.g. see][]{1999A&A...348..351D,2002A&A...387..383D}. In addition,
the instability correction due to subtraction of the $\mathrm{
div}(\vec{B})$ term was used in the force equation. Various
hydro-dynamical variables at the final time (e.g. $t=7$ in this case)
are shown. The black lines with error bars show the SPH-MHD result, the red lines
are the reference result obtained with Athena in a 1D setup. Shown are
(from upper left to the bottom right panel) the density, total energy
and pressure, the $x$- and $z$-component of the velocity field, the
$y$-component in the velocity field, the three components of the
magnetic field and the measure of the $\mathrm{div}(\vec{B})$ error,
obtained from equation (\ref{eqn:divb}).
Here we also switched back to the conventional formulation of 
the artificial viscosity as described by e.g. \citet{1992ARA&A..30..543M}, 
rather than that based on signal velocity as used in {\small GADGET-2}.
Although the SPH-MHD results in general follow the solution
obtained with Athena, there is a large scatter in the individual particle
values within the 3D volume elements, as well as some instability,
especially in the low-density part. But note that although the
mean values for the internal energy, as well as the velocity or magnetic
field, can locally show some systematic deviations from the
ideal solution, the total energy shows much better, nearly
unbiased, behaviour. This demonstrates the conservative nature of the
symmetric formulations in SPH-MHD.

Noticeable reduction of noise is obtained when using the
signal-velocity based artificial viscosity and
including the magnetic waves in the calculation of the signal
velocity. Therefore, the magnetic waves are directly captured for the
time step calculation and in the artificial viscosity, needed to capture
shocks. Also switching off the shear viscosity suppression again
leads to significant reduction in scatter. This can be seen in figure
\ref{fig:BrioWu2}, where the noise in the velocity as well
as in the magnetic field components is significantly reduced. Values of
$\mathrm{div}(\vec{B})$ are also reduced (by a factor of $\approx 2$)
compared to before.
In general, the SPH-MHD implementation gains from the new 
formulation of SPH, including the $\frac{\mathrm{
d}W}{\mathrm{ d}h}$ terms and the new way to determine the SPH smoothing
length, both contributing to a reduction of noise (and $\mathrm{div}(\vec{B})$) in the
general treatment of hydro-dynamics.
We will refer to this implementation of SPH-MHD as
{\it basic SPH-MHD} further on in this paper.

\subsection{The effect of regularization}

\begin{figure}
\begin{center}
  \includegraphics[width=0.45\textwidth]{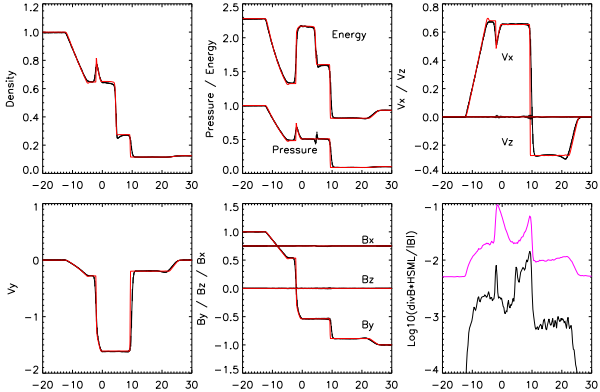}
\end{center}
\caption{As figure \ref{fig:BrioWu2}, but including
time dependent artificial magnetic dissipation as a regularization scheme.
No significance improvement is obtained. Note that here in the lower right panel the
artificial dissipation constant ($\alpha_{B}$) is shown. The effect of
suppressing the dissipation is clearly visible, and the maximum value
is only reached in peaks associated with the region of strong
shocks. However the improvement in the smearing of sharp features is
not very significant.}
\label{fig:BrioWu_tdd}
\end{figure}

As described in section (\ref{sec:reg}), there are several suggestions
for regularization of the magnetic field. Here we will show
results obtained by two regularization methods, namely smoothing the
magnetic field in regular intervals and including an artificial dissipation.

\begin{table*}
\begin{tabular}{*{9}{| c |}} 
\hline\hline
\multirow{2}{*}{ TEST Nr.  }  & \multicolumn{4}{| c |}{Left}  & \multicolumn{4}{ | c | }{Right} \\ 
  & $\rho$ & ${V}$ & ${B}$ & $P$  & $\rho$ & ${V}$ & ${B}$ & $P$   \\\hline\hline
| 1A | & 1.00 & $[10.0, 0.0  , 0.0]$   & $[5.0,5.0,0.0]/(4\pi)^{2} $             & 20.0  &  1.000 & $[-10.0,0.0,0.0]$ & $[5.0,5.0,0.0]/(4\pi)^{2} $            &  1.00   \\
| 1B | & 1.00 & $[ 0.0, 0.0  , 0.0]$   & $[3.0,5.0,0.0]/(4\pi)^{2} $             &  1.0  &  0.100 & $[  0.0,0.0,0.0]$ & $[3.0,2.0,0.0]/(4\pi)^{2} $            & 10.0   \\
| 2A | & 1.08 & $[ 1.2, 0.01 , 0.5]$   & $[2.0,3.6,2.0]/(4\pi)^{2} $             & 0.95  &  1.000 & $[  0.0,0.0,0.0]$ & $[2.0,4.0,2.0]/(4\pi)^{2} $            &  1.00   \\
| 2B | & 1.00 & $[ 0.0, 0.0  , 0.0]$   & $[3.0,6.0,0.0]/(4\pi)^{2} $              &  1.0  &  0.100 & $[ 0.0,2.0,1.0]$ & $[3.0,1.0,0.0]/(4\pi)^{2} $            & 10.0   \\
| 3A | & 1.00 & $[50.0, 0.0  , 0.0]$   & $-[0.0,1.0,2.0]/(4\pi)^{2} $           &  0.4  &  0.100 & $[  0.0,0.0,0.0]$ &  $[0.0,1.0,2.0]/(4\pi)^{2} $            &  0.20   \\
| 3B | & 0.10 & $[-1.0, 0.0  , 0.0]$   & $[0.0, 1.0   , 0.0]$                   &  1.0  &  1.000 & $[  1.0,0.0,0.0]$ & $[0.0,1.0,0.0]$                         &  1.00   \\
| 4A | & 1.00 & $[0.00, 0.0  , 0.0]$   & $[1.0, 1.0   , 0.0]$                   &  1.0  &  0.200 & $[  0.0,0.0,0.0]$ & $[1.0,0.0,0.0]$                         &  0.10   \\
| 4B | & 0.40 & $[-0.669,0.986,0.0]$   & $[1.3,0.0025293,0.0]$               & 0.5247 & 1.000 & $[0.0,0.0,0.0]$ & $[1.3,1.0,0.0]$                              &  1.00   \\
| 4C | & 0.65 & $[0.667,-0.257,0.0]$   & $[0.75, 0.55,0.0]$                     & 0.50  &  1.000 & $[ 0.4,-0.94,0.0]$ & $[0.75,0.00001,0.0]$                   & 0.75   \\
| 4D | & 1.00 & $[0.0,0.0,0.0]$        & $[7.0,0.001,0.0]$                      &   1.0 &  0.300 & $[  0.0,0.0,0.0]$ & $[7.0,1.0,0.0]$                         & 0.20    \\
| Brio Wu | & 1.00 & $[0.0,0.0,0.0]$   & $[0.75,1.0,0.0]$                       &   1.0 &  0.125 & $[  0.0,0.0,0.0]$ & $[0.75,-1.0,0.0]$                       & 0.10   \\
\hline\hline
\end{tabular}
\caption{Summary table with the initial conditions of the left and right side of the shock tubes.}
\label{TTable1}
\end{table*}

For the first method, the magnetic field is smoothed using the same
kernel as used for the normal SPH calculations. In this case,
there are two numerical parameters one can choose. One is $q$ in
equation (\ref{eqn:bsmooth}), which quantifies the weight with which
the smoothed component enters into the updated magnetic field. We
always use $q=1$ here, which means that we completely replace the
magnetic field by the smoothed value. The second is $T_\mathrm{BS}$,
which is the time interval at which the smoothing is done. Here we
use a value corresponding to a smoothing every 30$^{th}$ global time
step. This correspond to the SPH-MHD implementation used to study the
magnetic field in clusters and large scale structure within the local
universe, see \citet{2004JETPL..79..583D,dolag2005b}.
Figure \ref{fig:BrioWu_bsmooth} shows the result for the same shock-tube
test as before. Clearly, the noise in the individual
quantities is strongly reduced. Also the error in
$\mathrm{div}(\vec{B})$ is reduced by more than one order of
magnitude. Note that the error bars for the SPH-MHD implementation are
of the size of the line width or smaller in most of the cases and
therefore no longer clearly visible. However, one can notice
some small effect of smearing sharp features. Additinalally, some states
-- like the region with the negative $x$-component of the velocity behind
the the fast rarefaction wave propagating to the right --
converge to values which have small but systematic deviations from the
exact solution.

\begin{figure*}
\begin{center}
    \subfigure[Test 1A]{
    \label{fig:1A}
    \includegraphics[width=0.45\textwidth]{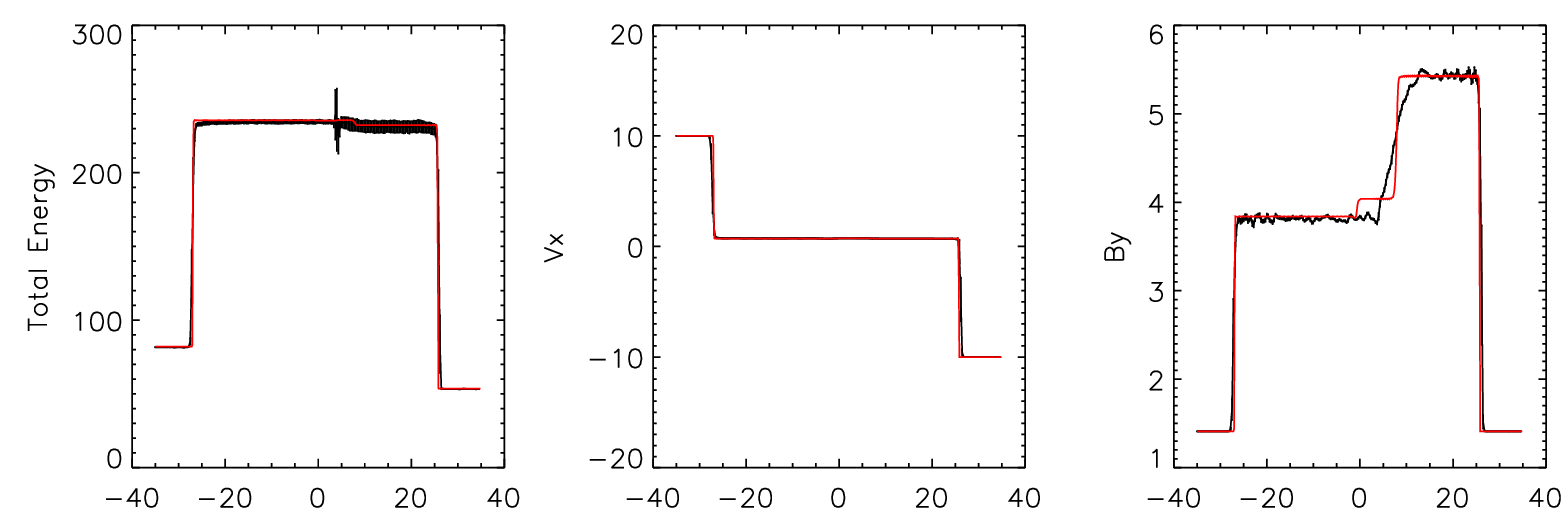}
    }
    \subfigure[Test 1B]{
    \label{fig:1B}
    \includegraphics[width=0.45\textwidth]{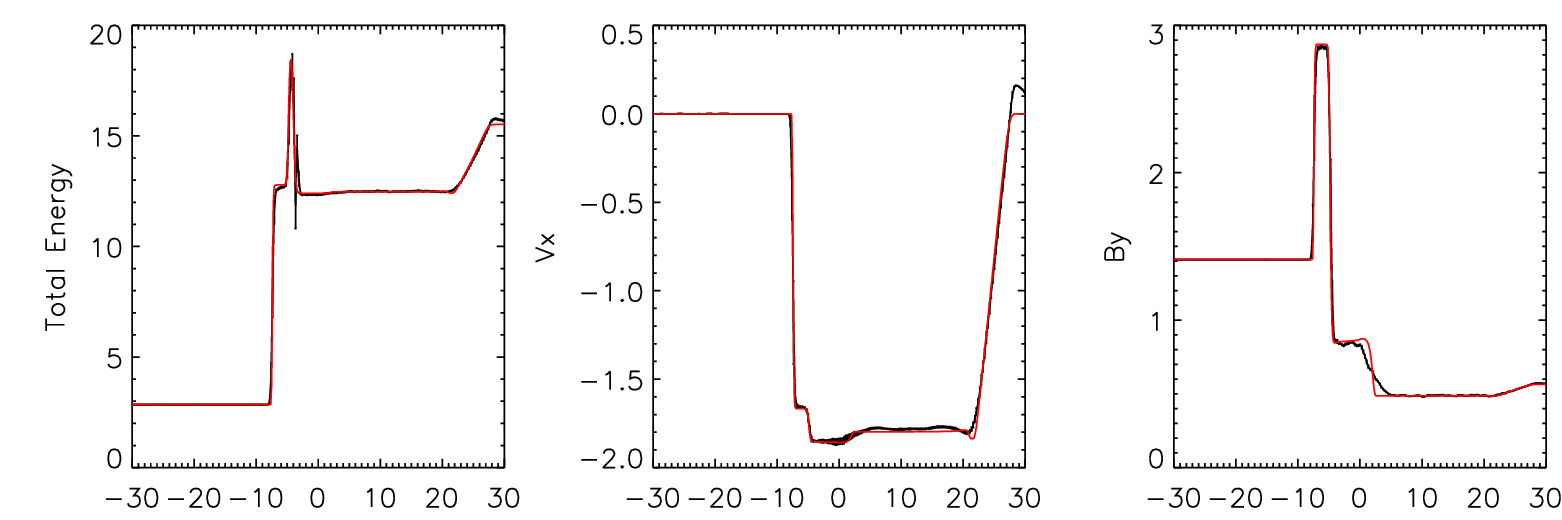}
    }
    \subfigure[Test 2A]{
    \label{fig:2A}
    \includegraphics[width=0.45\textwidth]{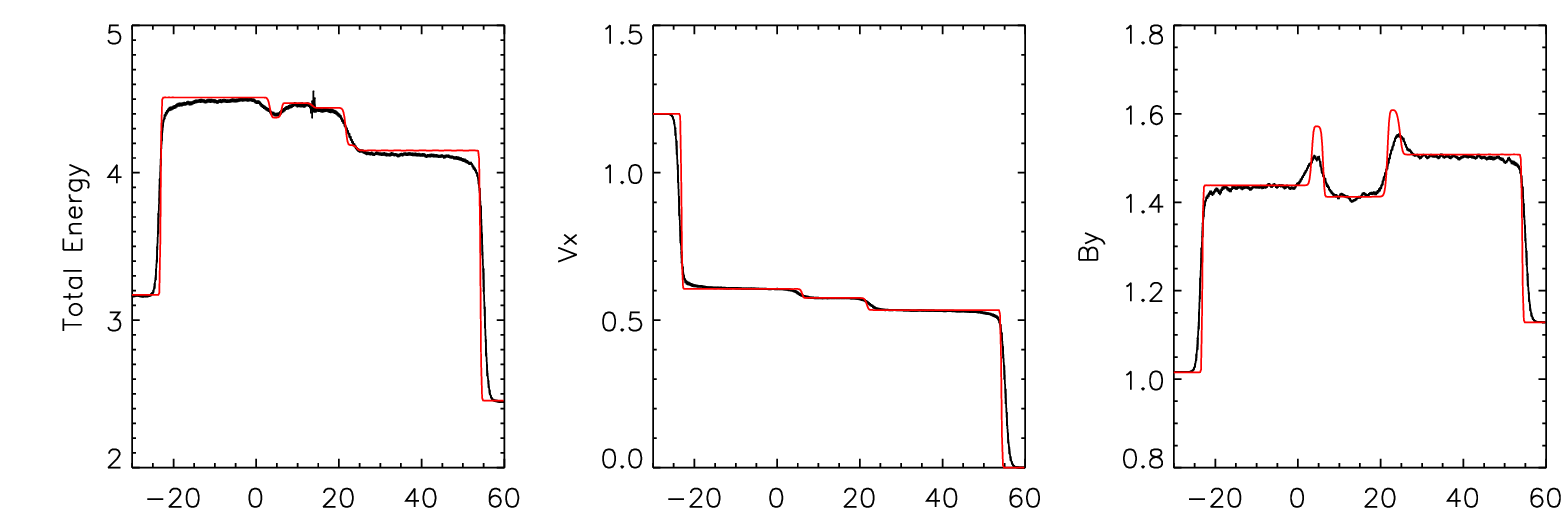}
    }
    \subfigure[Test 2B]{
    \label{fig:2B}
    \includegraphics[width=0.45\textwidth]{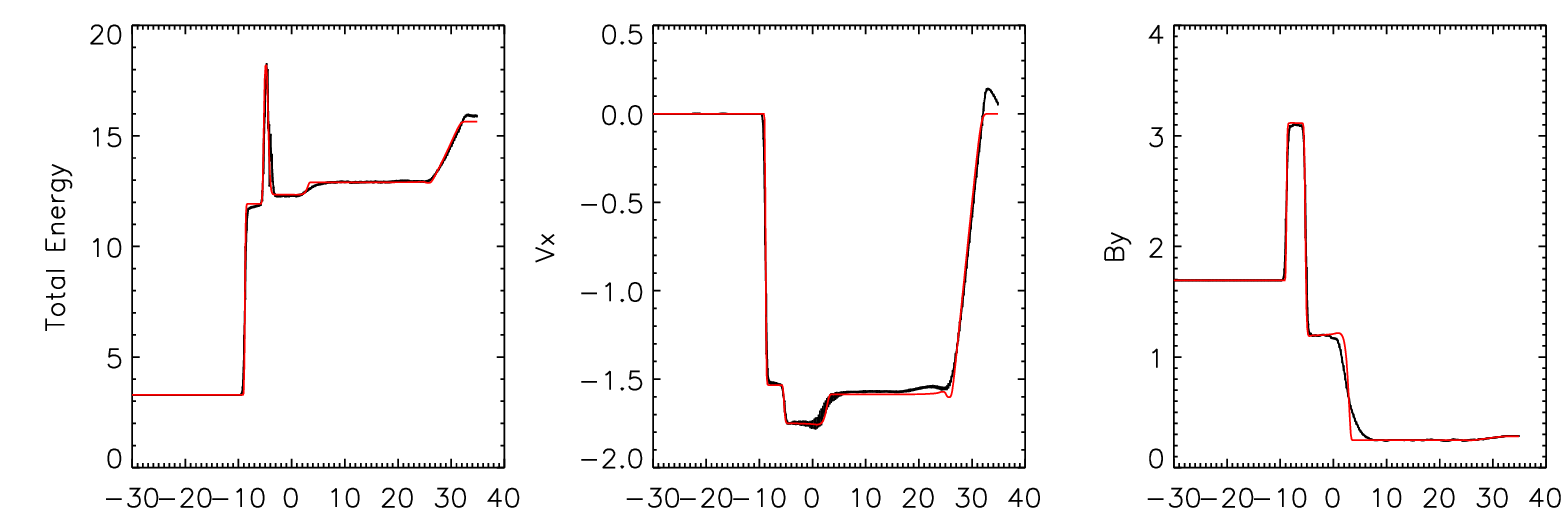}
    }
    \subfigure[Test 3A]{
    \label{fig:3A}
    \includegraphics[width=0.45\textwidth]{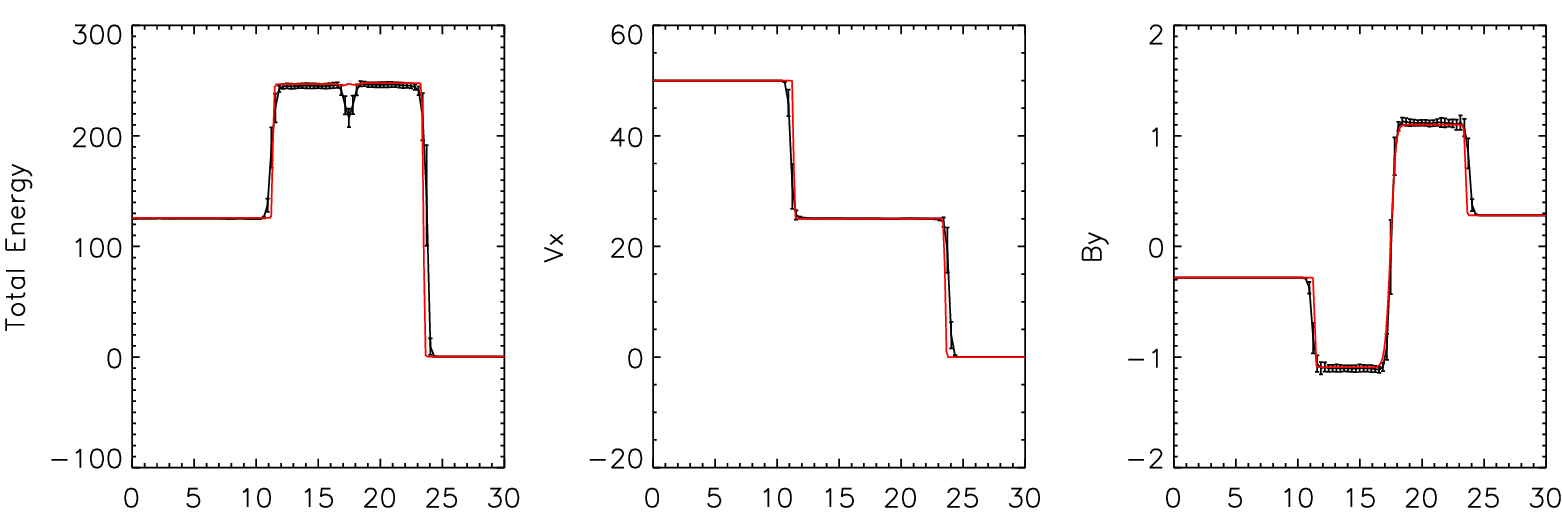}
    }
    \subfigure[Test 3B]{
    \label{fig:3B}
    \includegraphics[width=0.45\textwidth]{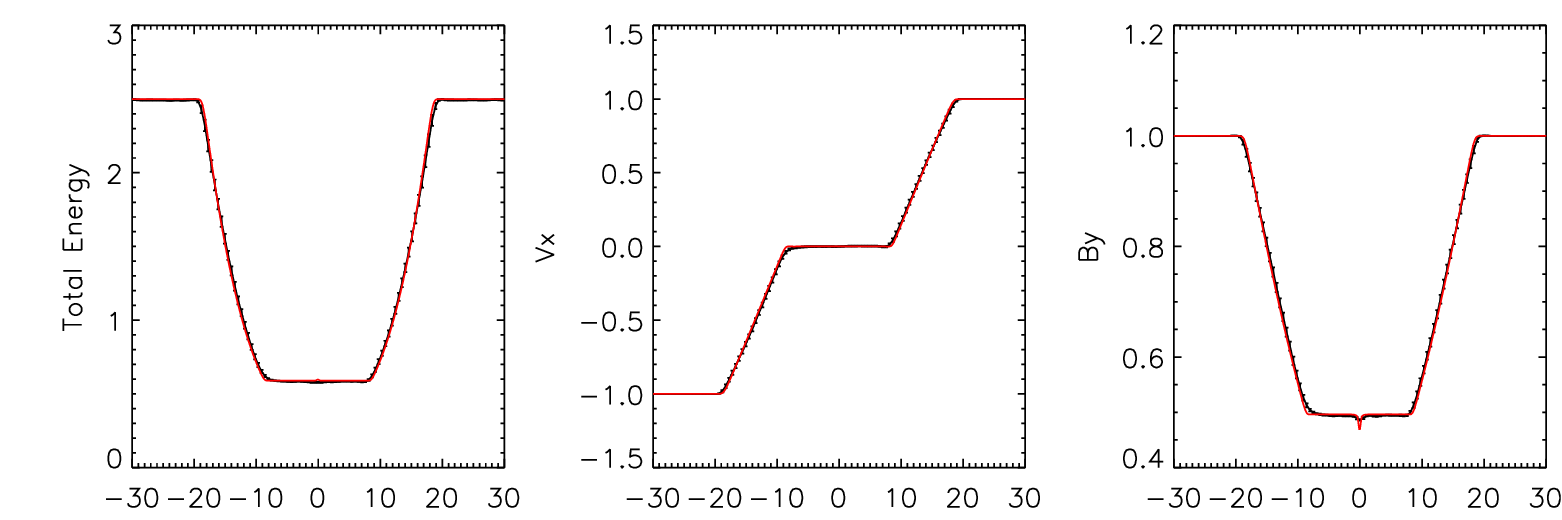}
    }
    \subfigure[Test 4A]{
    \label{fig:4A}
    \includegraphics[width=0.45\textwidth]{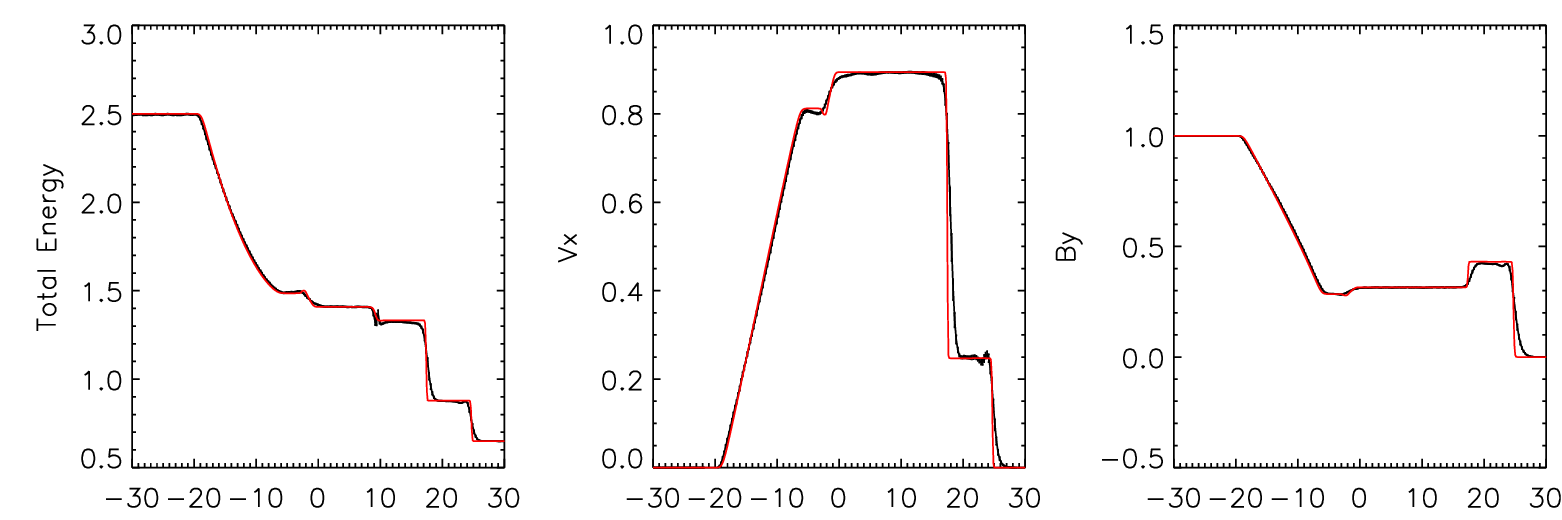}
    }
    \subfigure[Test 4B]{
    \label{fig:4B}
    \includegraphics[width=0.45\textwidth]{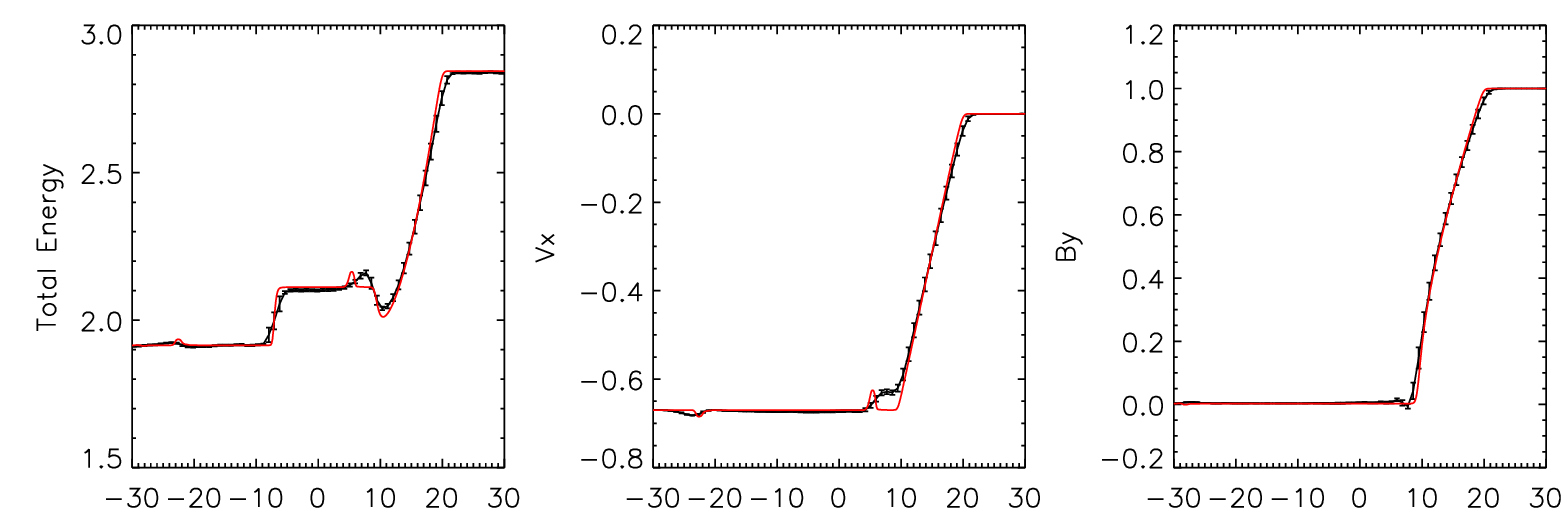}
    }
    \subfigure[Test 4C]{
    \label{fig:4C}
    \includegraphics[width=0.45\textwidth]{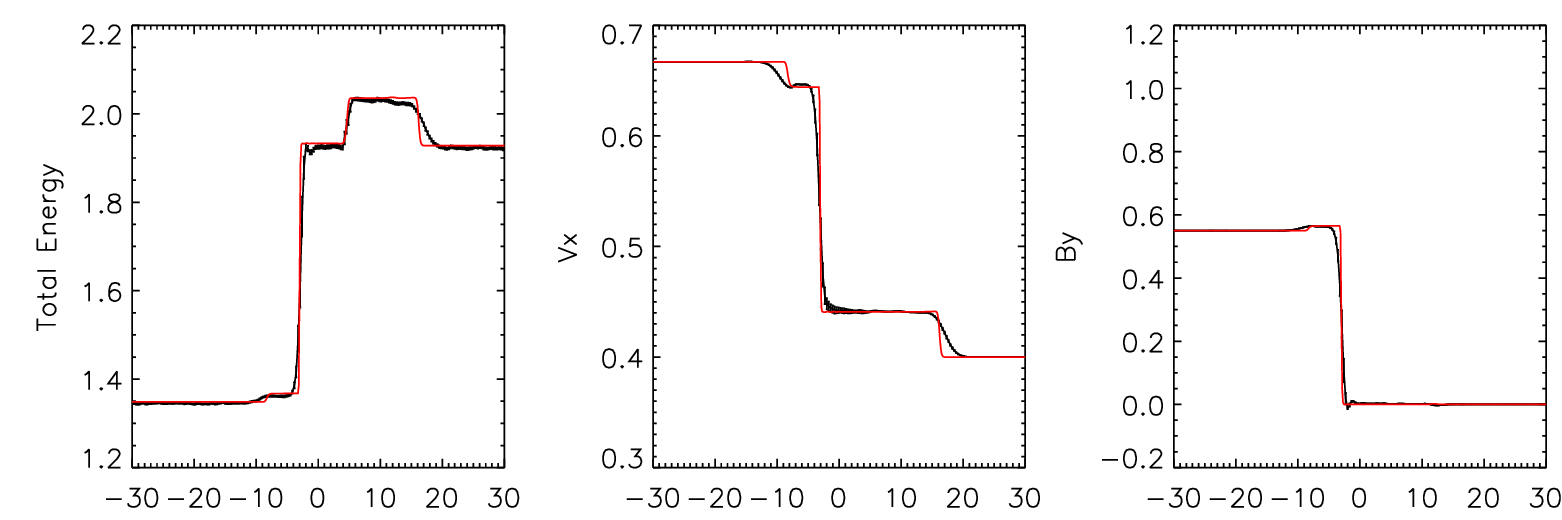}
    }
    \subfigure[Test 4D]{
    \label{fig:4D}
    \includegraphics[width=0.45\textwidth]{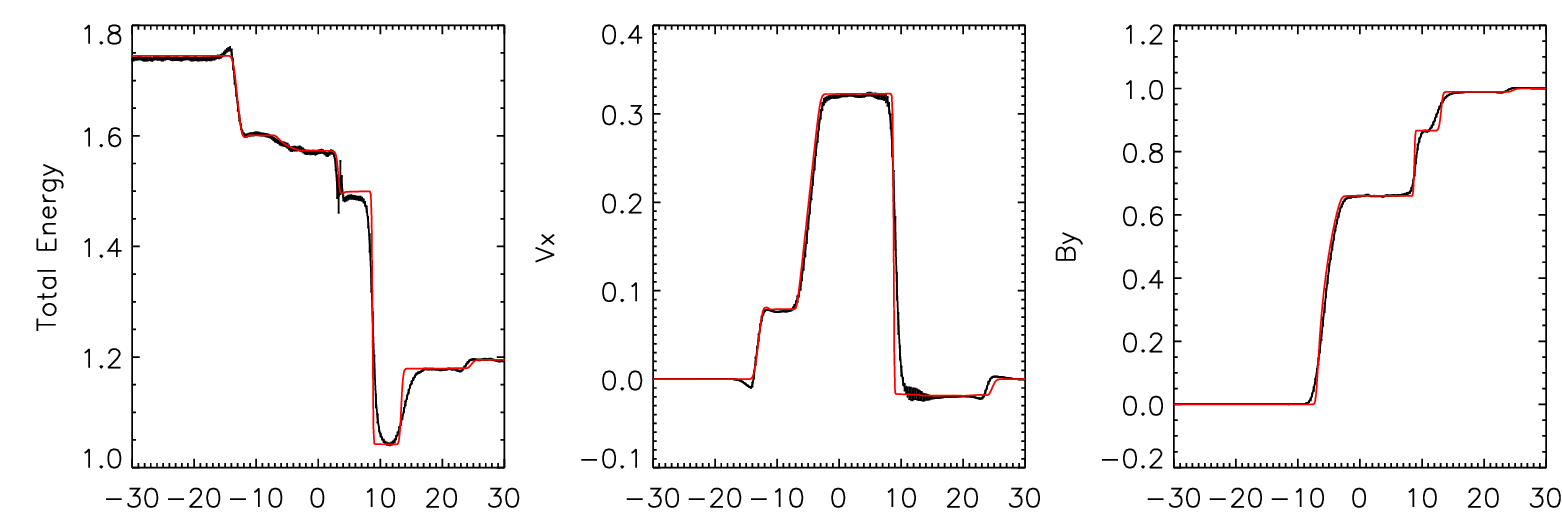}
    }
\end{center}
\caption{Representative plots of the additional 10 shock-tube tests
from \citet{1995ApJ...442..228R}. Shown for each test are the total energy
(left panels), the velocity along the $x$-direction (middle panels) and
the magnetic field along the $y$-direction (right panels). }
\label{fig:RyuJones}
\end{figure*}

In the second method, the magnetic field can be dissipated in the same way as
artificial dissipation works in the hydrodynamics. Here the numerical
parameter one has to chose is the strength of this artificial,
magnetic dissipation $\alpha_B$ in equation (\ref{eqn:induction_dis})
and (\ref{eqn:entropy_dis}). Figure \ref{fig:BrioWu_dissipation} shows the
result for the same shock-tube test as before using
$\alpha_B=0.1$. Similar to the first regularization method presented, the noise in
the individual quantities is strongly reduced and also the error in
$\mathrm{div}(\vec{B})$ is reduced by one order of magnitude. Again,
the error bars are smaller than the line width nearly
everywhere. Also, some small effects of smearing sharp features are
visible as well as some small but systematic deviations from the exact
solution. In general, this method works
slightly better than the smoothing of the magnetic field, but the
differences are generally small.

One idea to reduce the unwanted side effects of such regularization
schemes was presented in \citet{2005MNRAS.364..384P} and is based on a
modification of the artificial, magnetic dissipation constant
$\alpha_B$. Whereby every particle evolves its own numerical constant,
so that this value can decay where it is not needed and therefore
the effects of the artificial dissipation are suppressed here. 
Figure \ref{fig:BrioWu_tdd} shows the
same test as before, but this time where $\alpha_B$ evolves
for each particle, as shown in the lower right panel. Clearly,
the values are strongly reduced outside the regions associated
with sharp features (e.g. shocks), but the effect of smearing sharp
features and the small offset of some states are not significantly
reduced. This is because in the region in which these side effects
originate, the dissipation is still working with its
maximum numerical value. On the other hand, due to the suppression 
of the artificial magnetic dissipation constant outside the shock region, the 
regularization after the shock passes is nearly switched off. Therefore 
it is not as efficient as before in the post shock region, 
visible as increase in the $\mathrm{div}(\vec{B})$ error compare to the 
run with a constant, artificial magnetic dissipation.

\subsection{Shock tube problems}

As can be seen in figures \ref{fig:BrioWu_bsmooth} and
\ref{fig:BrioWu_dissipation}, the side effects of smoothing features
by the different regularization methods depend on the details of the underlying
structure of the shock-tube test. Even more interesting, the states where one can see small
deviations from the ideal solution are different for the two different
regularization methods. Therefore we performed the full set of
different shock-tube tests as presented in \citet{1995ApJ...442..228R} to test
the overall performance of the different implementations under
different circumstances. The four test families deal with different
complexities of velocity and magnetic field structures, leading to
different kinds of waves propagating. A summary of the results of these tests
can be found in figure \ref{fig:RyuJones}. Plotted are the
total energy (left panels), the velocity along the $x$-direction
(middle panels) and the magnetic field along the $y$-direction (right
panels). The red lines reflects the ideal solution obtained with
Athena, the black lines with error bars mark the results from the
SPH-MHD implementation using the magnetic field smoothing every
30th main time step. Note that the error bars in most cases are
smaller than the line width. The initial setups for the shock-tube
tests can be found in table \ref{TTable1}, which lists the state vector
of the left and right states for the different shock tube tests.

The first family of tests ({\it 1A/1B}) has no structure in the
tangential direction of the propagating shocks in magnetic field and
velocity, e.g. $B_z = v_z = 0$. As we expect, in the {\it 1A} test,
the strong shock (large jump in $v_x$) leads to some visible noise in
the magnetic field component $B_y$, also translating into significant
noise in the total energy. The regularization method here suppresses the
formation of the intermediate state in $B_y$ in the SPH-MHD
implementation, as can be seen in figure \ref{fig:1A}. The second
case, the {\it 1B} test, the weak shock is captured well. Again in
some regions some smearing of sharp features due to the regularization
method is clearly visible.

The second class of shocks ({\it 2A/2B}) involve three dimensional velocity
structures, where the plane of the magnetic field rotates. All features
(e.g. fast/slow shocks, rotational discontinuity and fast/slow
rarefaction wave, for details see \citet{1995ApJ...442..228R}), are well
captured, see figure \ref{fig:2A} and \ref{fig:2B}. Some of the
features are clearly smoothed by the regularization method.

The third class of tests ({\it 3A/3B}) shows handling of magnetosonic
structures. The first has a pair of magnetosonic shocks with zero
parallel field and the second are magnetosonic
rarefractions. Although there is slightly more noise present, all
states are captured extremely well, except the numerical feature left
at the position dividing the two states initially, see figure
\ref{fig:3A} and \ref{fig:3B}.

The fourth test family ({\it 4A/4B/4C/4D}) deals with the so-called
switch-on and switch-off structures. The tangential magnetic
field turns on in the region behind switch-on fast shocks and switch-on slow
rarefractions. Conversely, in the switch-off slow shocks and
switch-off fast rarefractions the tangential magnetic field turns off.
Again, all structures are captured well with the exception of one
feature in figure \ref{fig:4B} as well and maybe \ref{fig:4D} too, 
where clearly the regularization
leads to the washing out of a state. Otherwise the regularization leads to
smoothing of some structures similar to the tests presented before.

In general, figure \ref{fig:RyuJones} demonstrates that all these
different situations have to be included when trying to
measure the performance and quality of different implementations of
regularization methods.


\subsection{Finding optimal numerical parameters}

To optimize, we performed all these 11 shock-tube tests with various
different settings for the parameters in the regularization methods and
evaluated the quality of the result obtained with the SPH-MHD
implementation. To measure this, we used two estimators. First,
we have chosen the mean of all  $\mathrm{div}(\vec{B})$ errors within the
simulation region shown in the plots, as defined by
\begin{equation}
   \Delta_{\mathrm{div}(\vec{B})} = \left< \mathrm{div}(\vec{B})
   \frac{h}{|\vec{B}|} \right>_x.
\end{equation}
Second, we measured the discrepancy of the SPH-MHD result for the
magnetic field relative to the results obtain by Athena. Therefore
we calculate first
\begin{equation}
   \delta_{B^i}(x) = \frac{\left(\vec{B}_\mathrm{SPH}^i(x) -
   \vec{B}_\mathrm{Athena}^i(x)\right)^2}{\mathrm{RMS}_{B^i}^2(x)}
\end{equation}
for each component $i$ of the magnetic field $\vec{B}$ within each 3D
slab corresponding to the smoothing length. The RMS of $B^i$ reflects
the noise of $B_i$ within the chosen slab. We then calculate
\begin{equation}
   \hat{\Delta}_{B^i} = \left(\sum_x \delta_{B^i}(x)\right)
   \left(\sum_x \mathrm{RMS}_{B^i}^2(x)\right),
\end{equation}
for each component of the magnetic field. This includes both
contributions, the deviation of the SPH-MHD from the ideal solution as
well as the noise within each 3D slab of the SPH-MHD
implementation. To judge the improvement of the regulatization methods
we sum up all three components and further relate this measurement to
the value obtained with the {\it basic SPH-MHD} implementation, e.g.
\begin{equation}
   \Delta_B =
   \frac{\sum_i\hat\Delta_{B^i}}{\sum_i\hat\Delta_{B^i}^\mathrm{std}}
   - 1.  \label{eqn:quality}
\end{equation}
We will use these two error estimators,
$\Delta_{\mathrm{div}(\vec{B})}$ and $\Delta_B$, to measure the
quality of the individual SPH-MHD implementations.


\subsubsection{Regularization by smoothing the magnetic field}

\begin{figure*}
\begin{center}
  \includegraphics[width=0.45\textwidth]{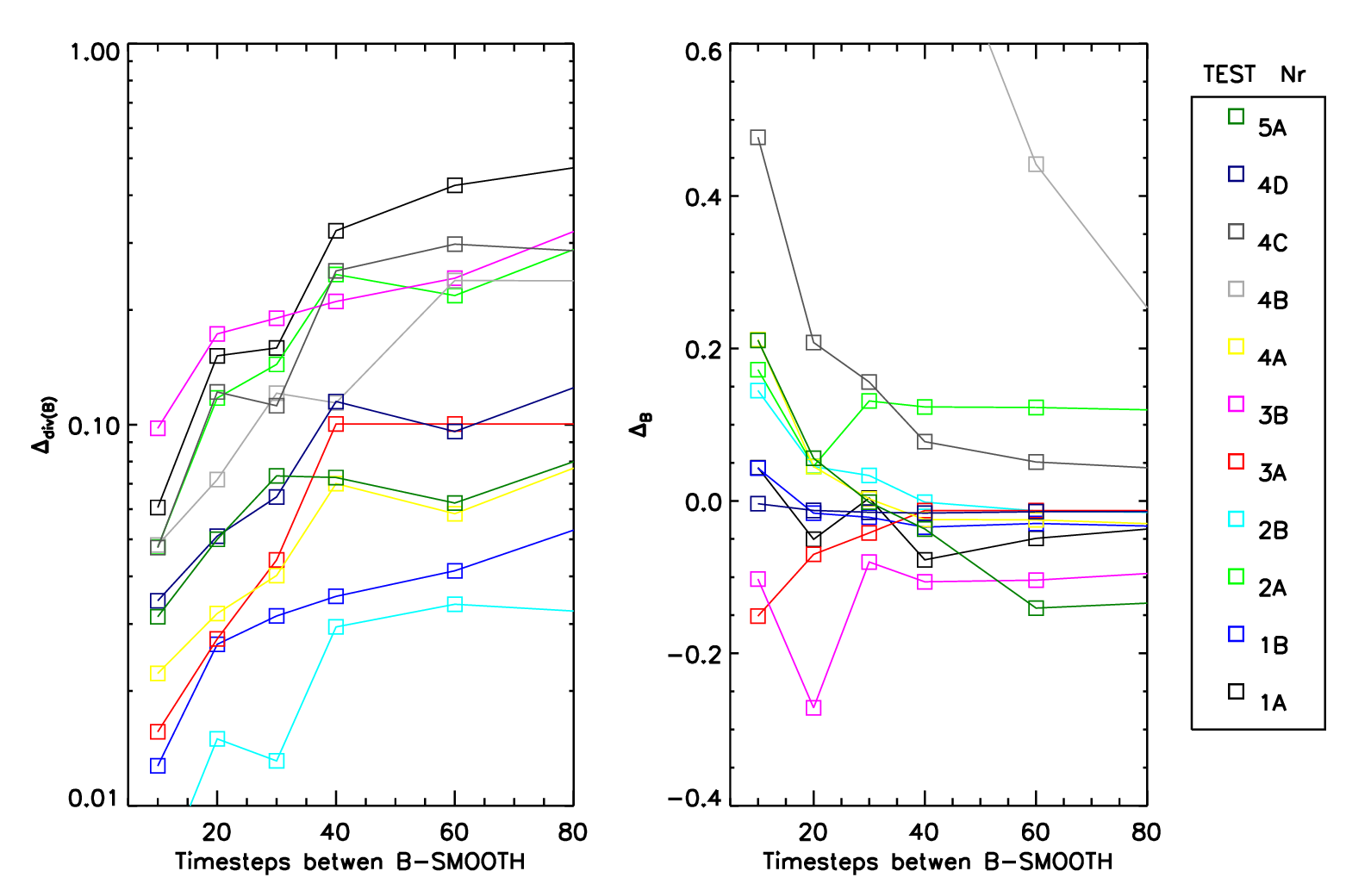}
  \includegraphics[width=0.45\textwidth]{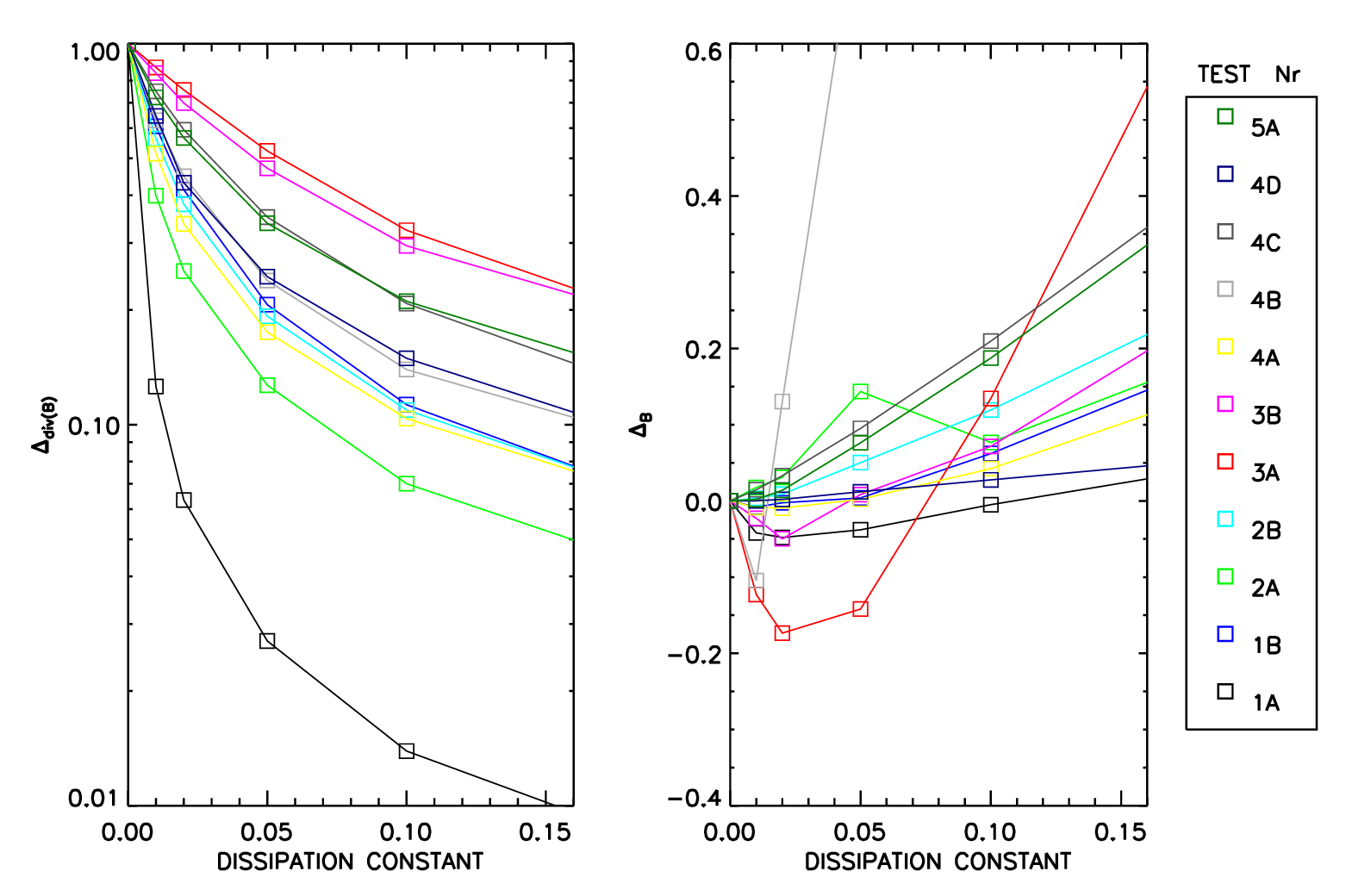}
\end{center}
\caption{Shown in the two left panels are the mean error in divergence 
(first panel on the left) and the measure of the quality (second panel from left) 
as defined in equation (\ref{eqn:quality})
obtained by the SPH-MHD implementation for different values of the
smoothing interval. The different lines are for the 11 different
shock-tube tests as indicated by the labels. The two right panels show the same 
quantities but for different values of the artificial magnetic dissipation constant $\alpha_B$.}
\label{fig:test_bsmooth}
\end{figure*}

Choosing the time interval between smoothing the magnetic field is a
compromise between reducing the noise in the magnetic field components,
as well as $\mathrm{div}(\vec{B})$) (by smoothing more often) and 
preventing sharp features from being smeared out. The left two panels of Figure
\ref{fig:test_bsmooth} show a summary of the results of the
individual shock-tube test computed with different smoothing
intervals. As expected, when using shorter smoothing intervals the error in
$\mathrm{div}(\vec{B})$ reduces. For the quality measure of the
SPH-MHD implementation the situation changes. Short smoothing
intervals generally increase the discrepancy, many of them
even to larger values than the {\it basic SPH-MHD} run. Specifically {\it 4B}
and {\it 4C} show strong deviations due to smearing of sharp
features. Note that the non monotonic behavior shown in some tests
usually relates to some residual resonances between the magnetic
waves and the smoothing intervals in the noise. Some tests show a
minimum in the differences at smoothing intervals around 20. The test
{\it 3A} seems to prefer even shorter smoothing intervals. 
In general is not clear how an optimal decision between such quality 
measure and the reduction in $\mathrm{div}(\vec{B})$ can be reached, given 
the different nature and amplitude of the two measures. However, ignoring
{\it 4B} which strongly suffers from smearing sharp features when smoothing the 
magnetic field, a good compromise seems to be for values around 20-30, where 
pro and con in the quality measure are small and compensating within the 
different tests but $\mathrm{div}(\vec{B})$ is still drastically reduced 
in all tests. We will 
refer to this as the {\it Bsmooth SPH-MHD} implementation in the rest of the paper.

\begin{figure*}
\begin{center}
  \includegraphics[width=0.45\textwidth]{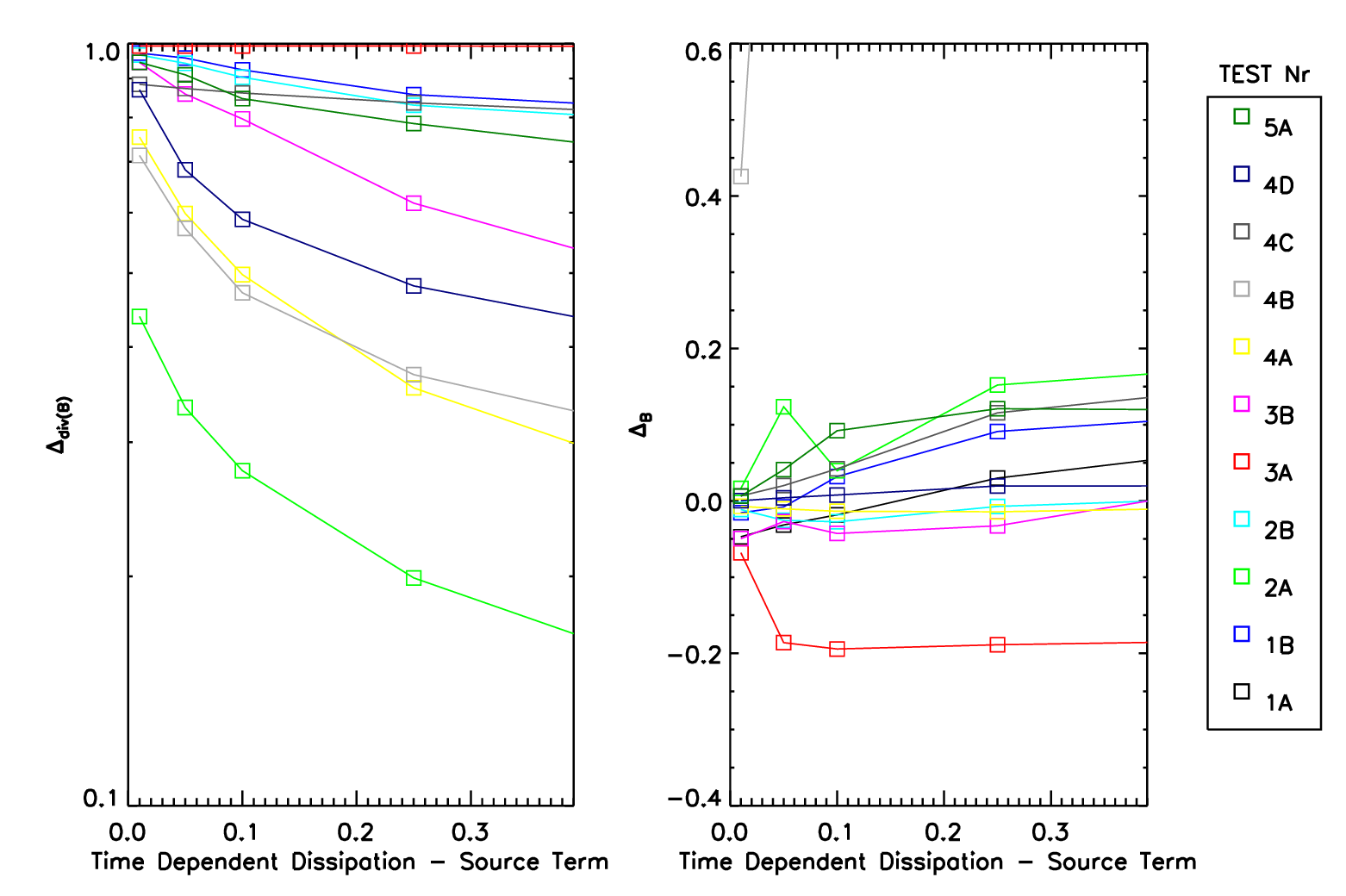}
  \includegraphics[width=0.45\textwidth]{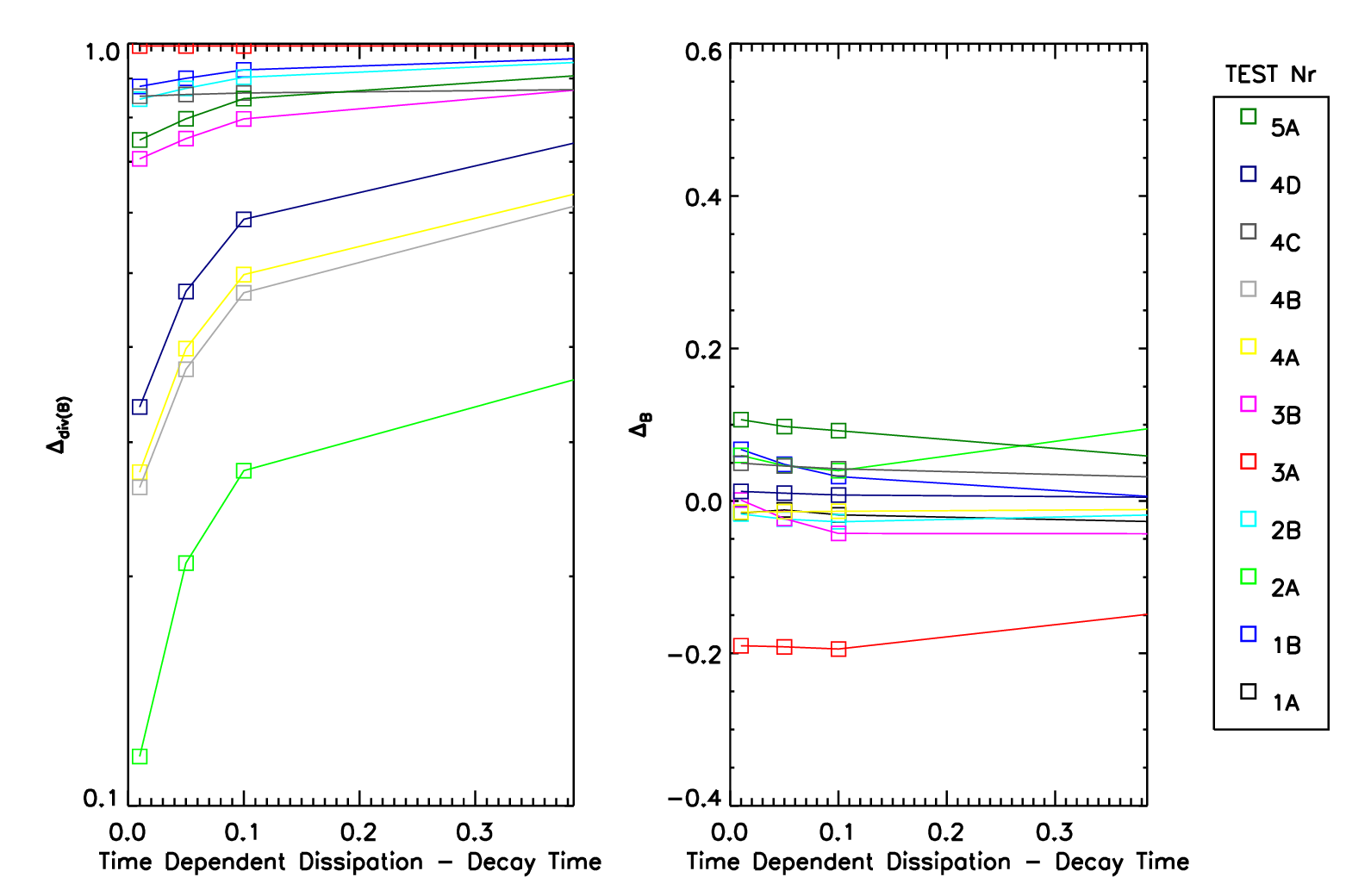}
\end{center}
\caption{Similar to figure \ref{fig:test_bsmooth} but for different
values of the source term $S_0$ (left panels) and the decay term $C$
(right panel) of the time dependent, artificial
magnetic dissipation.}
\label{fig:test_timedependent_dissipation}
\end{figure*}

\subsubsection{Artificial dissipation}

As before, choosing the value for the artificial magnetic dissipation
constant $\alpha_B$ is a compromise between reducing the noise in the 
magnetic field components
(as well as reducing $\mathrm{div}(\vec{B})$) and preventing sharp
features from smearing out due to the effect of the dissipation. 
The right two panels of Figure
\ref{fig:test_bsmooth} show a summary of the results of the
individual shock-tube tests computed with different values for the
artificial magnetic dissipation. As expected, using larger
values reduces the error in
$\mathrm{div}(\vec{B})$ significantly. Similar to before,
using larger values also generally results in an increase of the
discrepancy between the SPH MHD implementation and the true solution,
again usually to even larger values than in the {\it
basic SPH-MHD} run. As before, especially the shock-tube test {\it 4B}
and {\it 4C} show strong deviations due to smearing of sharp
features. Note that here less non-monotonic behavior is visible
(except for test {\it 4B}). The main reason is that
dissipation is a continuous process, so resonances between dissipation
and the magnetic waves cannot be very pronounced. As before, 
it is difficult to infer the best choice of parameter for all the tests.
Again, once ignoring {\it 4B}, a compromise for choosing $\alpha_B$ 
seems to be between 0.02 and 0.1. Choosing $\alpha_B$ close to the upper 
value of 0.1 might lead to a significant reduction in $\mathrm{div}(\vec{B})$ 
without to strong signature from smearing out sharp features. 
We will refer to this as the {\it dissipation SPH-MHD} implementation 
in the rest of the paper.


\subsubsection{Time dependent artificial dissipation}

One idea to reduce the effect of the artificial dissipation is to make
the artificial magnetic dissipation constant $\alpha_B$ time
dependent. The idea here is that, if the evolution of $\alpha_B$ is
properly controlled, dissipation will happen only at the places where
it is needed and it will be suppressed in all other parts of the simulation
volume. The evolution of $\alpha_B$ is controlled by the two parameters
$S_0$ (source term) and $C$ (decay term) where we have chosen
$\alpha_B^\mathrm{min}$ and $\alpha_B^\mathrm{max}$ as 0.01 and 0.5
respectively. Figure \ref{fig:test_timedependent_dissipation} shows
the result for varying these two parameters. As before, generally, the larger the
dissipation is (e.g. large source term or small decay time)
the smaller the noise and the error in  $\mathrm{div}(\vec{B})$
becomes. However, as soon as these parameters have values which drive
$\alpha_B$ in the shocks to the maximum allowed value, there is
marginally no gain in quality, although the values for $\alpha_B$
outside the shocks can still be quite small. Therefore, the time
dependent method does not improve the results significantly, as the regions in
which the artificial dissipation constant is suppressed do not
significantly contribute to the smearing of sharp features. 

\begin{figure*}
  \begin{center}
    \includegraphics[width=0.45\textwidth]{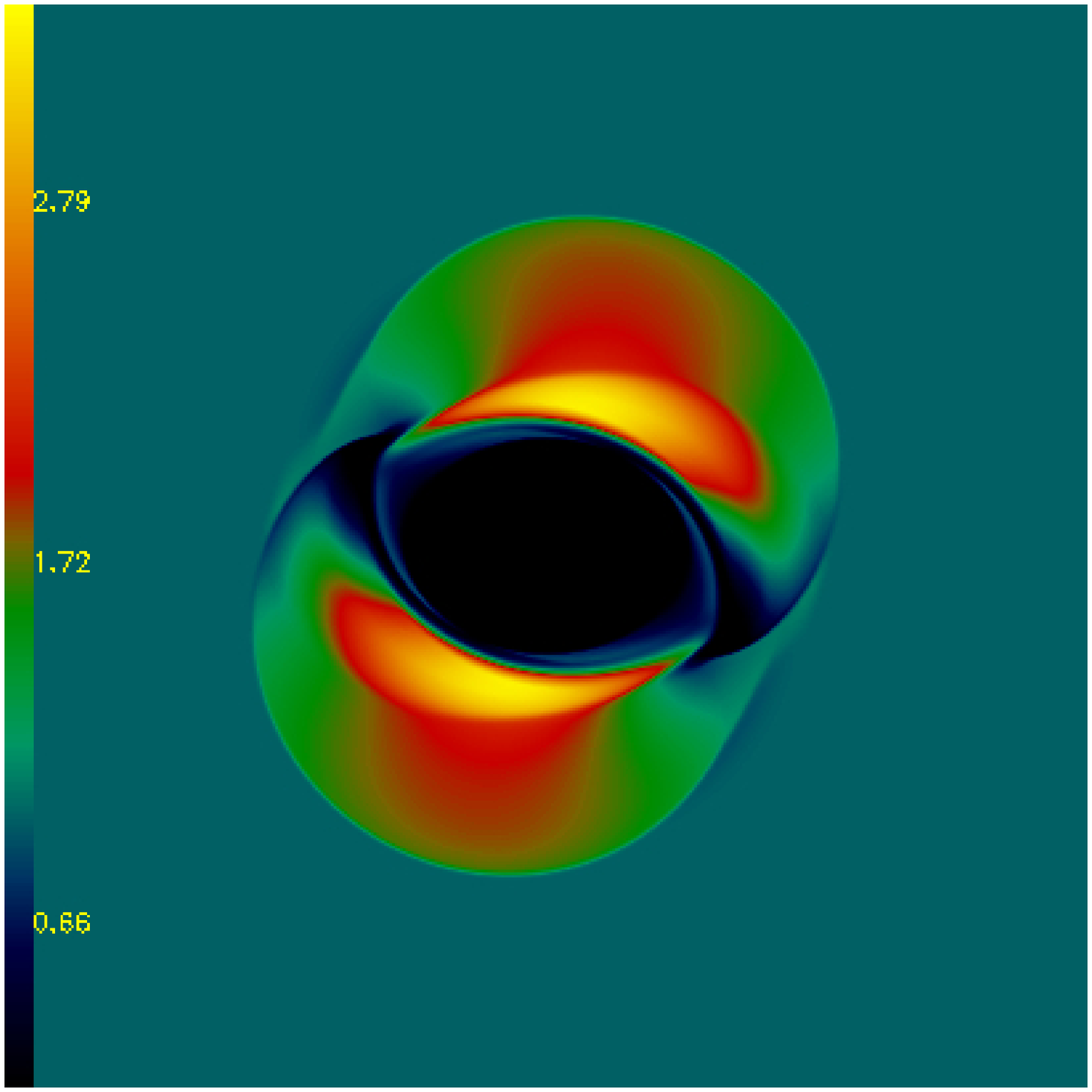}
    \includegraphics[width=0.45\textwidth]{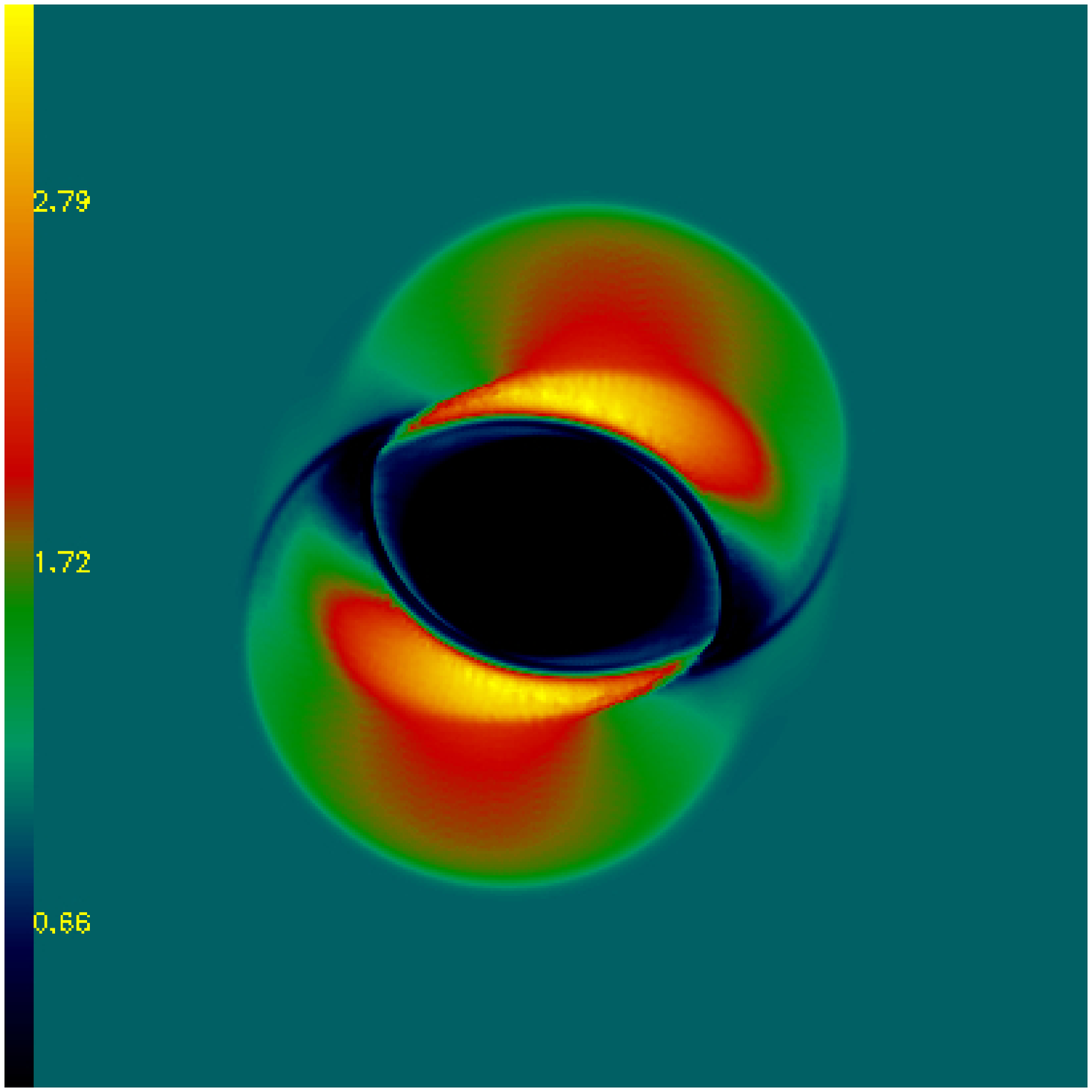} \\
    \includegraphics[width=0.45\textwidth]{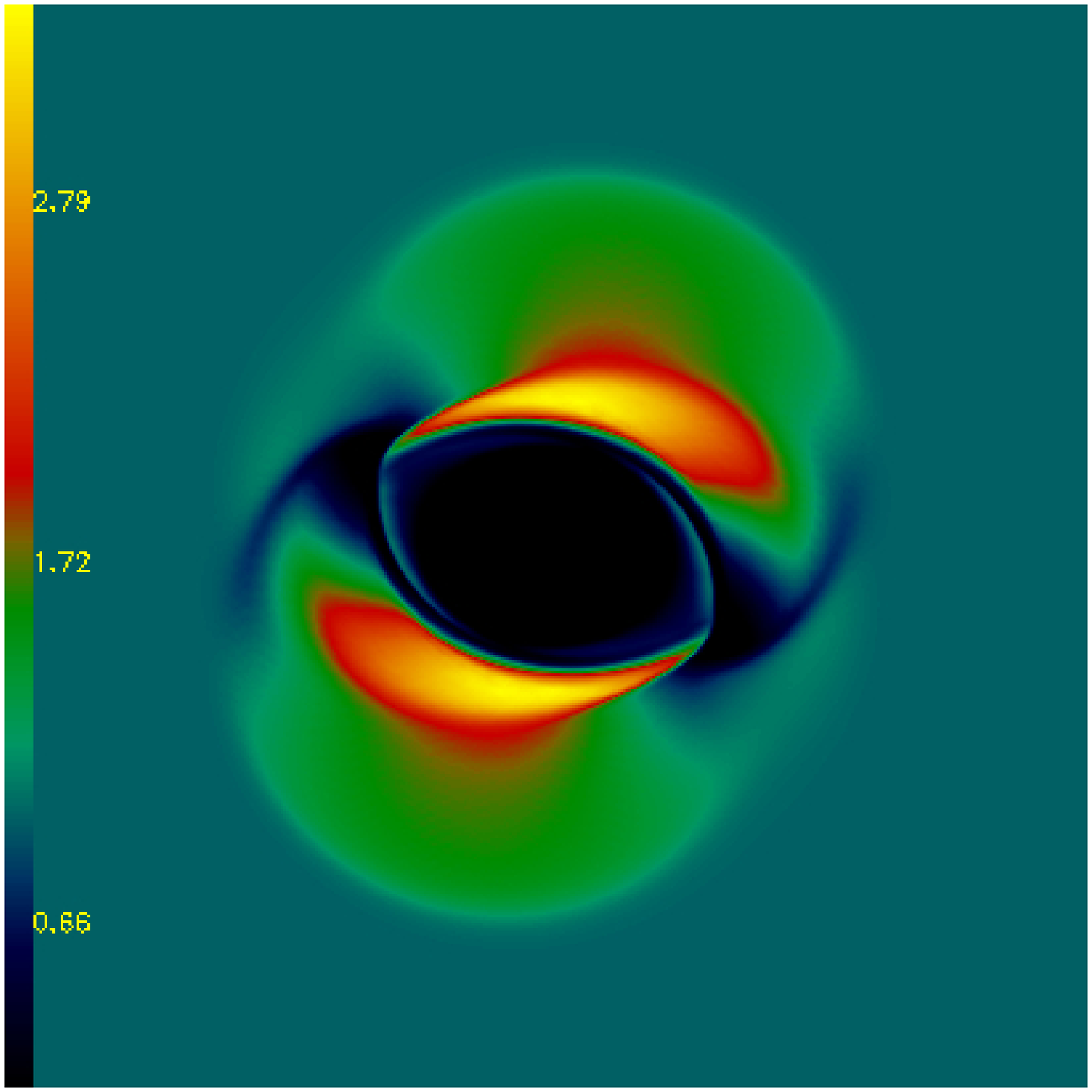}
    \includegraphics[width=0.45\textwidth]{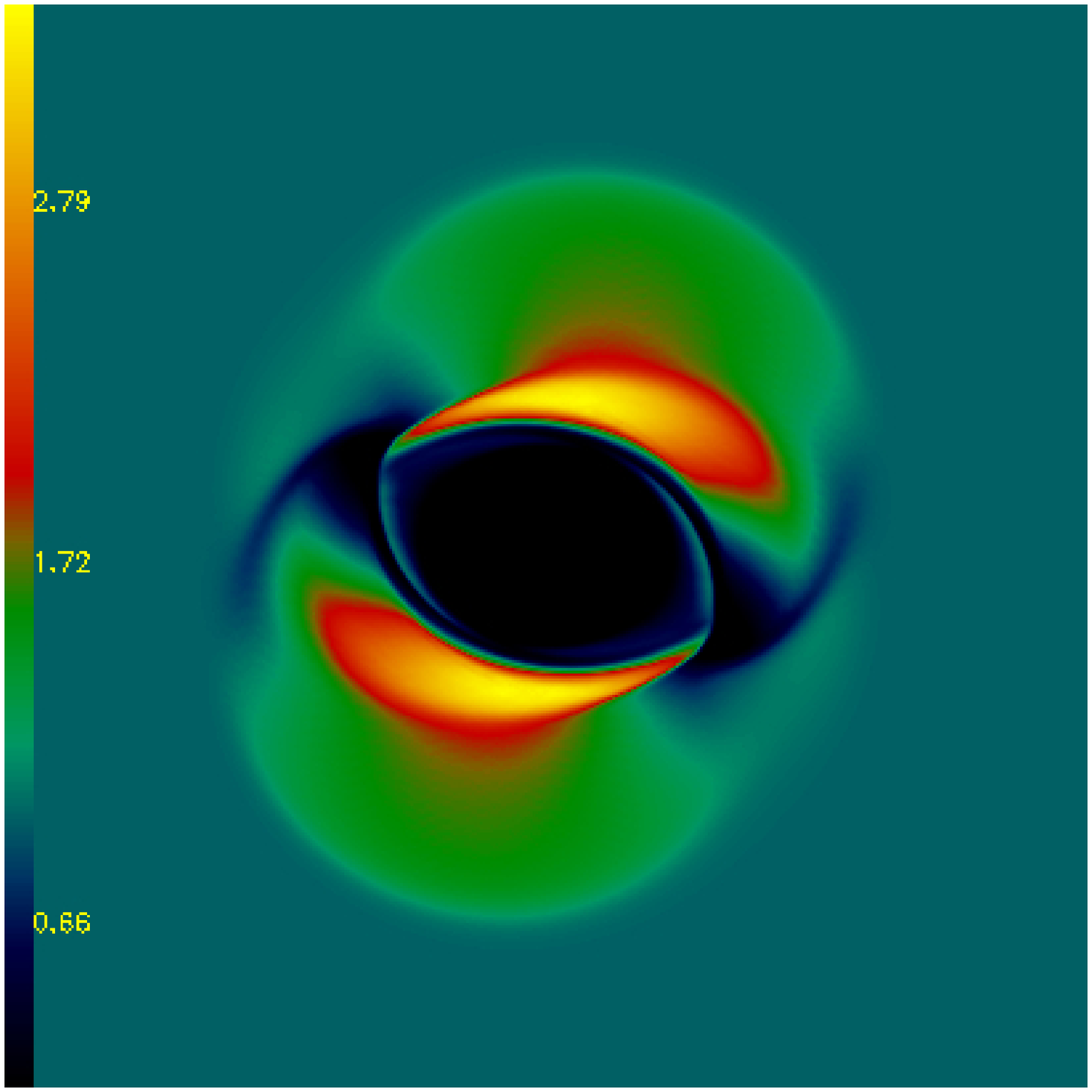}
    \end{center} \caption{The magnetic pressure ($B^2/2$) in
    the {\it Fast Rotor} test at $t=0.1$. The ATHENA solution of the test
    problem is shown in the upper left panel. The upper right panel
    shows the results obtained with the {\it basic SPH-MHD} implementation.
    The lower left and right panels shows the result obtained with the 
    {\it Bsmooth SPH-MHD} and the {\it dissipation SPH-MHD} implementation 
    respectively. All the main features are well reproduced in the GADGET runs. 
    The shape, positions and amplitudes
    correspond quite well, although the GADGET runs appear slightly more
    smoothed, depending on the regularization scheme used 
    (see also Figure \ref{fig:RotorCut}).}  \label{fig:Rotor}
\end{figure*}

\begin{figure}
  \begin{center}
    \includegraphics[width=0.45\textwidth]{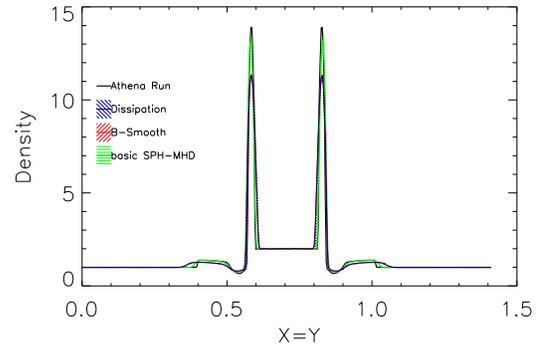}
    \end{center} \caption{Diagonal ($x=y$) cut through the {\it Fast Rotor}
    at $t=0.1$ showing the density. In black result obtained with
    ATHENA. The pink line with the red error bars show the GADGET
    solution using the {\it basic SPH-MHD} implementation. Also here
    the red error bars reflect the dispersion of
    the values among individual particles within a slab corresponding
    to the local smoothing length. In general, the SPH-MHD result
    shows an excellent agreement in all the features (peaks, valleys
    and edges), however there is a visible over-smoothing at the outer
    edges in the GADGET result.}
    \label{fig:RotorCut}
\end{figure}

\subsection{Multi dimensional Tests - Planar Tests}

Besides the one dimensional shock tube test described in the previous
section, two dimensional (e.g. planar) test problems are a good test-bed
to check code performance. Such higher dimensional tests include additional
interaction between the evolving components with non-trivial
solution. These can be quite complex (with several classes of waves
propagating in several directions) such as the Orszang-Tang Vortex or simple
(but with strong MHD discontinuities) such as Strong Blast or Fast Rotor.

\subsubsection{Fast Rotor}
This test problem was introduced by \citet{Balsara99}, to study
star formation scenarios, in particular the strong torsional Alfv\'en
waves, and is also commonly used to validate MHD implementations
\citep[for example see][]{Toth,2000ApJ...530..508L,2005MNRAS.364..384P,2006ApJ...652.1306B}.
The test consists of a fast rotating dense disk embedded in a
low density, static and uniform media, with a initial constant magnetic
field along the x-direction (e.g. $B_x=2.5\pi^{-1/2}$). In the
initial conditions, the disk with radius $r=0.1$, density $\rho=10$
and pressure $P=1$ is spinning with an angular velocity
$\omega=20$. It is embedded in a uniform background with $\rho = P
=1$. Again we setup the initial conditions by distributing the particles on a
glass like distribution in 3D, using $700\times700\times5$ particles
and periodic boundaries in all directions for the background
particles. The disk is created by removing all particles which fall
inside the radius of the disk and replacing this space with a denser
representation of particles of the same mass. 
As an ideal solution to compare with, we again used the result of a
simple, two dimensional ATHENA run with $400\times400$ cells. A visual 
impression of the results can be obtained from the maps presented in figure 
\ref{fig:Rotor}.

Figure \ref{fig:RotorCut} presents another quantitative
comparison. Shown is a diagonal cut through the {\it Fast Rotor} at $t=0.1$
showing the density. The different lines show the result
obtained with ATHENA (black line) and for the three different SPH-MHD 
implementations in GADGET (colored lines). The very small, red error bars
reflect the RMS of the values held by the individual particles within
the 3D slab through the three dimensional simulations corresponding to
the local smoothing length. The results show remarkable agreement
between the simulations and also compare well with results
quoted in the literature \citep[e.g.][]{2000ApJ...530..508L}. Here
the smoothing of sharp features in the two implementations with 
regularization is quite clear visible and leads to a less good match than the
{\it basic SPH-MHD} implementation.

Note that although we perform our calculations in three dimensions
and without a regularization scheme, the implementation produce a
result, which has the same quality as other schemes in two dimensions
with regularization \citep[e.g.][]{2005MNRAS.364..384P,2006ApJ...652.1306B}.

\begin{figure*}
  \begin{center}
    \includegraphics[width=0.45\textwidth]{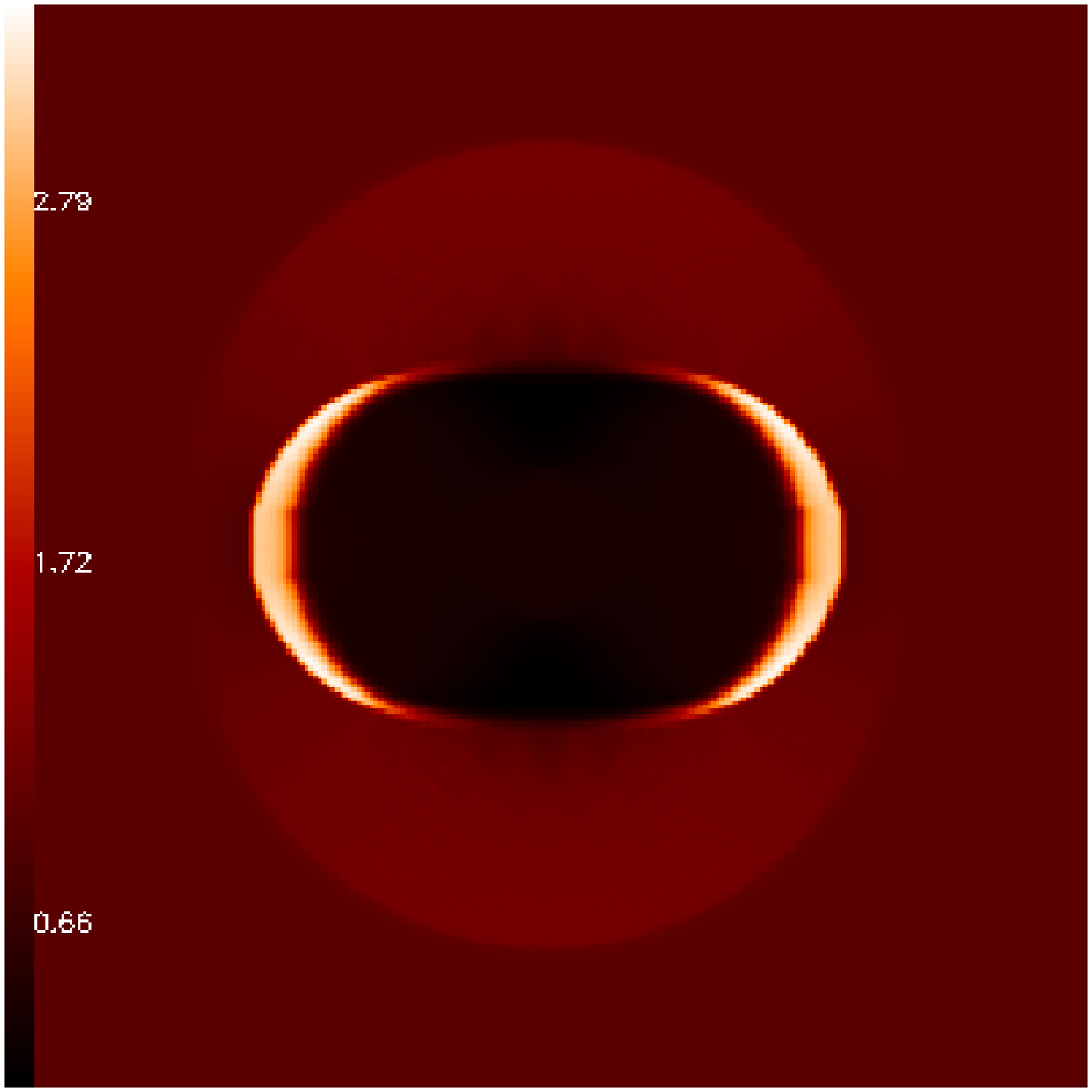}
    \includegraphics[width=0.45\textwidth]{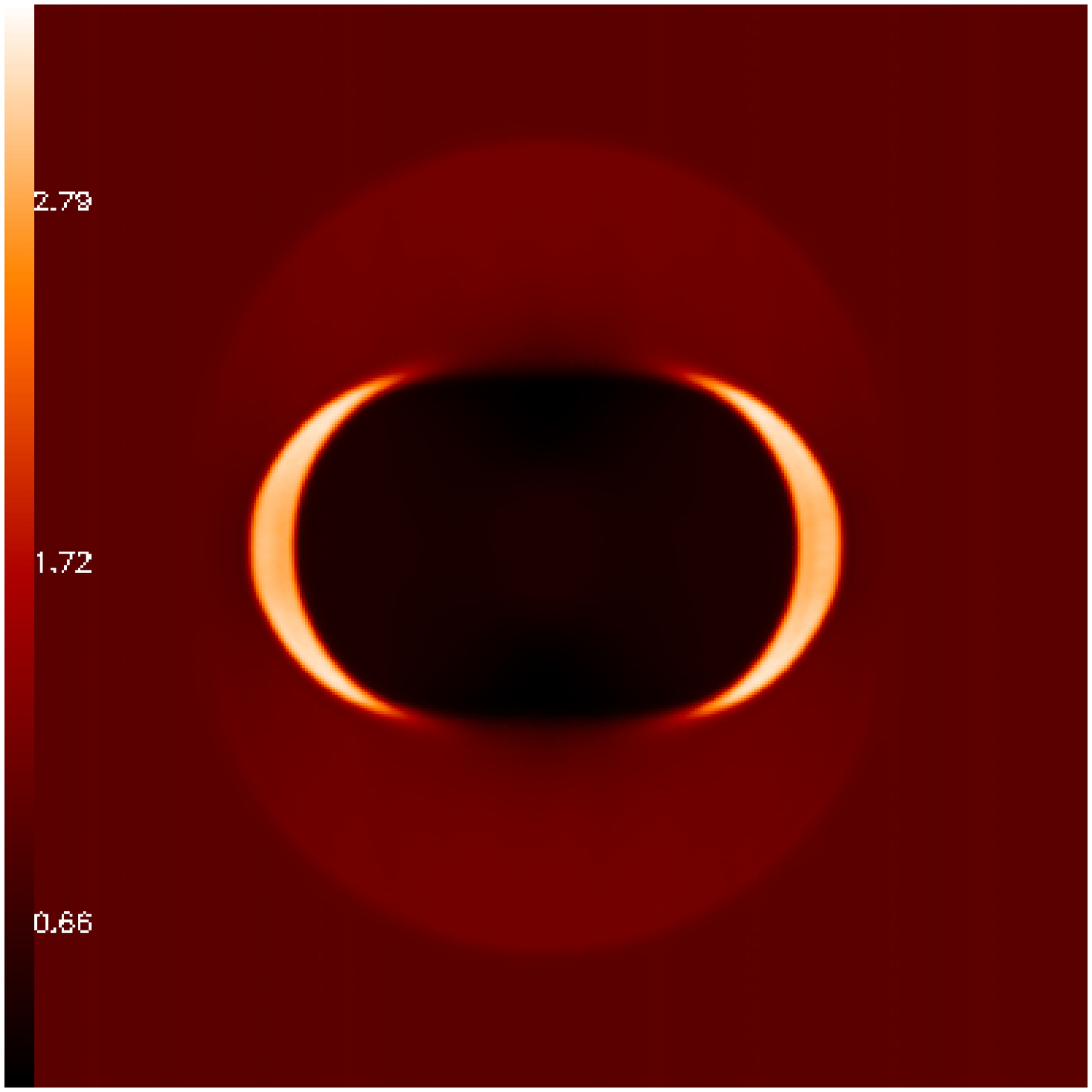}\\
    \includegraphics[width=0.45\textwidth]{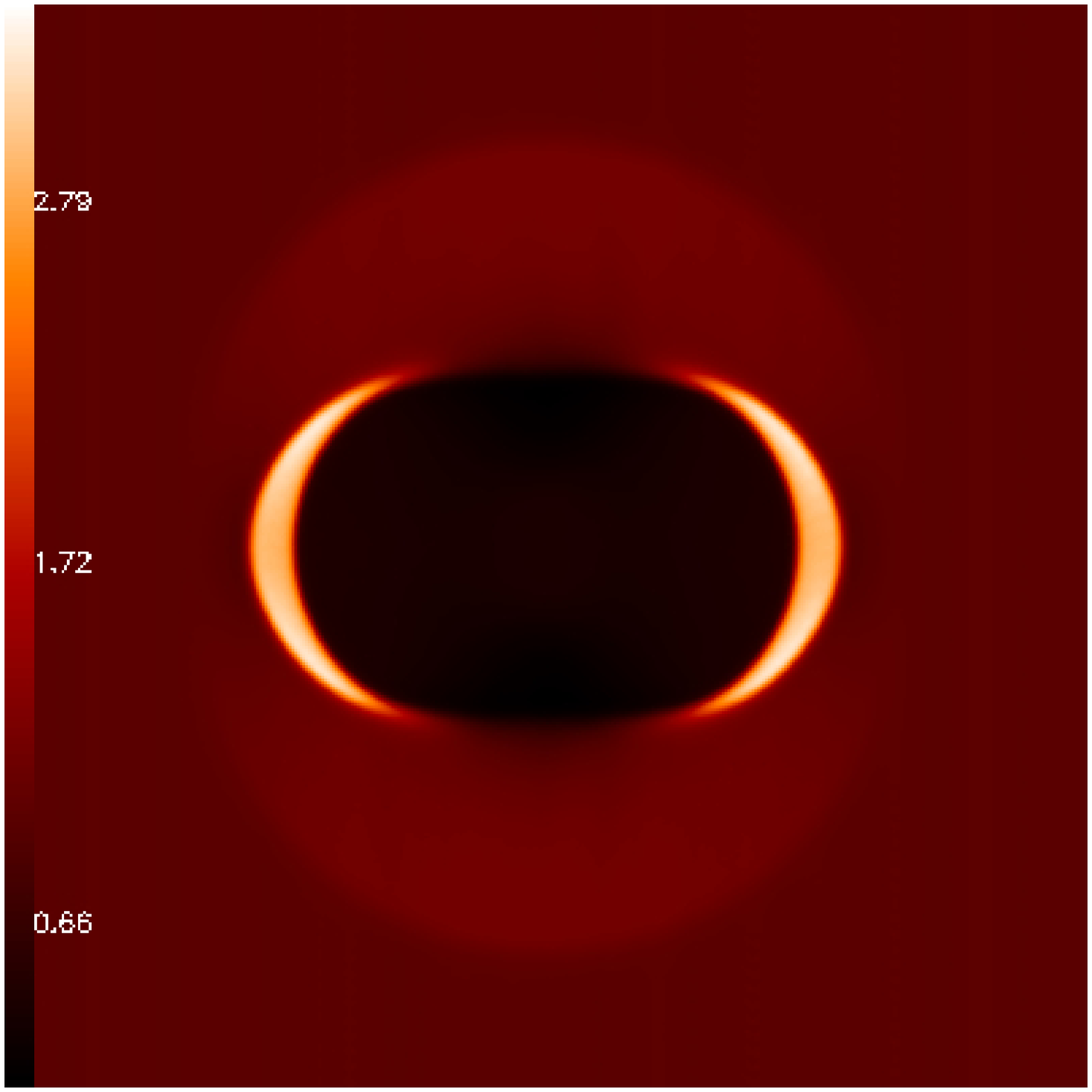}
    \includegraphics[width=0.45\textwidth]{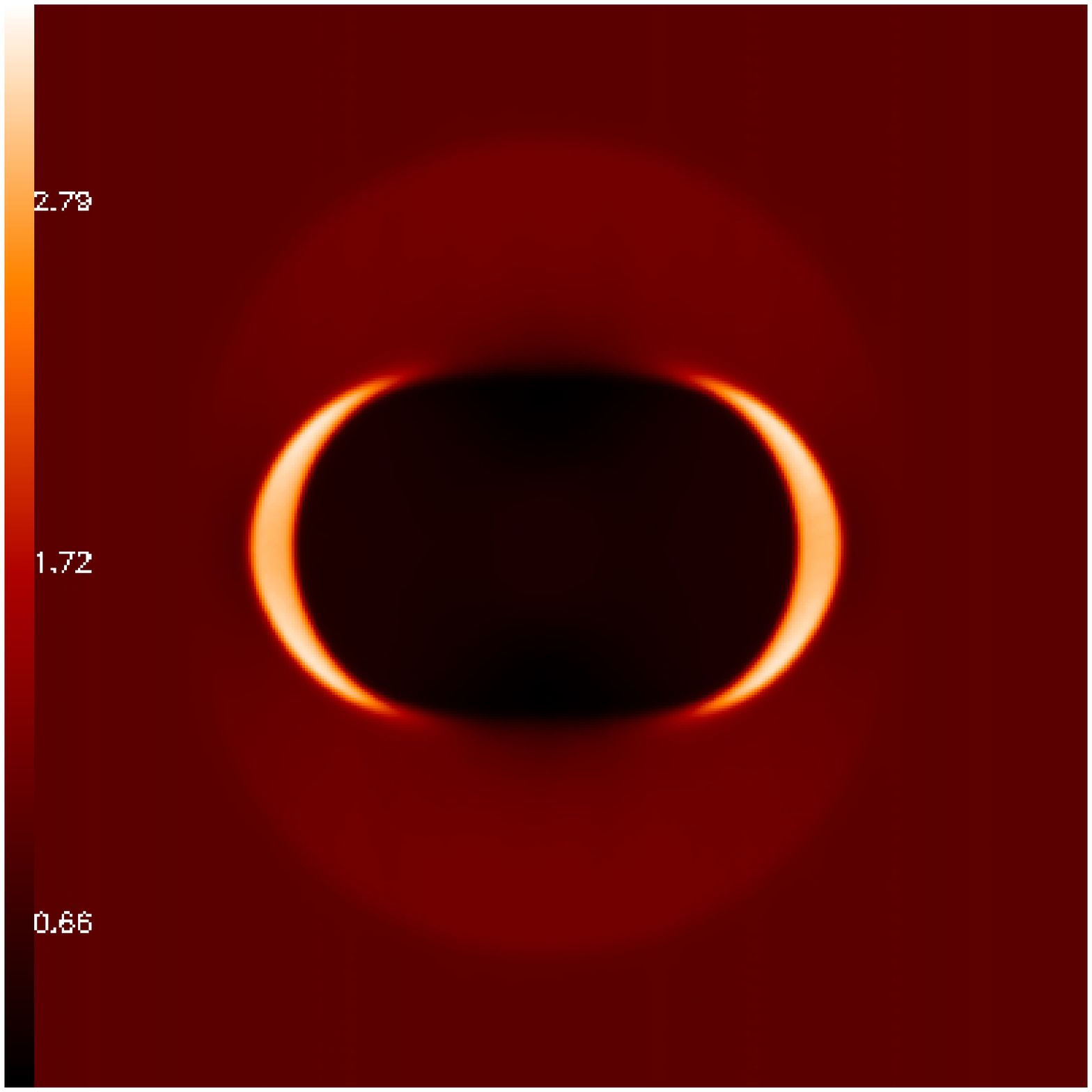}
    \end{center} \caption{Shown is the resulting density distribution
    for the {\it Strong Blast} test at $t=0.02$. The upper left panel 
    shows the result obtained with ATHENA, whereas the upper right panel
    shows the that from the {\it basic SPH-MHD} implementation.
    The lower left and right panels shows the results obtained with the 
    {\it Bsmooth SPH-MHD} and the {\it dissipation SPH-MHD} implementation 
    respectively. There are
    no visible difference between the results (see also figure
    \ref{fig:BlastCut}), indicating that all SPH-MHD implementations are 
    capable to handle such test situation.}
    \label{fig:Blast}
\end{figure*}

\begin{figure}
  \begin{center}
    \includegraphics[width=0.45\textwidth]{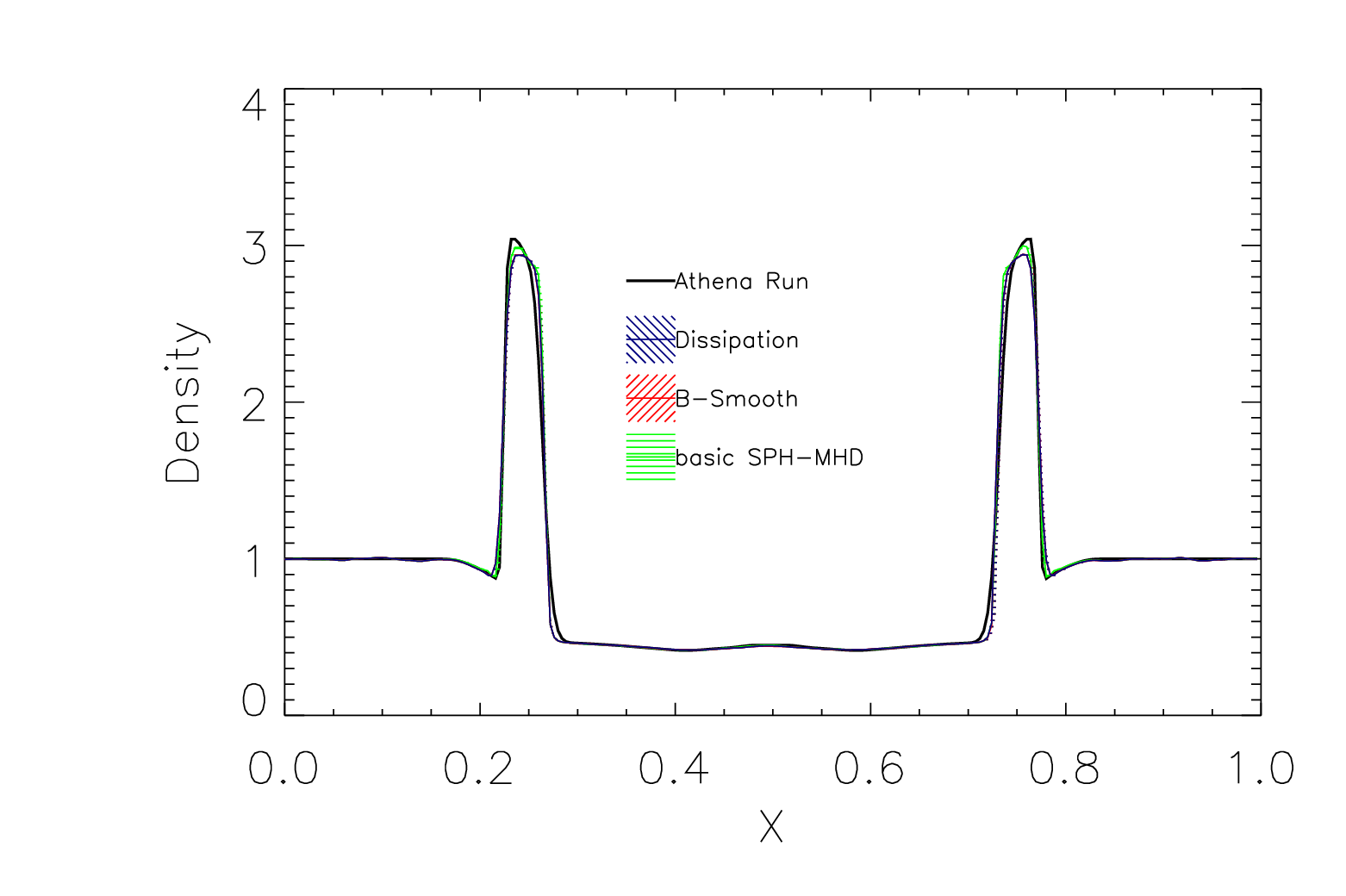}
    \end{center} \caption{Horizontal cut through the {\it Strong Blast} test
    ($x=[0.0;1.0],y=0.5$) showing the density. The black line is from
    the ATHENA simulation, the pink line with the red error bars (see
    figure \ref{fig:RotorCut}) reflects the GADGET result. The overall
    behavior is excellent, with only very small differences between the
    two solutions.}
    \label{fig:BlastCut}
\end{figure}

\subsubsection{Strong Blast}

The {\it Strong Blast} test consists of the explosion of a
circular hot gas in a static magnetized medium and is also regularly
used for MHD code validation \citep[see for example][]{2000ApJ...530..508L,Balsara99}.
The initial conditions consist of a constant density
$\rho=1$ where a hot disk of radius $r_0=0.125$ is embedded, which is
a hundred times over-pressured, e.g the pressure in the disk is set to
$P_d=100$ whereas the pressure outside the disk is set to
$P_o=1$. In addition there is initially an overall homogeneous
magnetic field in the $x$-direction, with a strength of $B_x=10$.
The system is evolved until time $t=0.02$ and an outgoing shock wave is
visible which, due to the presence of the magnetic field, is no longer
spherical but propagates preferentially along the field lines. Figure
\ref{fig:Blast} shows the density at the final time, comparing the ATHENA
results with the results from the three different SPH-MHD
implementations in GADGET. Although the setup is a strong blast wave,
there is no visible difference of the SPH-MHD implementation with the
ATHENA results. This is quantitatively confirmed in figure
\ref{fig:BlastCut} which shows a horizontal cut (at $y=0.5$) of the
density through the {\it Strong Blast} test, comparing the ATHENA
(black line) with the GADGET (colored lines with error bars)
results. Besides very small variations there is no significant
difference between the two results and all features are well
reproduced by the SPH-MHD implementation. Note that the error bars
of the GADGET results again are almost in all cases smaller than the
shown line width.

\subsubsection{Orszang-Tang Vortex}

\begin{figure*}
  \begin{center}
    \includegraphics[width=0.45\textwidth]{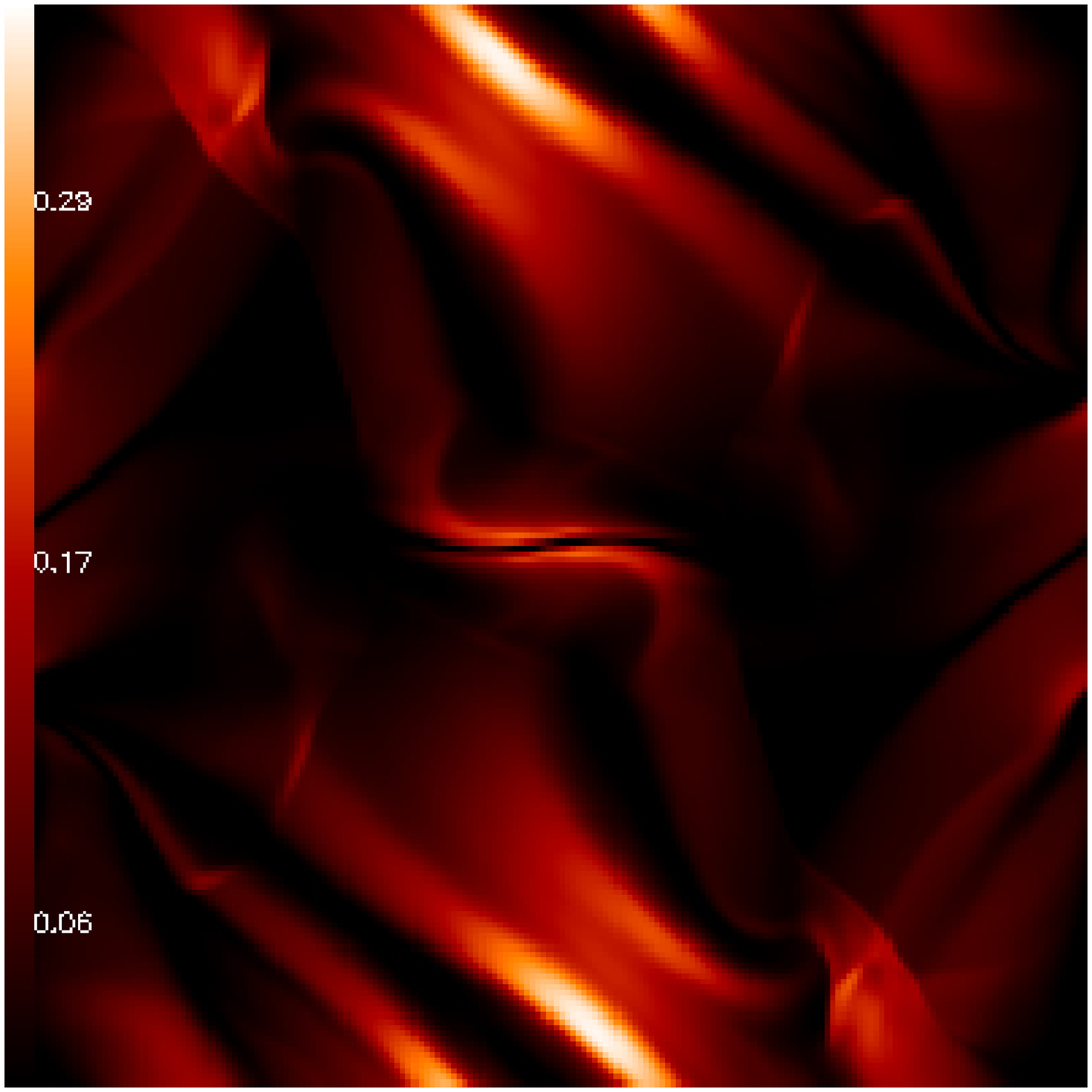}
    \includegraphics[width=0.45\textwidth]{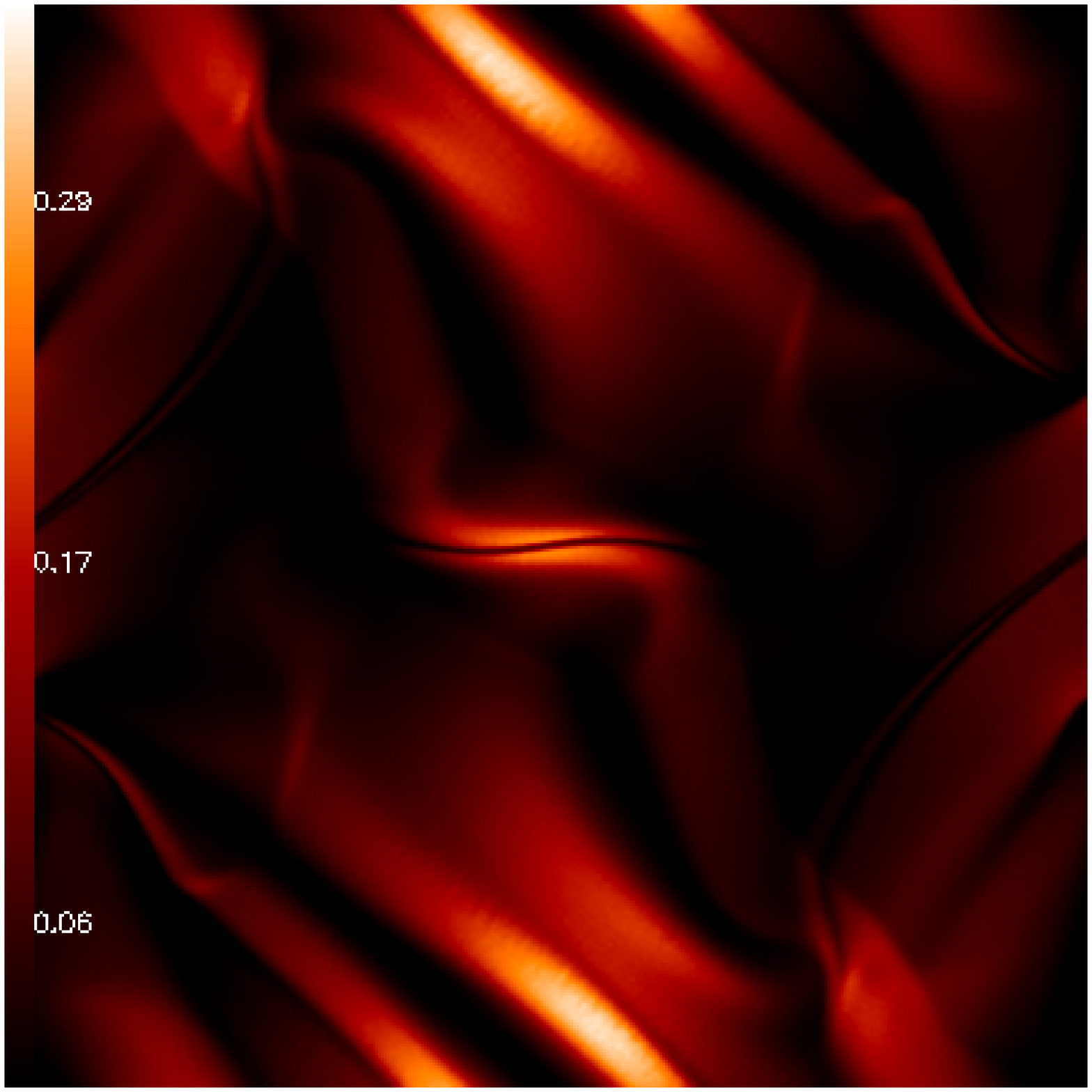}\\
    \includegraphics[width=0.45\textwidth]{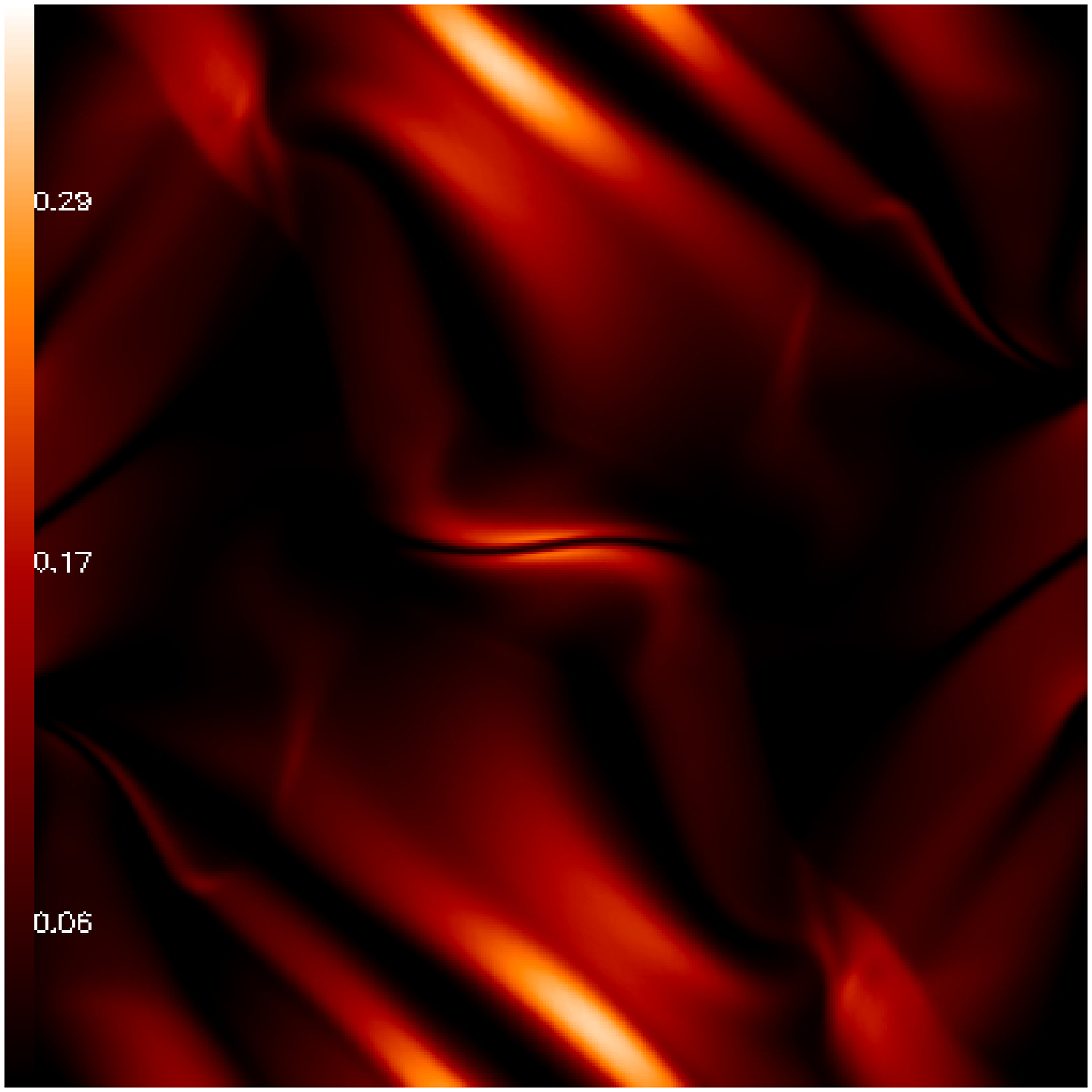}
    \includegraphics[width=0.45\textwidth]{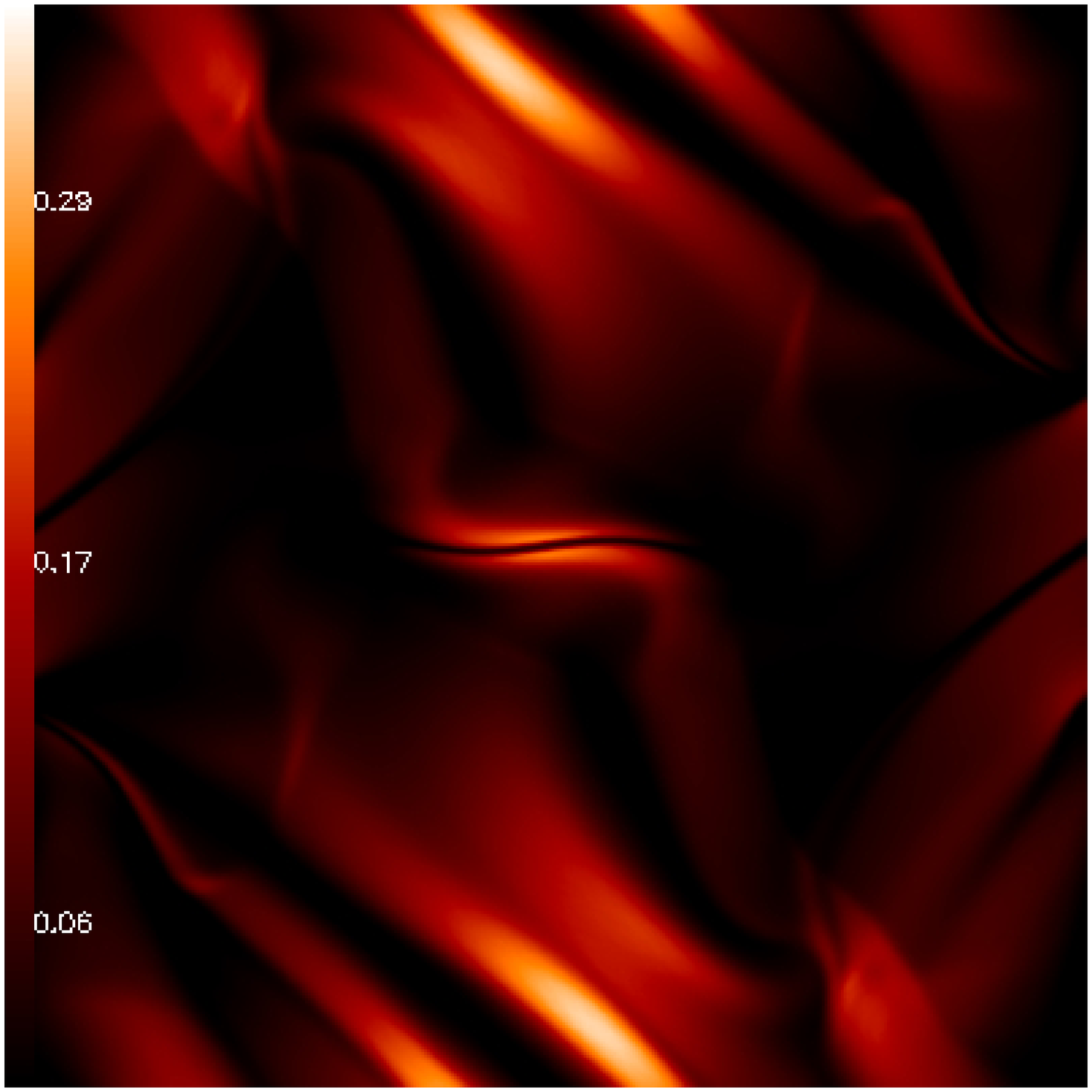}
    \end{center} \caption{The magnetic pressure $B^2/2$
    distribution in the
    {\it Orszang-Tang Vortex} at $t=0.5$. The upper left panel 
    shows the result obtained with ATHENA, while the upper right panel
    shows those from the {\it basic SPH-MHD} implementation.
    The lower left and right panels shows the results obtained with the 
    {\it Bsmooth SPH-MHD} and the {\it dissipation SPH-MHD} implementations 
    respectively. Some of the sharp features are more smoothed in the GADGET
    runs, depending on the choice of regularization scheme, but overall the results 
    compare very well (see also figure \ref{fig:VortexCut})
}  \label{fig:Vortex}
\end{figure*}

\begin{figure*}
  \begin{center}
    \includegraphics[width=0.45\textwidth]{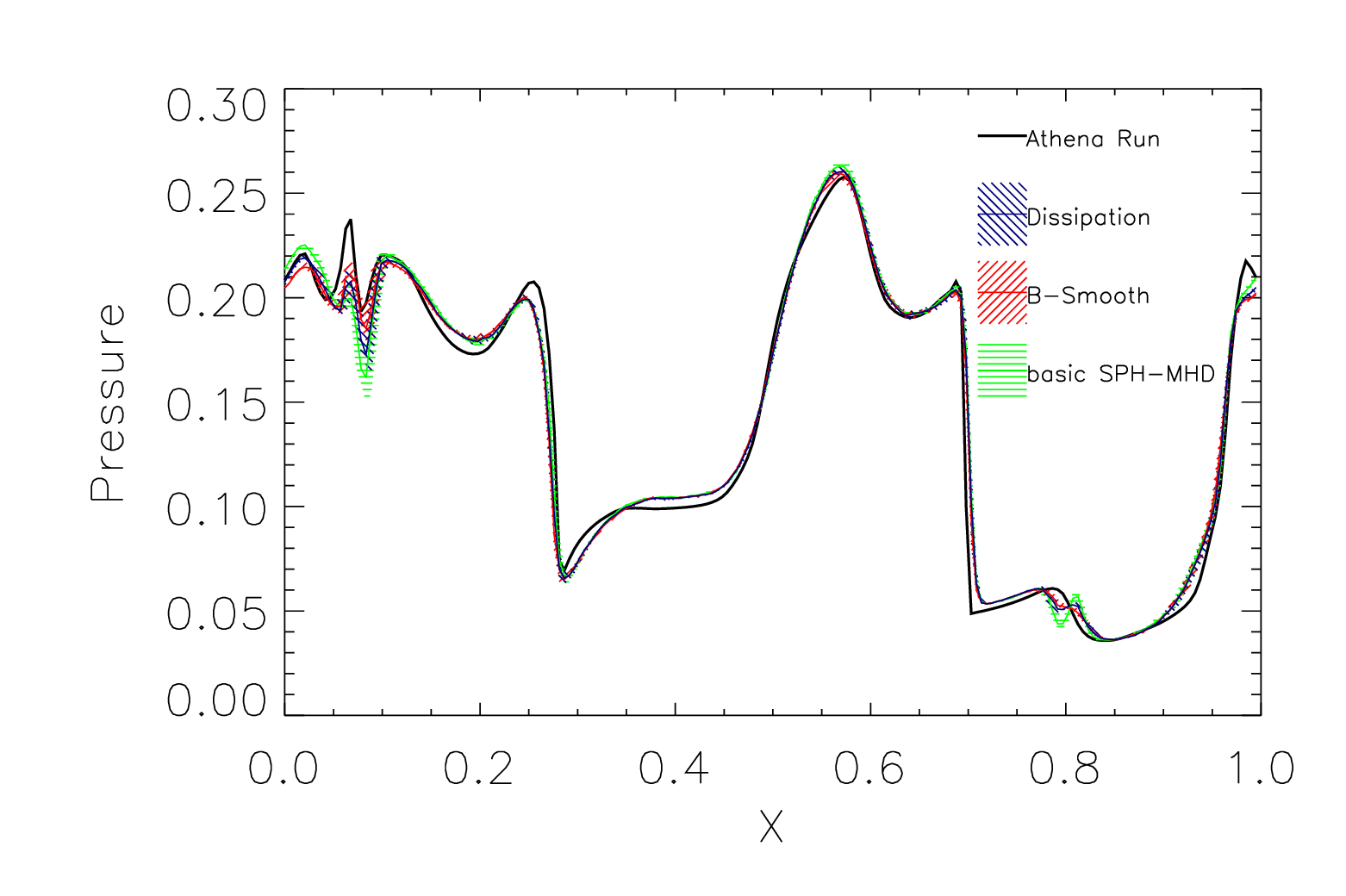}
    \includegraphics[width=0.45\textwidth]{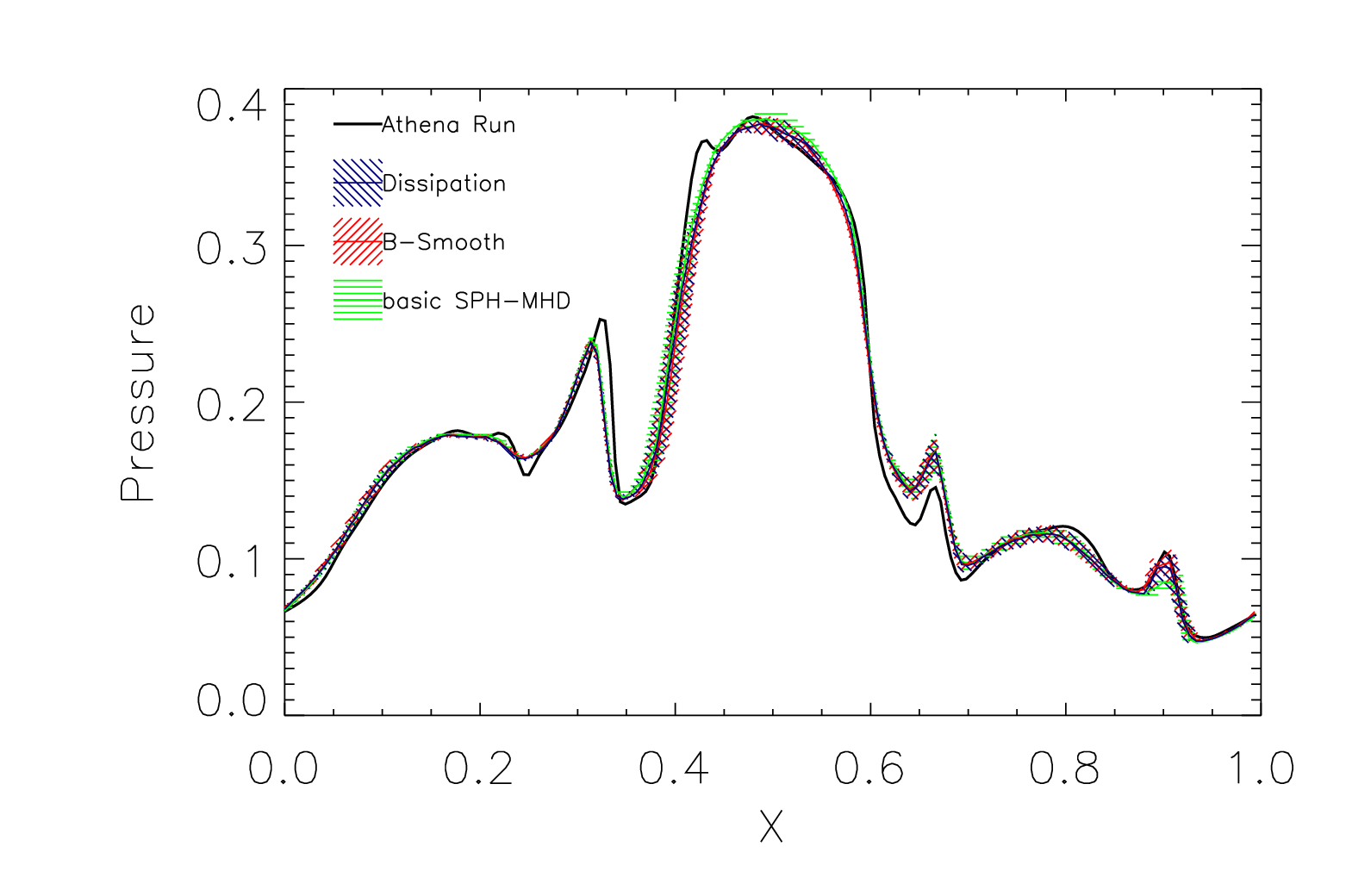}
    \end{center} \caption{Two $t=0.5$ cuts throught the pressure
    in the {\it Orszang-Tang Vortex} at $y=0.3125$ and at
    $y=0.4277$ (left and right panels, respectively).
    As before, the black line reflect the results obtained with
    ATHENA and the pink line with the red error bars is obtained with
    the {\it basic SPH-MHD} implementation in
    GADGET. The cuts are choosen for comparison with results from
    the literature, e.g. \citet{2006ApJ...652.1306B}.
}  \label{fig:VortexCut}
\end{figure*}

This planar test problem, introduced by \citet{Orzang}, is well known
to study the interaction between several classes of shock waves (at
different velocities) and the transition to MHD turbulence. Also, this
test is commonly used to validate MHD implementations \citep[for
example
see][]{1994JCoPh.115..485D,1991PhFlB...3...29P,2000ApJ...530..508L,2005MNRAS.364..384P,2006ApJ...652.1306B}.
The initial conditions for an ideal gas with $\gamma=5/3$ are
constructed within a unit-length domain (e.g. $x=[0,1],y=[0,1]$) with
periodic boundary conditions. The velocity field is defined by
$v_x=-\sin(2\pi y)$ and $v_y=\sin(2\pi x)$. The
initial magnetic field is set to be $B_x=B_0 v_x$
and $B_y=B_0 sin(4\pi x)$. The initial density is $\rho=\gamma P$ and
the pressure is set to $P=\gamma B_0^2$. This system is evolved until
$t=0.5$. Figure \ref{fig:Vortex} shows the final result for the
magnetic pressure for the ATHENA run and the 
three different SPH-MHD implementations in GADGET. Visually the
results are quite comparable, however the GADGET results look slightly
more smeared, which is the imprint of the underlying SPH and regularization methods. This
impression is confirmed in figure \ref{fig:VortexCut}, which shows two
cuts ($y=0.3125$ and $y=0.4277$) through the two simulations. Again,
the black line shows the ATHENA result, the colored lines with the 
error bars are showing the GADGET results. In general there is a reasonable
agreement, however the SPH-MHD results clearly show smoothing of 
features. The adaptive nature of the SPH-MHD implementation should allow
the central density peak to be resolved whereas in ATHENA is can only
be resolved by
increasing the number of grid cells. Never-the-less the SPH-MHD implementation
seems to converge slower when increasing the resolution (see Appendix).

\subsection{General performance of SPH-MHD}

In summary, as shown in the previous sections, an MHD implementation in SPH
is able to reliable reproduce the results of standard, one and two dimensional 
MHD test problems. We want to stress the point that all tests for the SPH-MHD
implementation where performed in a fully three dimensional setup to test the code 
under realistic circumstances. Regularization schemes in general are able to further 
suppress the numerical driven growth of $\mathrm{div}(\vec{B})$. Although some optimal 
numerical values for the regularization schemes can be inferred when comparing a suit of
different shock tube tests, such regularization schemes always introduce small
dissipative effects, which lead to a slight smearing of sharp features. This has to 
be kept in mind when applying the different SPH-MHD implementations to cosmological
applications.


\section{Cosmological Application} \label{sec:sim}

The cluster used in this work is part of a galaxy cluster sample
\citep{2008arXiv0808.3401D} extracted from a re-simulation
of a Lagrangian region selected from a cosmological, lower resolution
DM-only simulation \citep{2001MNRAS.328..669Y}. This parent simulation
has a box--size of $684$ Mpc, and assumed a flat $\Lambda$CDM cosmology
with $\Omega_m=0.3$ for the matter density parameter, $H_0=70$ for the
Hubble constant, $f_{bar}=0.13$ for the baryon
fraction and $\sigma_8=0.9$ for the normalization of the power
spectrum. The cluster has a final mass of $1.5\times10^{14} M_\odot$
and was re-simulated at 3 different particle masses for the high
resolution region. Using the ``Zoomed Initial Conditions'' (ZIC) technique
\citep{1993ApJ...412..455K,tormen97}, these regions were re-simulated with higher mass and
force resolution by populating their Lagrangian volumes with a larger
number of particles, while appropriately adding additional
high--frequency modes drawn from the same power spectrum. To optimize
the setup of the initial conditions, the high resolution region was
sampled with a $16^3$ grid, where only sub-cells are re-sampled at
high resolution to allow for quasi abritary shapes of the high
resolution region. The exact shape of each high--resolution region was iterated by
repeatedly running dark-matter only simulations, until the targeted
objects are free of any lower--resolution boundary particle out to
3-5 virial radii. The initial particle distributions, before adding
any Zeldovich displacement, were taken from a relaxed glass configuration
\citep{1996clss.conf..349W}. The three resolutions used correspond to
a mass of the dark matter particles of $1.6\times10^9
M_\odot$, $2.5\times10^8 M_\odot$ and $1.6\times10^8M_\odot$ for the
{\it 1x}, {\it 6x} and {\it 10x} simulation. The gravitational
softening corresponds to $7$, $3.9$ and $3.2$ kpc respectively.
For simplicity we assumed an initially homogeneous magnetic field of
$10^{-11}$ G co-moving as also used in previous work
\citep{1999A&A...348..351D,2002A&A...387..383D}. Furthermore we applied the
regularization by smoothing
the magnetic field in the same way as we did in previous work
\citep{2004JETPL..79..583D,dolag2005b} but also tested the effects of
regularization by artificial dissipation for varying values of $\alpha_B$.

\begin{figure*}
\begin{center}
  \includegraphics[width=0.99\textwidth]{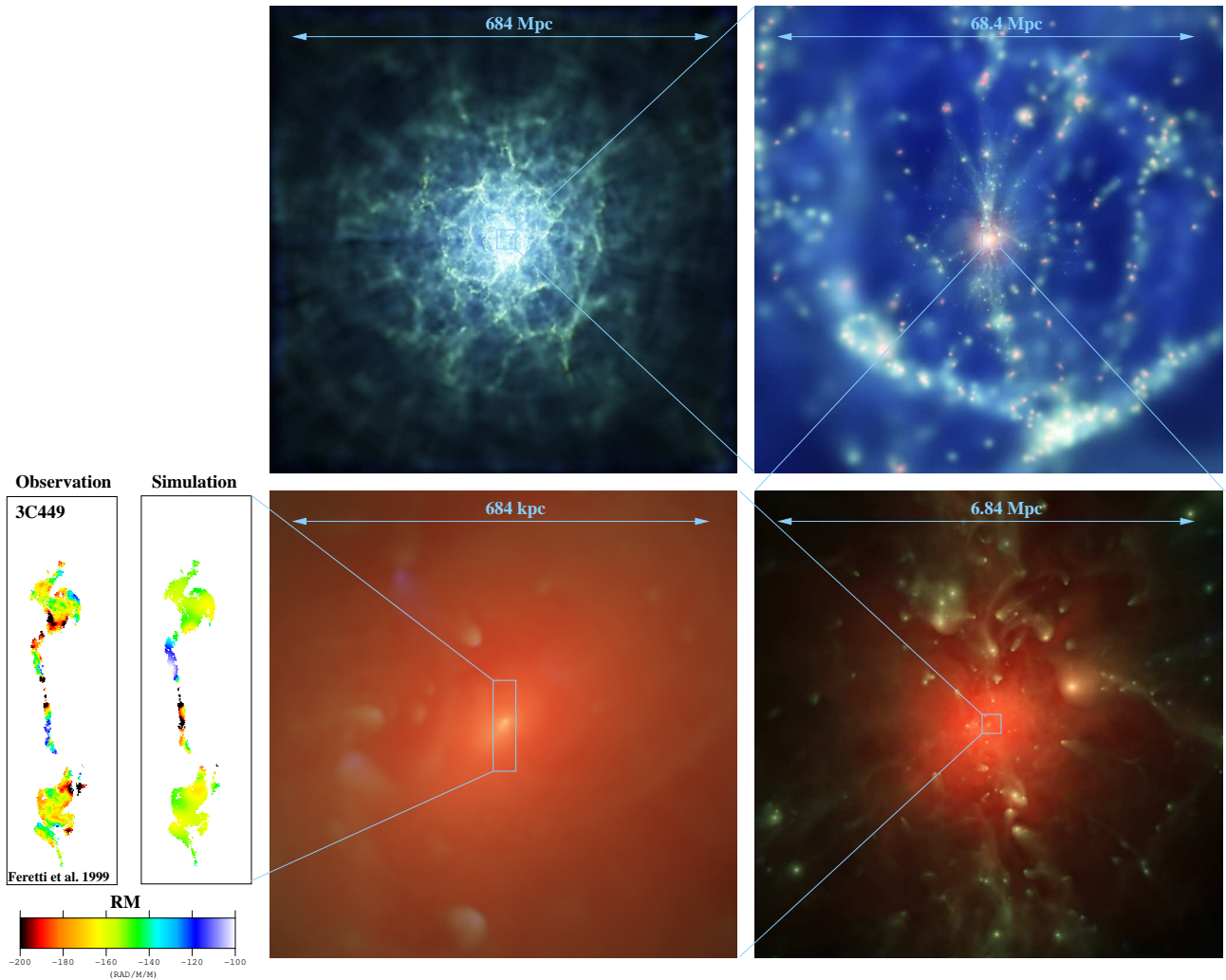}
\end{center}
\caption{Zoom into the cluster simulated within the
cosmological box. Clockwise, each panel displays a factor 10 increase
in imaging magnification, starting from the full box (684 Mpc) down to the cluster center
(680 kpc). On the very large scale, the density of the dark matter
particles are shown, whereas in the high resolution region the
temperature of the gas is rendered to emphasize the presence and
dynamics of the substructure. The last zoom extracts a region of the
same size of an observed radio jet (3C449) with measured rotation
measure \citep{1999A&A...341...29F}. Both, the simulated and the observed
map are displayed using a linear color-scale based on the minimum and
maximum values in the maps. The synthetic RM map is clipped to the
shape of the observations. Clearly, the simulations are still
lacking in resolution, however they do come quite close.}
\label{fig:Zoom}
\end{figure*}

Figure \ref{fig:Zoom} shows a zoom-in from the full cosmological box
down to the cluster. The structures in the outer parts get less
pronounced due to the decrease in resolution, which is designed to capture 
only the very largest scales of the simulation volume.
Each panel shows (in clockwise order) a zoom-in by a
factor of ten. Finally the elongated box in the lower left panel marks the
size of the observational frame shown on the left. For comparison we
produced a synthetic Faraday Rotation map from the simulation and
clipped it to the shape of the actual observations to give an
indication of the structures resolved by such simulations. 
We used the same linear colorscale for both the observed and 
the simulated RM map, using the highest resoluton simulation 
({\it 10x}). Note that we added a constant, galactic foreground signal 
to the simulated RM map to account for the non zero mean in the 
observed RM map. The dynamical range of the simulation
spans more than five orders of magnitudes in spatial dimension, and
the size of the underlying box is 6 and 5 times larger than
the AMR simulations presented in \citet{2008A&A...482L..13D} and
\citet{2005ApJ...631L..21B}, respectively. Still the resolution of the underlying
dark matter distribution is, respectively, 2 and 5 times better than these
AMR simulations and the cluster is resolved with more than one million
dark matter particles within the virial radius at the {\it 10x} resolution. To
perform the simulation, the {\it 10x} resolution run
needed $\approx 730$ CPUh on an AMD Opteron
cluster. This again is demonstrating the advantages of the underlying
SPH scheme in making large, cosmological zoomed simulations possible.

\begin{figure}
\begin{center}
  \includegraphics[width=0.49\textwidth]{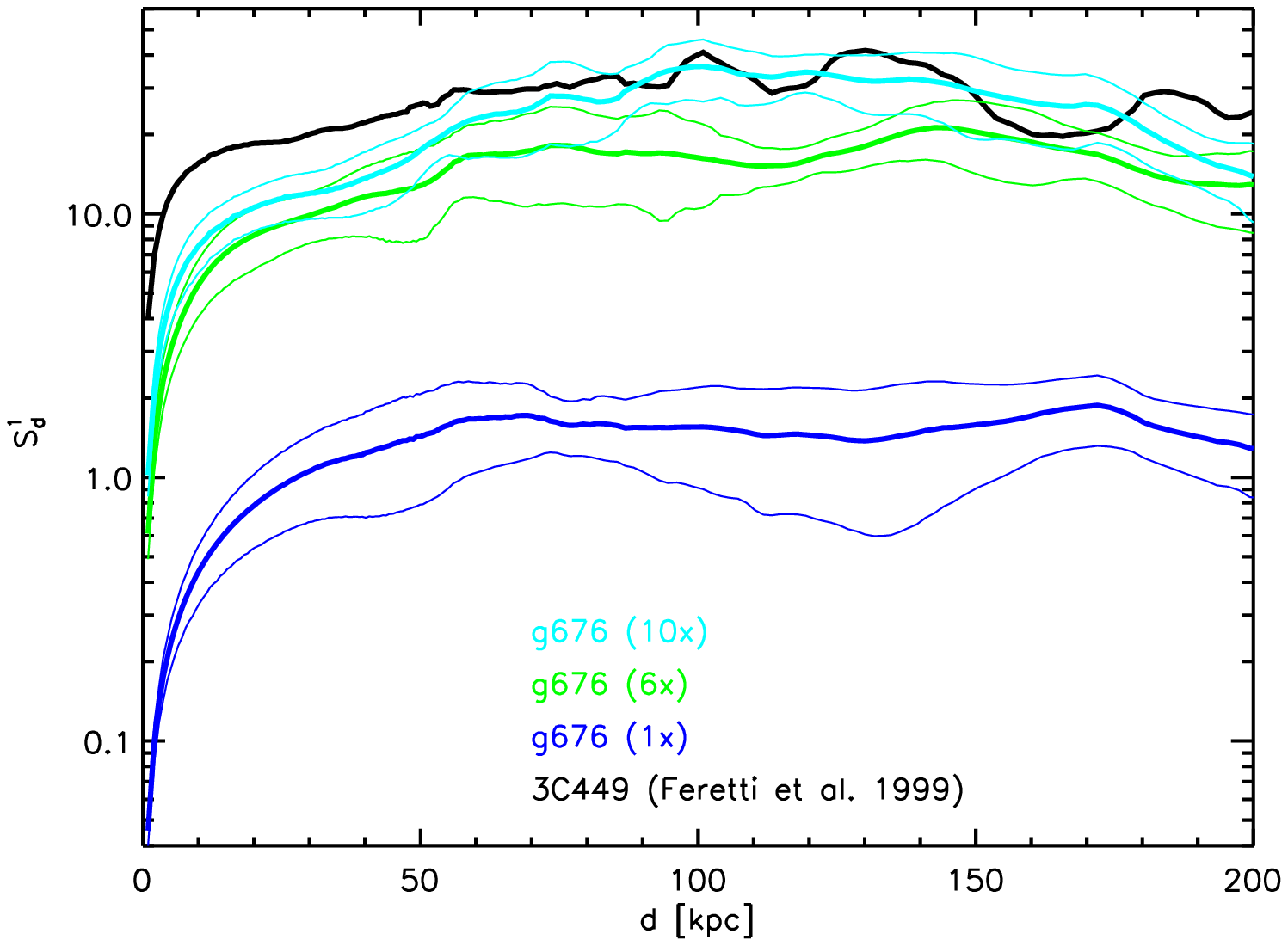}
\end{center}
\caption{Shown is the structure function averaged in radial bins 
(for details see text) calculated 
from the observed (black line) and from the synthetic Faraday Rotation maps (colored lines). 
The different colors correspond to different resolutions ({\it 1x}, {\it 6x} and {\it 10x}). 
The thick lines correspond to the mean calculated over 9 realizations of the maps, whereas 
the thin lines mark the RMS scatter between the different maps (see text for details).}
\label{fig:correl}
\end{figure}

To obtain a more quantitative comparison we calculated the 
projected structure function $S^{(1)}(d)$ from both the observed 
and synthetic Faraday Rotation maps. 
\begin{eqnarray}
S^{(1)}(\Delta x,\Delta y) &= \left<\left| RM(x,y) - RM(x+\Delta x, y+\Delta y)\right|\right>,
\end{eqnarray}
with $\Delta x$ and $\Delta y$ being the offsets from a pixel at position $(x,y)$.
The resulting matrix is then averaged in radial bins 
$d=\sqrt{\Delta x^2+\Delta y^2}$ to obtain
the structure function. Figure \ref{fig:correl} shows a comparison
of the obtained structure function from the observations (black line) and
the simulations. For each simulated cluster we calculated the
synthetic rotation measure maps, clipped accordingly to the shape of the
observed map. To obtain different realizations of the same simulation
we produced nine different maps where we shifted the clipped region by
$\pm20$ kpc in both spacial directions within the original, cluster
centered maps. The thick lines mark the mean structure function over
these maps, whereas the thin lines show the RMS scatter between the
different maps. It is clear that, due to the additional resolved 
turbulent velocity field, when increasing the resolution the same 
initial seed fields are amplified more. For our {\it 10x} simulation
the chosen magnetic seed field gets amplified to the observed level 
in our example cluster. Although increasing the resolution resolve smaller 
scales in the RM maps, the slope of the structure function at small separations, 
even the {\it 10x} resolution simulation, gets not as steep as the 
observed one. This confirms the visual impression from the lower left 
panel in Figure \ref{fig:Zoom} which also indicates more structure at 
small scales in the observed than the simulated RM map.
Pushing the mass resolution by one more order of magnitude would probably
result in a spacial resolution which should be sufficient to reproduce the
observed small scale structure in the RM maps, however in 
such a case we would expect to have to start from even smaller seed 
fields to avoid overestimating the amplitude of the Rotation Measure in the 
simulations. Such a study lies outside of the scope of this paper, 
as it ultimately would lead to questions of the role of magnetic 
dissipation and viscosity within the real intra cluster medium.

\begin{figure*}
\begin{center}
  \includegraphics[width=0.49\textwidth]{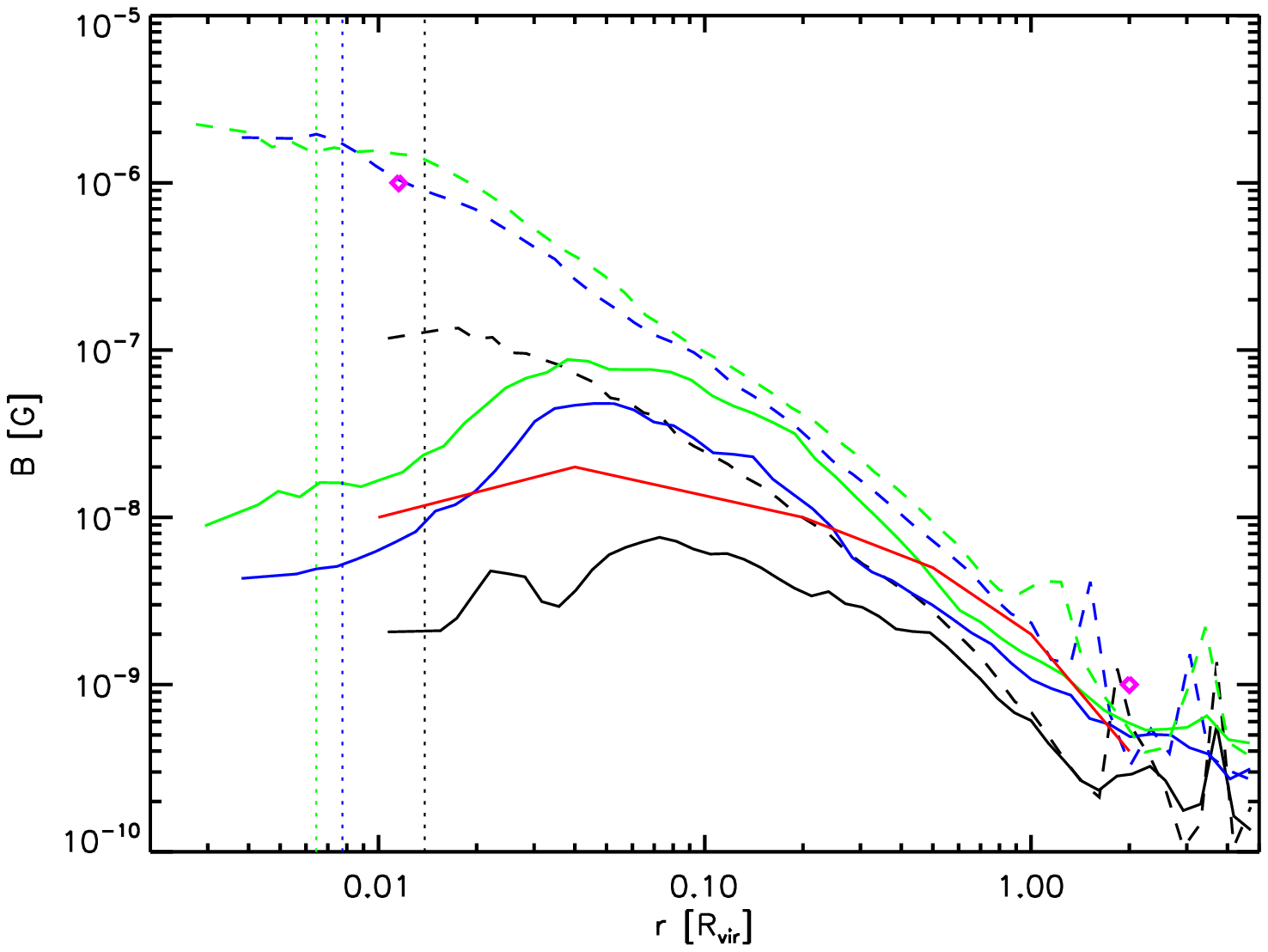}
  \includegraphics[width=0.49\textwidth]{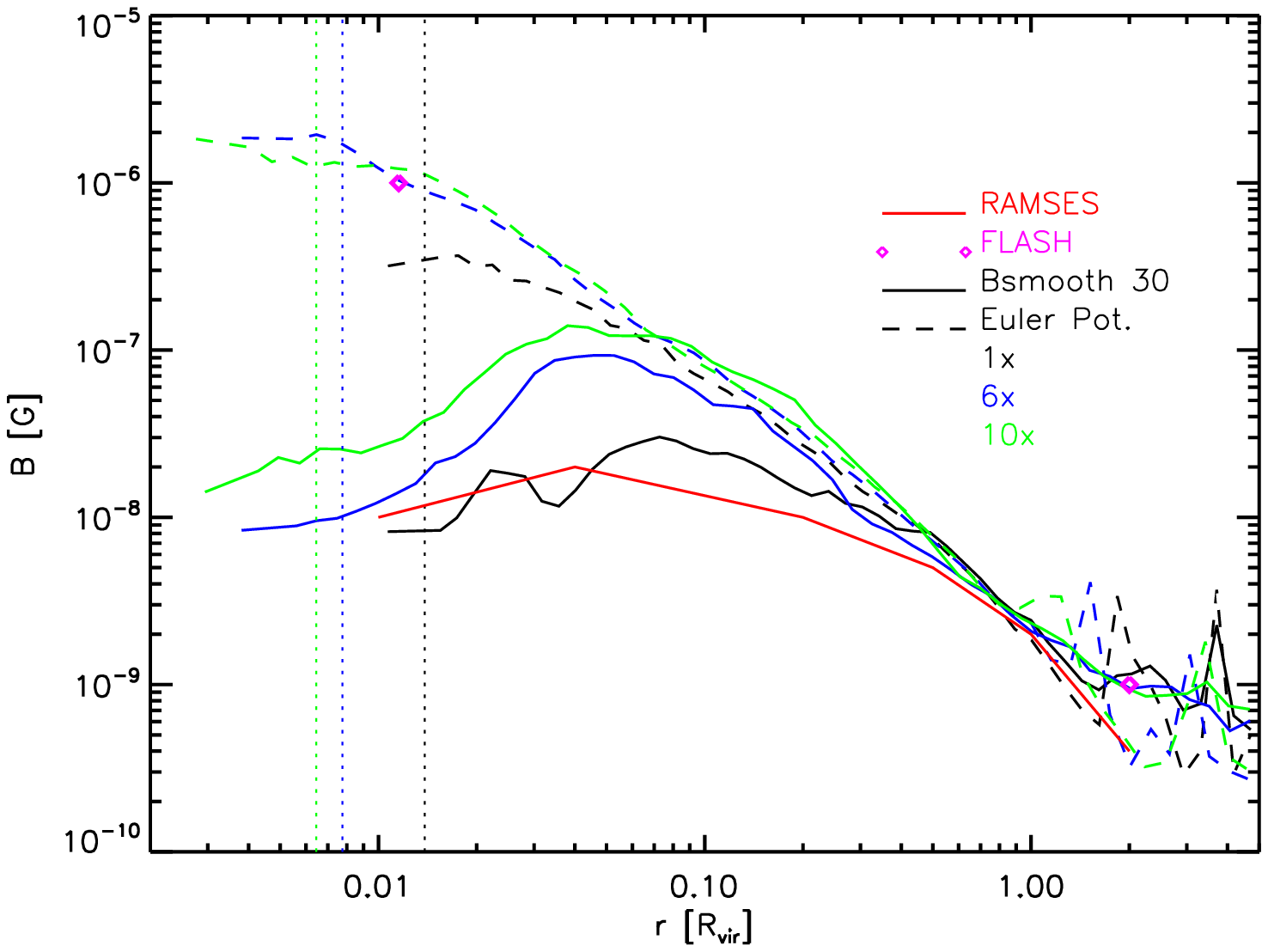}
\end{center}
\caption{The magnetic field profiles for a resolution study
of our cluster. Solid lines are obtained with the {\it Bsmooth SPH-MHD}
implementation, dashed lines are for using {\it Euler Potentials}. The
different line colors indicating different resolution. Right panel
shows the same but normalizing all the profiles in the outer part of 
the cluster. Results taken from \citet{2008A&A...482L..13D} 
(red line) and \citet{2005ApJ...631L..21B} (diamonds) are also shown.}
\label{fig:bprof}
\end{figure*}

In figure \ref{fig:bprof}, the radial magnetic field profiles are
shown for the three resolutions comparing the results obtained with
the normal configuration for cosmological simulations with results
where we just used the {\it Euler Potential} to follow the evolution
of the magnetic field ignoring back reactions. As already noted in
earlier work \citep{2002A&A...387..383D}, the left panel shows the 
dependence of the amplification
of magnetic fields with resolution. In addition, the solution
obtained with the {\it Euler Potential} agrees nicely with the 
{\it Bsmooth SPH-MHD} runs in
the outer part of the profiles. It is important to note that 
the increase in amplification of the magnetic field with increasing  
resolution when using the {\it Euler Potential} clearly demonstrates that 
this effect originates from resolving more velocity structure,
especially driven by the increased amount of substructure in the
underlying dark matter representation. Thus the result reflects the increased
complexity of the structures in the density and velocity fields, 
once the resolution of a cosmological simulation 
is increased. The {\it  Euler Potential} implementation
provides an unique possibility to study these effects as they reflect the
result of integrating the wind--up of the complex flows within galaxy
clusters, revealing information which can not easily be obtained in Eulerian
schemes. In the central parts, the {\it Bsmooth SPH-MHD}
simulation falls below the solution obtained using the {\it Euler
Potential}. This is easy to understand, because the {\it Euler
Potential} are free from any numerical magnetic dissipation. Additionally the
magnetic field is strongest in the cluster core and therefore,
including the magnetic force in the normal runs will lead to a
suppression of the amplification. In general the comparison of the two
methods demonstrate that the amplification of the magnetic field in
the {\it bsmooth SPH-MHD} implementation is not significantly influenced by
the non-zero $\mathrm{div}(\vec{B})$. In addition, although the absolute value
of the amplification is not converged with resolution, the shape of
the predicted magnetic field profile appears to be converged. This is
shown in the right panel of figure \ref{fig:bprof}, where the
profiles are normalized artificially at large radii to demonstrate the
self similar shapes. Note that
this convergence, as usual for all hydro-dynamical quantities, is
only reached at radii significant larger than the size of the
gravitational softening, indicated as dashed lines for the lowest
(e.g. {\it 1x}), medium (e.g. {\it 6x}) and highest (e.g. {\it 10x}) resolution runs.
In both panels of Figure \ref{fig:bprof} we also show the 
results from a cluster simulation using RAMSES, taken from 
\citet{2008A&A...482L..13D}, and FLASH, taken from 
\citet{2005ApJ...631L..21B}. This comparison is over-simplistic, 
as results are based on simulations of different objects and 
can only be compared with some care. Nevertheless,
the shape of the radial profiles obtained with RAMSES indicate 
slightly more dissipative effects compared to our {\it Bsmooth SPH-MHD} 
implementation, whereas the steeper profile obtained with FLASH 
resembles our results using the {\it Euler Potential} implementation.

\begin{figure*}
\begin{center}
  \includegraphics[width=0.49\textwidth]{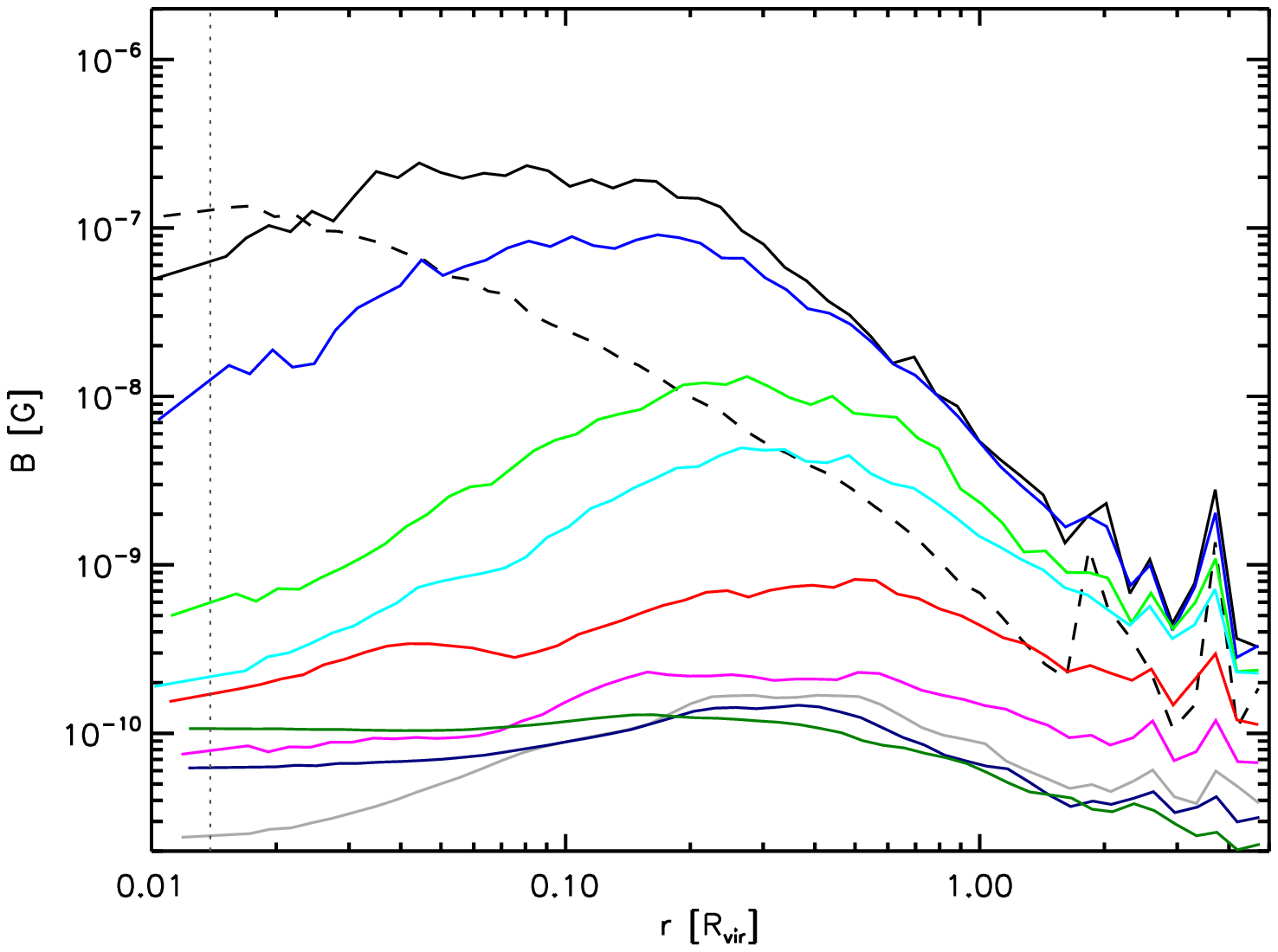}
  \includegraphics[width=0.49\textwidth]{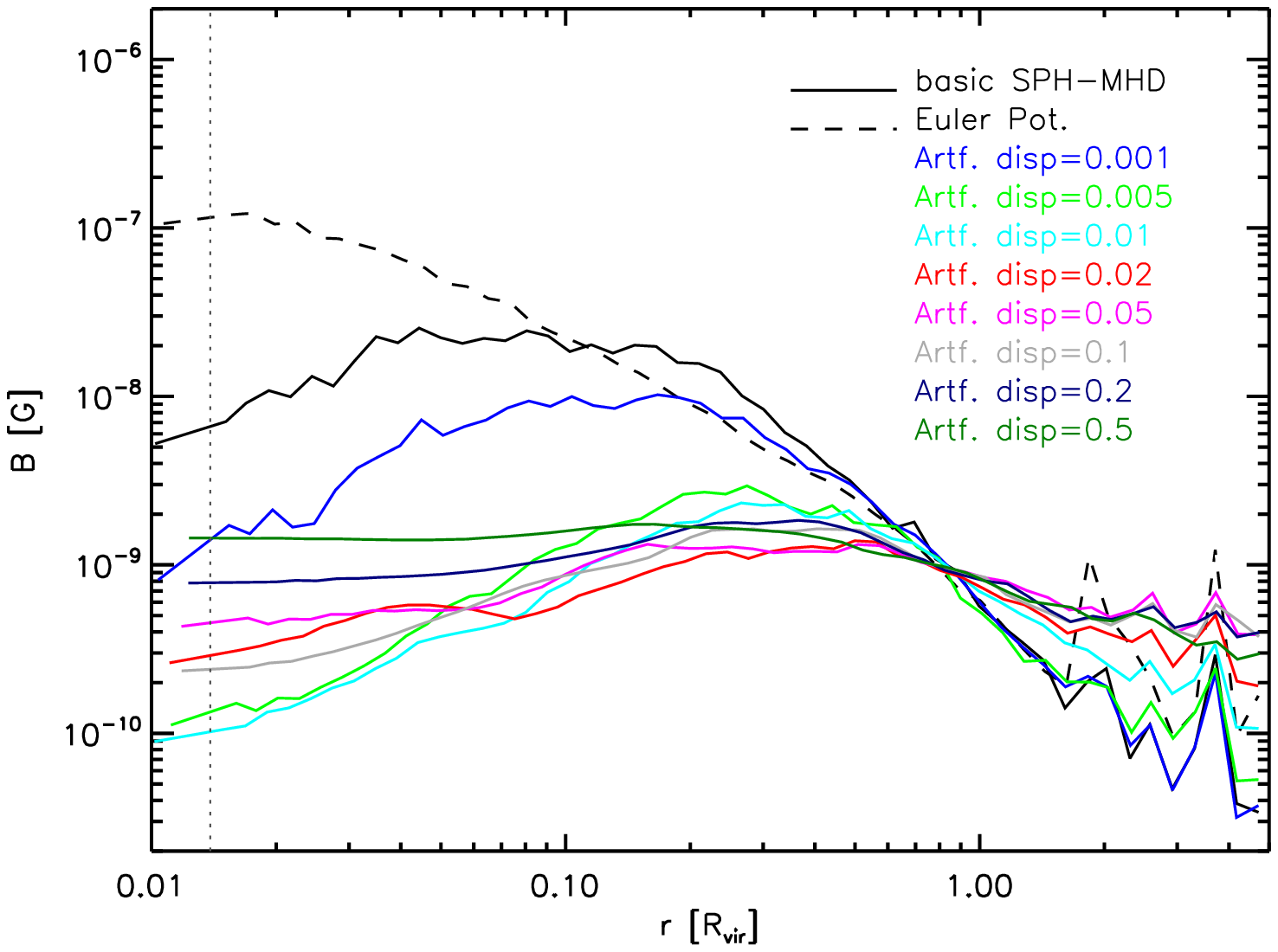}
\end{center}
\caption{The magnetic field profiles obtained for the galaxy
cluster using different values of the artificial dissipation. 
In addition the results using the {\it Euler Potentials} 
and the {\it Basic MHD} implementation are shown. Right panel
shows the same with profiles normalized in the outer part of the cluster.}
\label{fig:bprof_dis}
\end{figure*}

\begin{figure*}
\begin{center}
  \includegraphics[width=0.49\textwidth]{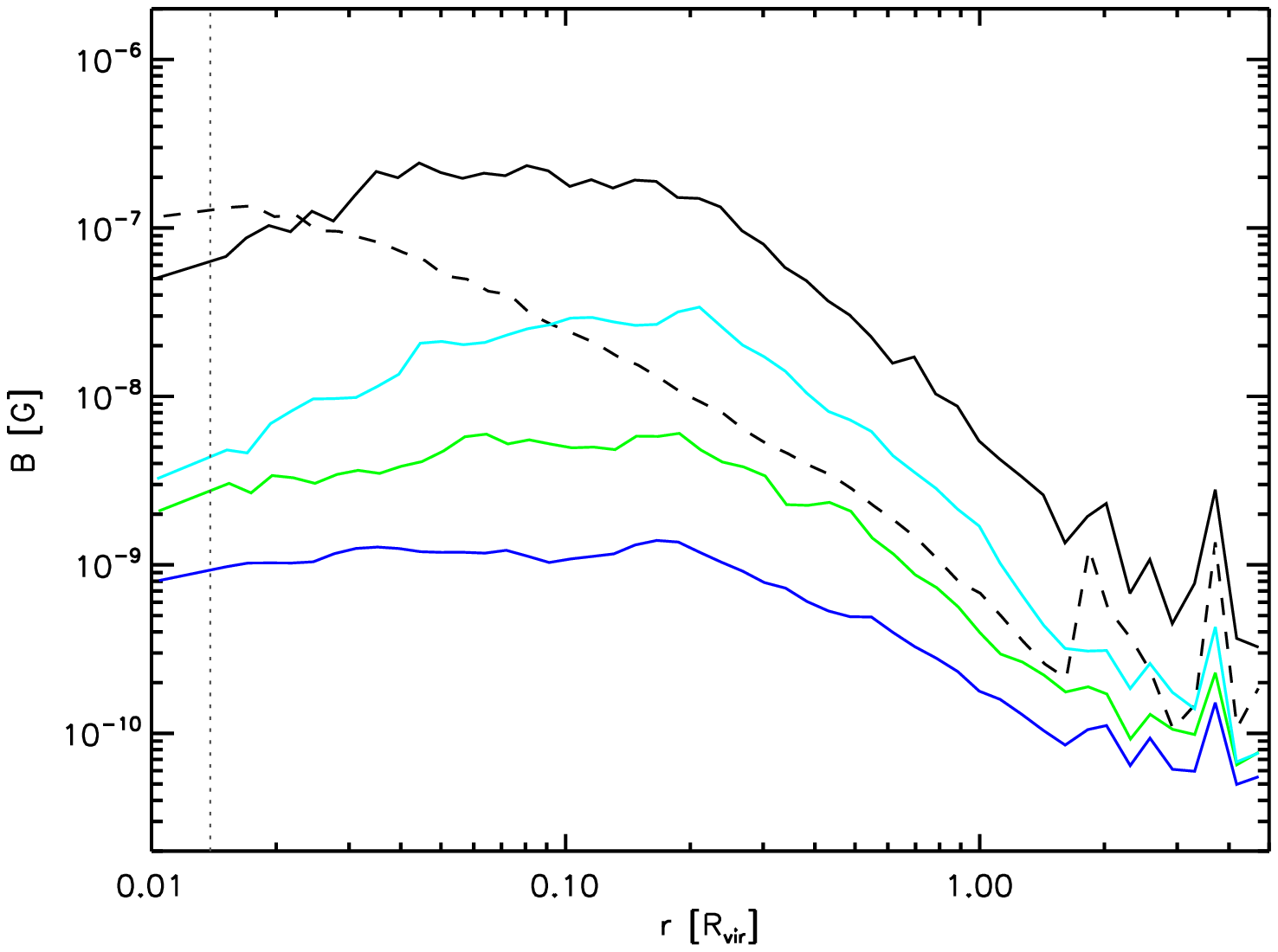}
  \includegraphics[width=0.49\textwidth]{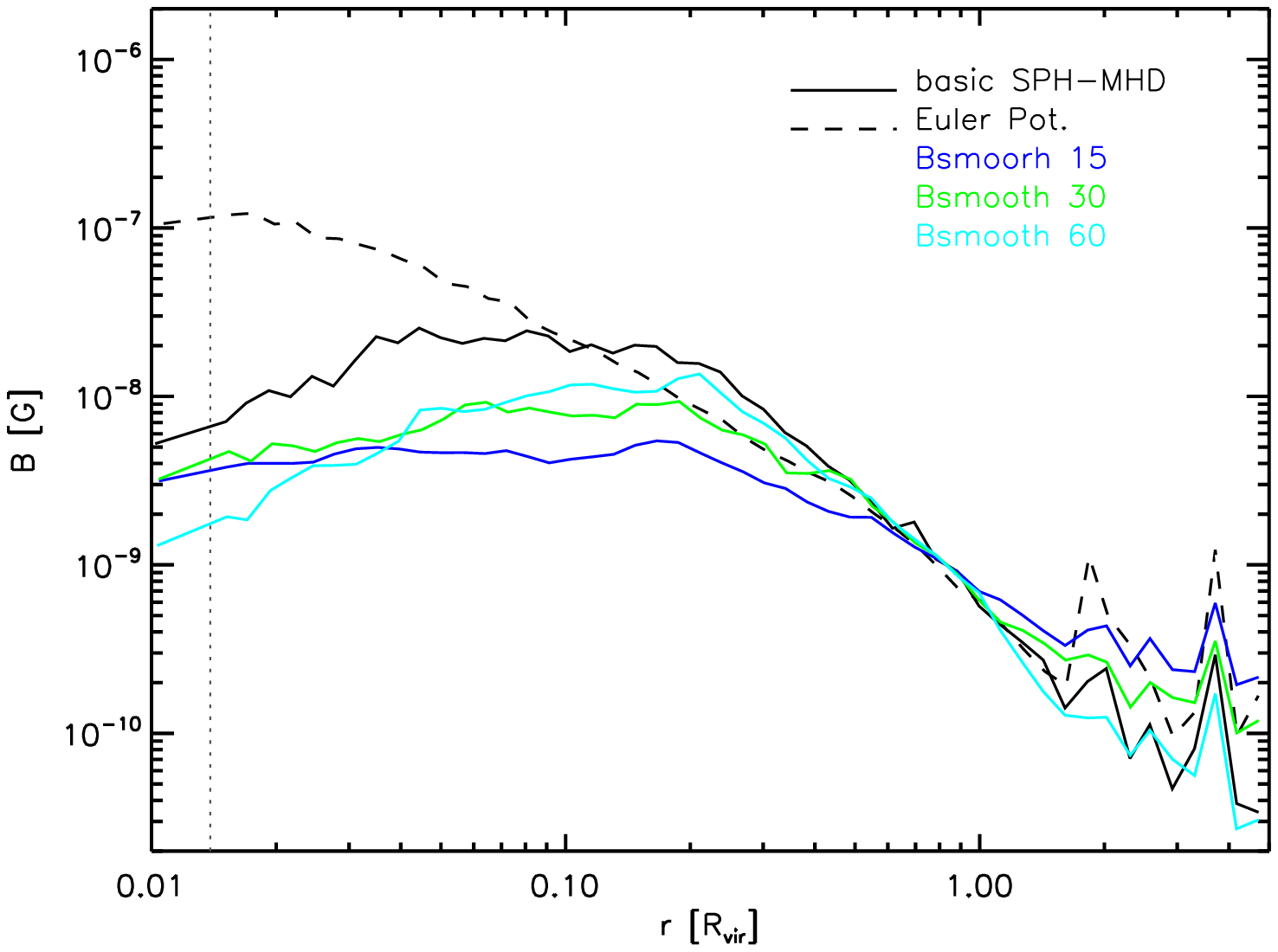}
\end{center}
\caption{The magnetic field profiles obtained for the galaxy
cluster using different time intervals for the smoothing procedure. 
In Addition the results for using the {\it Euler Potentials} 
and the {\it Basic MHD} implementation are shown. Right panel
shows the same with profiles normalized in the outer part of the cluster.}
\label{fig:bprof_bsm}
\end{figure*}

The situation changes when using artificial magnetic dissipation, as
shown in figure \ref{fig:bprof_dis}. The left panel shows the
magnetic field profiles for several values of $\alpha_B$ compared with
the profiles for the {\it basic SPH-MHD} run and that using {\it Euler
Potentials}. Clearly a normal value for artificial magnetic
dissipation leads to a large dissipation of magnetic field over the
simulation time (e.g. close to the Hubble time). The right panels show
the profiles artificially normalized at large radius. Clearly the self
similarity of the profiles is lost. Therefore it appears
that the use of artificially dissipation as a regularization
scheme is not a good choice for cosmological simulations. Additionally
it points out that true physical dissipation might play an important role
in determining the shape of the magnetic field profile in galaxy
clusters. Here, transport processes, cosmic rays, turbulence 
(especially at unresolved scales) and reconnection of 
magnetic field lines are not well understood, especially within the ICM. 
As the micro-physical origin of most of them are far outside the scales 
which can be ever reached by cosmological simulations, future work will 
have to include them as approximative, sub-grid models, possibly motivated by 
small scale numerical experiments.


\section{Conclusions} \label{sec:conc}

We presented the implementation of MHD in the cosmological, SPH code
GADGET. We performed various test problems and discussed several
instability correction and regularization schemes. We also
demonstrated the application to cosmological simulations, the role of
resolution and the role the regularization schemes play in
cosmological simulations.

Our main findings are:

\begin{itemize}

\item The combination of many improvements in the SPH implementation,
like the correction terms for the variable smoothing length
\citep{springel02} as well as the usage of the signal velocity in the
artificial viscosity \citep{1997JCoPh..136....298S} together with its
generalization to the MHD case \citep{2004MNRAS.348..123P} improve the
handling of magnetic fields in SPH significantly.

\item Correcting the instability by explicitly subtracting the
contribution of a numerical non-zero divergence of the magnetic field
to the Lorenz force from the Maxwell tensor as suggested by
\citet{2001ApJ...561...82B}  seems to perform well. Specifically in three
dimensional setups where it seems to work much better than other suggestions
in the literature.

\item The SPH-MHD implementation performs very well on simple shock
tube tests as well as on planar test problems. We performed all tests
in a fully three-dimensional setup and find excellent agreement of
the results obtained with the SPH-MHD implementation compared to the
results obtained with ATHENA in one or two dimensions.

\item With a convergence study we demonstrate that the SPH-MHD results
when increasing the resolution are converging to the true
solution, especially in the sharp features. However, in some regions it
seems that small but systematic differences converge only very slowly
to the correct solution.

\item Regularization schemes help to further suppresses
noise and $\mathrm{div}(\vec{B})$ errors in the test simulations,
however one has to carefully select the numerical parameters to
avoid too strong smoothing of sharp features. Performing a full set of
individual shock tube tests allows one to tune the numerical schemes and to
determine optimal values. However they reflect an optimal choice for
problems where the local timescales are mostly similar to the global 
timescale of the problem. For cosmological simulations it turns out that
regularization by artificial dissipation leads to questionable results, 
whereas the regularization by smoothing the magnetic field (which is applied 
on global timescales) produces reasonable results.

\item The SPH-MHD implementation allows us to perform challenging
cosmological simulations, covering a large dynamical range in
length-scales. For galaxy clusters, only the shape of the predicted magnetic
profiles is, (with the exception of the central part of clusters) converged
in resolution and in good agreement with previous studies. Also the
structures obtained in synthetic Faraday Rotation maps are in
good agreement with previous findings and compare well with observations.

\end{itemize}

The results obtained with artificial dissipation in
cosmological simulations indicate that physical dissipation could
play a crucial role in determining the exact shape of the predicted,
magnetic field profiles in galaxy clusters. Future work, especially when
including more physical processes at work in galaxy cluster -- as can be
done easily with our SPH-MHD implementation -- will reveal an interesting
interplay between dynamics of the cluster atmosphere and amplification
of magnetic fields. Thus having the potential to shed light on many,
currently unknown aspects of cluster magnetic fields, their structure and
their evolution.


\section*{acknowledgements}
KD acknowledges the financial support by the ``HPC-Europa
Transnational Access program'' and the hospitality of CINECA and ``Istituto di Radioastronomia'' (IRA) in bologna, where
part of the work was carried out. FAS acknowledges the support of the European Union's ALFA-II programme, through LENAC,
the Latin American European Network for Astrophysics and Cosmology.
We want to thank Stuart Sim for carefully reading and inmproving the manuscript.
We also want to thank James M. Stone for making ATHENA -- which we used for
obtaining our reference solutions -- publicly available as well as
Romain Teyssier for sending us the results for the Orszang-Tang Vortex
test obtained with RAMSES. We further thank Daniel Price for many
enlightening discussions and Volker Springel to always granting access to
the developer version of GADGET. We also want to thank Luigina Feretti
for providing the RM map of 3C449.

\bibliographystyle{mn2e}
\bibliography{master3}

\appendix

\section{Convergence}

\begin{figure}
\begin{center}
  \includegraphics[width=0.49\textwidth]{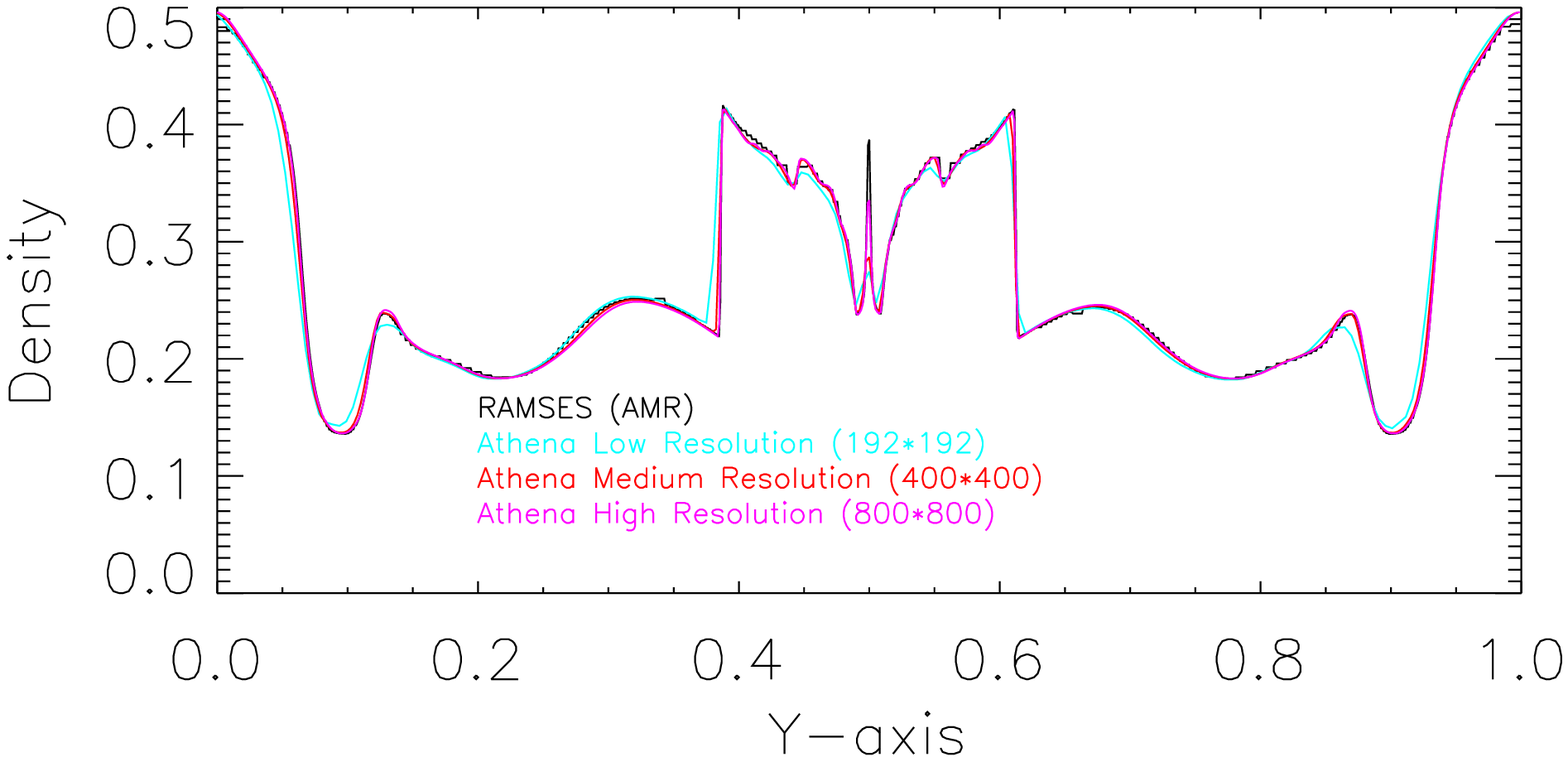}
\end{center}
\caption{A cut throgh the density for the Orszang-Tang Vortex
test (see Figure 13/14). Shown in black is the result obtained with
Ramses, compared to the results obtained with Athena using 3 different resolutions.}
\label{fig:convergence_ath}
\end{figure}

\begin{figure}
\begin{center}
  \includegraphics[width=0.49\textwidth]{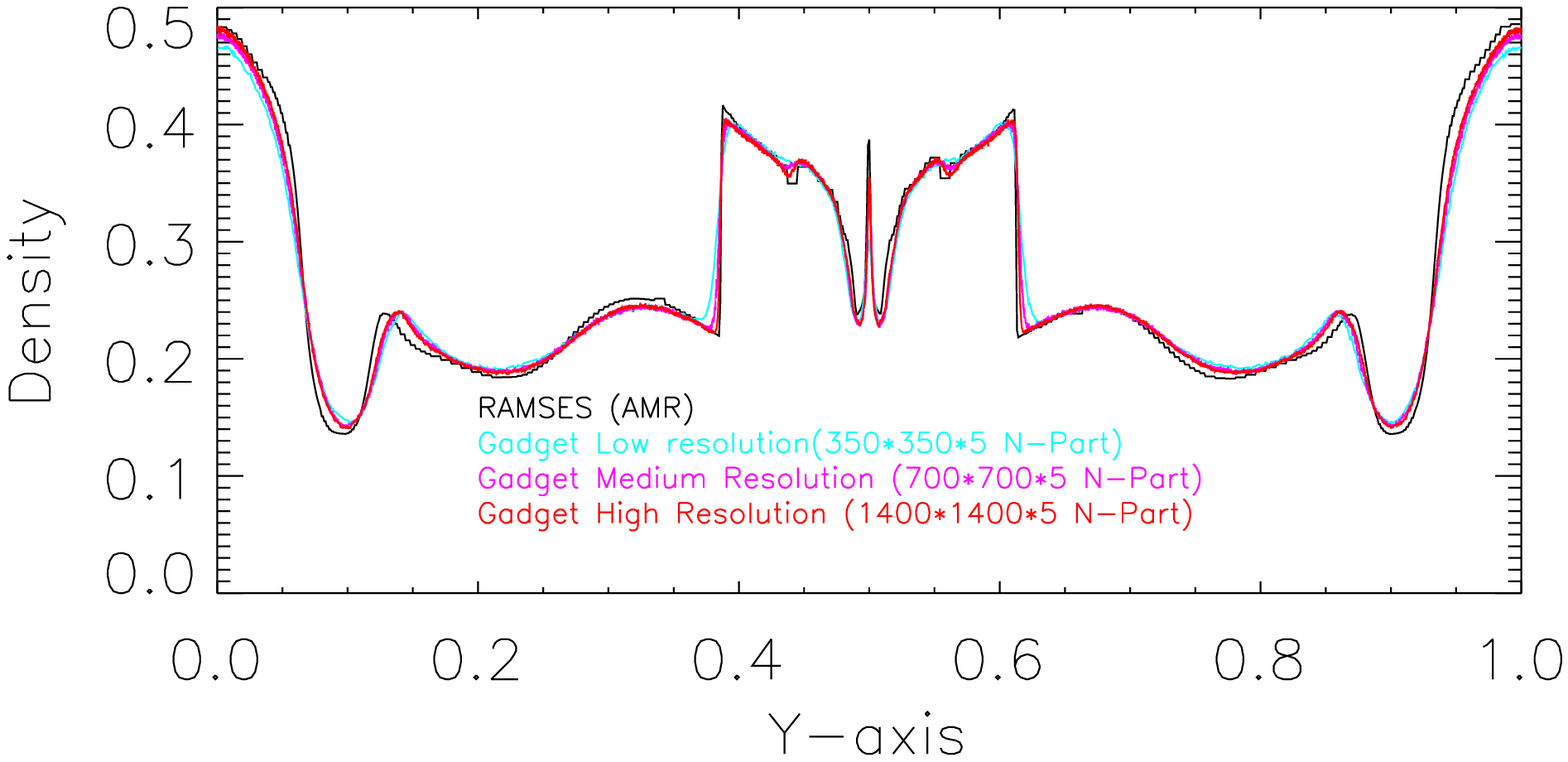}
\end{center}
\caption{Same than figure \ref{fig:convergence_ath}, but showing the
results obtained with the {\it basic SPH-MHD} implementation at two
different resolutions compared to the results obtained with Ramses.}
\label{fig:convergence_gad}
\end{figure}

Numerical experiments are normally restricted by the resolution one
can technically (in terms of computing/memory requirements) achieve.
Therefore tests as presented in section 3 are usually at
nominally better resolution than can be obtained in relevant (in this
case cosmological) simulations. Never-the-less an interesting question
is, how good do the numerical methods used converge if one further
increase the resolution? Figure \ref{fig:convergence_ath}
and \ref{fig:convergence_gad} show this for Athena and the {\it
basic SPH-MHD} implementation respectively. We repeated the
Orszang-Tang Vortex test problem with Athena on a $192^2$, $400^2$ and
$800^2$ grid. Figure \ref{fig:convergence_ath} shows a cut through
the density of the Orszang-Tang Vortex, comparing with the result
obtained with the AMR code Ramses \citep{2002A&A...385..337T}. Clearly, the results
obtained with Athena when increasing the resolution approaches the results obtained
with Ramses. Figure \ref{fig:convergence_gad} shows the same for
setups with $350^2\times5$, $700^2\times5$ and $1400^2\times5$
particles. The SPH-MHD implementation also converges towards the
Ramses results with increasing resolution. However, although the central
feature is better resolved in the SPH-MHD implementation than in the
Athena run with comparable resolution, some other features can be
seen to converge slower in the SPH-MHD implementation when
increasing the resolution. Specifically, in some very smoothed features 
there are small but systematic differences between the
SPH and the true solution. Here the SPH results seem to converge only
extremely slowly (if at all). 

\end{document}